\documentclass[fleqn,usenatbib]{mnras}

\usepackage{newtxtext,newtxmath}
\usepackage[T1]{fontenc}

\DeclareRobustCommand{\VAN}[3]{#2}
\let\VANthebibliography\thebibliography
\def\thebibliography{\DeclareRobustCommand{\VAN}[3]{##3}\VANthebibliography}

\usepackage{graphicx}
\usepackage{amsmath}
\usepackage{amsfonts}
\usepackage{amsbsy}
\usepackage{enumitem}
\usepackage{physics}
\usepackage{siunitx}
\usepackage{afterpage}
\usepackage{stmaryrd}
\usepackage{xcolor}
\usepackage{array}
\usepackage{hyphenat}

\DeclareMathOperator*{\argmax}{argmax}
\DeclareMathOperator*{\argmin}{argmin}

\title[CARPool]{CARPool: fast, accurate computation of large-scale structure statistics by pairing costly and cheap cosmological simulations}

\author[Chartier et al.]{
\parbox{\textwidth}{
\LARGE
Nicolas Chartier,$^{1, 2}$\thanks{E-mail: \href{mailto:nicolas.chartier@phys.ens.fr}{nicolas.chartier@phys.ens.fr}}
Benjamin Wandelt,$^{2,3}$
Yashar Akrami$^{1,4}$
and Francisco Villaescusa-Navarro$^{5,3}$
}
\vspace{0.4cm}
\\
$^{1}$Laboratoire de Physique de l'\'{E}cole Normale Sup\'{e}rieure, ENS, Universite PSL, CNRS, Sorbonne Universit\'{e}, Universit\'{e} de Paris, F-75005 Paris, France\\
$^{2}$Sorbonne Universit\'{e}, CNRS, UMR 7095, Institut d'Astrophysique de Paris, 98 bis bd Arago, 75014 Paris, France\\
$^{3}$Center for Computational Astrophysics, Flatiron Institute, 162 5th Avenue, New York, NY 10010, USA\\
$^{4}$Observatoire de Paris, Universit\'{e} PSL, Sorbonne Universit\'{e}, LERMA, 75014 Paris, France\\
$^{5}$Department of Astrophysical Sciences, Princeton University, Princeton, New Jersey, NJ 08544, USA
}

\date{Accepted 2021 February 9. Received 2021 February 9; in original form 2020 October 29}
\pubyear{2020}

\begin{document}
\label{firstpage}
\pagerange{\pageref{firstpage}--\pageref{lastpage}}
     
\maketitle
\begin{abstract}
    To exploit the power of next-generation large-scale structure surveys, ensembles of numerical simulations are necessary to give accurate theoretical predictions of the statistics of observables. High-fidelity simulations come at a towering computational cost. Therefore, approximate but fast simulations, \textit{surrogates}, are widely used to gain speed at the price of introducing model error. We propose a general method that exploits the correlation between simulations and surrogates to compute fast, reduced-variance statistics of large-scale structure observables \textit{without model error} at the cost of only a few simulations. We call this approach Convergence Acceleration by Regression and Pooling (CARPool). In numerical experiments with intentionally minimal tuning, we apply CARPool to a handful of \texttt{GADGET-III} $N$-body simulations paired with surrogates computed using COmoving Lagrangian Acceleration (COLA). We find $\sim 100$-fold variance reduction even in the non-linear regime, up to $k_\mathrm{max} \approx 1.2$ $h {\rm Mpc^{-1}}$ for the matter power spectrum. CARPool realises similar improvements for the matter bispectrum. In the nearly linear regime CARPool attains far larger sample variance reductions. By comparing to the 15,000 simulations from the \textit{Quijote} suite, we verify that the CARPool estimates are unbiased, as guaranteed by construction, even though the surrogate misses the simulation truth by up to $60\%$ at high $k$. Furthermore, even with a fully configuration-space statistic like the non-linear matter density probability density function, CARPool achieves unbiased variance reduction factors of up to $\sim 10$, without any further tuning. Conversely,  CARPool can be used to remove model error from ensembles of fast surrogates by combining them with a few high-accuracy simulations.
\end{abstract}


\section{Introduction}
\label{sec:intro}

The next generation of galaxy surveys will provide a detailed chart of cosmic structure and its growth on our cosmic light cone. These include the {\it Euclid} space telescope \citep{2011arXiv1110.3193L,2020A&A...642A.191E}, the Dark Energy Spectroscopic Instrument (DESI) \citep{2016arXiv161100036D,2016arXiv161100037D}, the Rubin Observatory Legacy Survey of Space and Time (LSST) \citep{2019ApJ...873..111I,2009arXiv0912.0201L,2018arXiv180901669T}, the Square Kilometre Array (SKA) \citep{2015MNRAS.450.2251Y,2020PASA...37....7S},
the Wide Field InfraRed Survey Telescope (WFIRST) \citep{2015arXiv150303757S}, the Subaru Hyper Suprime-Cam (HSC) and Prime Focus Spectrograph (PFS) surveys \citep{2018PASJ...70S...4A,2016SPIE.9908E..1MT} and the Spectro-Photometer for the History of the Universe, Epoch of Reionization, and Ices Explorer (SPHEREx) \citep{2014arXiv1412.4872D,2018arXiv180505489D}. These data sets will provide unprecedented statistical power to constrain the initial perturbations, the growth of cosmic structure, and the cosmic expansion history. To access this information requires accurate theoretical models of large-scale structure statistics, such as power spectra and bispectra. While analytical work, such as standard perturbation
theory \citep[SPT,][]{Jain:1993jh,1986ApJ...311....6G}, Lagrangian perturbation theory \citep[LPT,][]{Bouchet:1994xp,Matsubara:2007wj}, renormalised perturbation theory \citep{Crocce:2005xy} and effective field theory~\citep[EFT,][]{Carrasco:2012cv,Vlah:2015sea,2016arXiv161009321P}, has made great strides (see also \citet{Bernardeau:2001qr,Desjacques:2016bnm} for reviews), the reference models for large-scale structure are based on computationally intensive $N$-body simulations that compute the complex non-linear regime of structure growth. In recent years, the BACCO simulation project \citep{2020arXiv200406245A}, the Outer Rim Simulation \citep{2019ApJS..245...16H}, the Aemulus project I \citep{DeRose_2019}, the ABACUS Cosmos suite \citep{2018ApJS..236...43G}, the Dark Sky Simulations \citep{2014arXiv1407.2600S}, the MICE Grand Challenge \citep[MICE-GC,][]{10.1093/mnras/stv1708}, the Coyote Universe I \citep{2010ApJ...715..104H} and the Uchuu simulations \citep{2020arXiv200714720I}, among others, involved generation of expensive $N$-body simulations.

While analytical methods compute expectation values of large-scale structure statistics, a simulation generates a single realisation and its output therefore suffers from sample variance. Reducing this variance to a point where it is subdominant to the observational error therefore requires running ensembles of simulations. 

Computational cosmologists have been tackling the challenge of optimising $N$-body codes and gravity solvers for a growingly larger number of particles. Widely used codes include the parallel Tree Particle-Mesh (TreePM or TPM) codes \texttt{GADGET-II} by \citet{2005MNRAS.364.1105S} and \texttt{GreeM} by \citet{2009PASJ...61.1319I}, the adaptive treecode \texttt{2HOT} by \citet{2013arXiv1310.4502W}, the GPU-accelerated \texttt{ABACUS} code originated from \citet{phdAbacus}, the Hardware/Hybrid Accelerated Cosmology Code (\texttt{HACC}) developed by \citet{2016NewA...42...49H} and the distributed-memory and GPU-accelerated \texttt{PKDGRAV3}, based on Fast Multipole Methods and adaptive particle timesteps, from \citet{2017ComAC...4....2P}.
The memory and CPU time requirements of such computations are a bottleneck for future work on new-generation cosmological data sets. As an example, the 43,100 runs in the \textit{Quijote} simulations from \cite{villaescusanavarro2019quijote}, of which the data outputs are public and used in this paper, required $35$ million CPU-core-hours.

The search for solutions has led to alternative, fast and approximate ways to generate predictions for large-scale structure statistics. The COmoving Lagrangian Acceleration (COLA) solver of \citet{Tassev_2013} is a PM code that solves the particle equations of motion in an accelerated frame given by LPT. Particles are nearly at rest in this frame for much of the mildly non-linear regime. As a consequence, much larger timesteps can be taken, leading to significant time savings.
The $N$-body solver \texttt{F\textsubscript{AST}PM} of \citet{2016MNRAS.463.2273F} operates on a similar principle, using modified kick and drift factors to enforce the Zel'dovich approximation in the mildly non-linear regime. The  spatial COLA (sCOLA) scheme  \citep{2015arXiv150207751T} extends the idea of using LPT to guide the solution in the spatial domain. \citet{2020arXiv200304925L} have carefully examined and implemented these ideas to allow splitting large $N$-body simulations into many perfectly parallel, independently evolving small simulations.

In a different family of approaches, but still using LPT, \citet{2013MNRAS.433.2389M} proposed a parallelised implementation of the PINpointing Orbit Crossing-Collapsed HI-erarchical Objects (PINOCCHIO) algorithm from \citet{pino}.
\citet{2015MNRAS.446.2621C} developed a physically motivated enhancement of the Zel'dovich approximation called EZmocks.
Approximate methods and full $N$-body simulations can also be jointly used. For instance, \citet{2012JCAP...04..013T} proposed a statistical linear regression model of the non-linear matter density field using the density field given by perturbation theory, for which the random residual error is minimised.

Recently,  so-called {\it emulators} have been of great interest: they predict  statistics in the non-linear regime based on a generic mathematical model whose parameters are trained on simulation suites covering a range of cosmological parameters. An emulator is trained by \citet{2020arXiv200406245A} on the BACCO simulations; similarly, the Aemulus project contributions II, III and IV \citep{McClintock_2019,Zhai_2019, 2019arXiv190713167M} respectively construct an emulator for the halo mass function, the galaxy correlation function and the halo bias using the Aemulus I suite \citep{DeRose_2019}. Not only do emulators that map cosmological parameters to certain outputs  need large numbers of simulations for training,  they also do not guarantee unbiased results with respect to full simulation codes, especially outside the parameter range used during training.

Recent advances in deep learning have allowed training emulators that reproduce particle positions or density fields starting from initial conditions, therefore essentially emulating the full effect of a low-resolution cosmological $N$-body code---these include the Deep Density Displacement Model ($\mathrm{D}^3\mathrm{M}$) of \citet{He_2019} stemming from the U-NET architecture \citep{2015arXiv150504597R}. \citet{Kodi_Ramanah_2020} describe a complementary deep learning tool that increases the mass and spatial resolution of low-resolution $N$-body simulations using a variant of Generative Adversarial Networks \citep[GAN,][]{2014arXiv1406.2661G}.

None of these fast approximate solutions exactly reproduce the results of more computationally intensive codes. They trade computational accuracy for computational speed, especially in the non-linear regime.
In this vein, the recent series of papers \citet{2019MNRAS.482.1786L}, \citet{2019MNRAS.485.2806B} and \citet{2019MNRAS.482.4883C} compare the covariance matrices of clustering statistics given by several low-fidelity methods to those of full $N$-body codes and find statistical biases in the parameter uncertainties by up to 20\%.

A different approach to this problem is to reduce the stochasticity of the initial conditions, thereby  modifying the statistics of the observables in such a way as to reduce sample variance. This is the spirit of the method of fixed fields invented and first explored by \citet{2016PhRvD..93j3519P} and \citet{2016MNRAS.462L...1A}. They found in numerical experiments that a large variety of statistics retain the correct mean, and analytically showed that pairing and fixing, while changing the initial distributions, only impact a measure-zero set of correlations when the errors are not smothered by the large number of available modes. While this approach does not guarantee that any given statistic will be unbiased, the numerical study by \citet{Villaescusa_Navarro_2018} showed that ``fixing'' succeeds in reducing variance for several statistics of interest with no detectable bias when comparing to an ensemble of hundreds of full simulations and at no additional cost to regular simulations. Still, it is clear that other statistics must necessarily be biased, for example, the square of any variance-reduced statistic, such as four-point functions. Still in the family of variance reduction methods, \citet{2019MNRAS.486.1448S} built a composite model of the matter power spectrum and managed to cancel most of the cosmic variance on large scales, notably by using the ratio of matched phase initial conditions.
 
In this paper, we show that it is possible to get the best of both worlds: the speed of fast surrogates \textit{and} the guarantee of full-simulation accuracy.\footnote{As a jargon reminder, the accuracy and precision of an estimate refer, respectively, to the trueness of its expectation (in terms of the statistical bias) and the confidence in the expectation (standard errors, confidence intervals).} We take inspiration from \textit{control variates}, a classical variance reduction technique that directly and optimally minimises the variance of any random quantity (see \citet{Lavenberg1981APO} for a review, and \citet{GORODETSKY2020109257} and \citet{doi:10.1137/15M1046472} for related recent applications), to devise a way to combine fast but approximate simulations (or \textit{surrogates}) with computationally intensive accurate simulations to vastly accelerate convergence while \textit{guaranteeing} arbitrarily small bias with respect to the full simulation code. We call this Convergence Acceleration by Regression and Pooling (CARPool).\footnote{We will consider surrogates to be much faster than simulations, so that we only need to consider the number of simulations to evaluate computational cost.}

The paper is organised as follows. In Section~\ref{sec:2sec}, we explore the theory of univariate and multivariate estimation with control variates and highlight some differences in our setting for cosmological simulations. In Section~\ref{sec:3sec}, we briefly discuss both the $N$-body simulation suite and our choice of fast surrogates we use in the numerical experiments presented in Section~\ref{sec:results}. We conclude in Section~\ref{sec:conclusions}.

Table~\ref{table:notations} lists  mathematical notation and definitions used throughout this paper.

\begin{table}
    \centering
    \caption{Mathematical notation and definitions}
    \begin{tabular}{| m{12em} | m{15em}|}
    \hline
    Notation & Description \\
    \hline \hline
     $\mathcal{S}_{N} = \left\{ r_1, \dots, r_N  \right\}$ & Set of $N$ random seeds $r_n$ of probability space \\
     \hline
     $\boldsymbol{y}(r_n)\equiv\boldsymbol{y}_n$ & Random column vector of size $p$ at seed $r_n$\\
     \hline
     $\mathbb{E}\left[ \boldsymbol{y} \right]\equiv\boldsymbol{\mu_y}$ & Expectation value of random vector $\boldsymbol{y}$ \\
     \hline
     $\llbracket m,n \rrbracket$ & Set of integers from $m$ to $n$\\
     \hline
     $\boldsymbol{M}^{\boldsymbol{T}}$ & Transpose of real matrix $\boldsymbol{M}$ \\
     \hline
     $\boldsymbol{M}^{\boldsymbol{\dagger}}$ & Moore-Penrose pseudo-inverse of matrix $\boldsymbol{M}$ \\
     \hline
     $\det\left( \boldsymbol{M}\right)$ & Determinant of matrix $\boldsymbol{M}$ \\
     \hline
     $\mathbb{E} \left[\left( \boldsymbol{x} - \mathbb{E} \left[ \boldsymbol{x}\right]\right) \left(\boldsymbol{x} - \mathbb{E} \left[\boldsymbol{x}\right]\right)^{\boldsymbol{T}} \right]$ $\equiv \boldsymbol{\Sigma_{\boldsymbol{xx}}}$ & Variance-covariance matrix of random vector $\boldsymbol{x}$\\
     \hline
     $\mathbb{E} \left[ \left( \boldsymbol{y} - \mathbb{E} \left[ \boldsymbol{y} \right] \right) \left( \boldsymbol{x} - \mathbb{E} \left[ \boldsymbol{x}\right] \right) ^{\boldsymbol{T}} \right]$ $\equiv \boldsymbol{\Sigma_{\boldsymbol{yx}}}$ & Cross-covariance matrix of random vectors $\boldsymbol{y}$ and $\boldsymbol{x}$\\
     \hline
     $\sigma_{y}^2$ & Variance of scalar random variable $y$ \\
     \hline
     $\boldsymbol{0}_{p,q}$ and $\boldsymbol{0}_p$ & Null matrix in $\mathbb{R}^{p \times q}$ and null vector in $\mathbb{R}^{p}$\\
     \hline
     $\boldsymbol{I}_p$ & Square $p \times p$ identity matrix \\
     \hline
    \end{tabular}
    \label{table:notations}
\end{table}

\section{Methods}\label{sec:2sec}
Let us consider a set of observables  $y_i$ we would like to model (e.g., power spectrum or bispectrum bins) and collect them into a random vector $\boldsymbol{y}$ with values in $\mathbb{R}^p$. The standard estimate of the theoretical expectation of  $\boldsymbol y$, $\mathbb{E}\left[ \boldsymbol{y}\right] =\boldsymbol{\mu}$, from a set of \textit{independent and identically distributed} realisations $\boldsymbol{y}_n$, $n=1,\dots N$,  is the \textit{sample mean}
\begin{equation}
\boldsymbol{\bar{y}}=\frac{1}{N}\sum_{n=1}^{N} \boldsymbol{y}_n.
\end{equation}
Then the standard deviation $\sigma_{i}$ of each element $\bar{y}_i$  decreases as $\mathcal{O}(N^{-\frac{1}{2}})$, under mild regularity conditions (principally that $\sigma_{i}$ exists).

Our goal is to find a more precise---i.e. lower-variance---and unbiased estimator of $\mathbb{E}\left[ \boldsymbol{y}\right]$  with a much smaller number of  simulations  $\boldsymbol{y}_n$. The means by which we achieve this is to construct another set of quantities that are fast to compute such that 1) their means are small enough to be negligible, and 2) their errors are anti-correlated with the errors in the $\boldsymbol{y}_n$,\footnote{The intuition behind this principle is that for two random scalars $a$ and $b$, we have $\sigma_{a+b}^2 = \sigma_{a}^2 + \sigma_{b}^2+2\mathrm{cov}(a,b)$.} and add some multiple of these to $\boldsymbol{\bar{y}}$ to cancel some of the error in the $y_n$. This is the \textit{control variates} principle.

\subsection{Theoretical framework}
In what follows we will use the word \textit{simulation} to refer to costly high-fidelity runs and \textit{surrogate} for fast but low-fidelity runs. 

\subsubsection{Introduction with the scalar case}\label{sec:unicv}
Let us consider a scalar simulated observable $y$, such that $\mathbb{E} \left[ y \right] = \mu $, and a surrogate $c$ of $y$ with $\mathbb{E} \left[ c \right] = \mu_c$. Note that $\mu_c\ne\mu$ in general. For any $\beta \in \mathbb{R}$, the quantity
\begin{equation}\label{eq:scalarCV}
    x(\beta) = y - \beta \left( c - \mu_c \right)
\end{equation}
is an unbiased estimator of $\mu$ by construction. The optimal value for $\beta$ is determined by minimising the variance of the new estimator,
\begin{equation}\label{eq:varScal}
    \sigma_{x(\beta)}^2 = \beta^2 \sigma_c^2 - 2\beta \mathrm{cov}(y, c) + \sigma_y^2\,.
\end{equation}
The function \eqref{eq:varScal} of $\beta$ has a strict global minimum point at
\begin{equation}\label{eq:betaStar}
    \beta^{\star} =\argmin_{\beta \in \mathbb{R}} \sigma_{x(\beta)}^2= \frac{\mathrm{cov}(y,c)}{\sigma_c^2}\,.
\end{equation}
Plugging equation \eqref{eq:betaStar} into equation \eqref{eq:varScal} allows us to express the variance reduction ratio of control variates as
\begin{equation}
    \frac{\sigma_{x(\beta)}^2}{\sigma_y^2} = 1 - \rho_{y,c}^2\,,
\end{equation}
with $\rho_{y,c}$ the Pearson correlation coefficient between $y$ and $c$.
The latter result shows that no matter how biased the surrogate $c$ might be, the more correlated it is with the simulation $y$, the better the variance reduction. For the classical control variates method, the choice of $c$ is restricted to cases where $\mu_c$ and $\beta$ are known \textit{a priori}.
In section \ref{sec:InPractice} below, we will consider the more general case, typically encountered in practice,  where $\beta$ is not known  and we must estimate it from data.

\subsubsection{Multivariate control variates} \label{sec:mvcv}
Let $\boldsymbol{y}$ be an unbiased and costly simulation statistic of expectation ${\boldsymbol{\mu} \in \mathbb{R}^p}$, and  $\boldsymbol{c}$ an approximate realisation with $\mathbb{E} \left[ \boldsymbol{c} \right] = \boldsymbol{\mu_c} \in \mathbb{R}^q$. \\
Similarly to the scalar case, for any $\boldsymbol{\beta} \in \mathbb{R}^{p \times q}$ the control variates estimator is
\begin{equation}\label{eq:mvCV}
    \boldsymbol{x}(\boldsymbol{\beta}) = \boldsymbol{y} - \boldsymbol{\beta} \left( \boldsymbol{c} - \boldsymbol{\mu_c} \right)\,.
\end{equation}
$\boldsymbol{\Sigma_{xx}}$, the covariance matrix of the random vector $\boldsymbol{x(\beta)}$, is expressed as a function of $\boldsymbol{\beta}$,
\begin{equation}\label{eq:varMV}
    \boldsymbol{\Sigma_{xx}}(\boldsymbol{\beta}) = \boldsymbol{\beta}\boldsymbol{\Sigma_{cc}} \boldsymbol{\beta^T} - \boldsymbol{\beta}\boldsymbol{\Sigma_{yc}^T} - \boldsymbol{\Sigma_{yc}} \boldsymbol{\beta^T} + \boldsymbol{\Sigma_{yy}}\,.
\end{equation}

Optimising variance reduction here means minimising the confidence region associated to $\mathbb{E}\left[ \boldsymbol{x}(\boldsymbol{\beta})\right]$ and represented by the generalised variance $\det\left(\boldsymbol{\Sigma_{xx}}(\boldsymbol{\beta})\right)$. Appendix \ref{app:bayesDer} presents a Bayesian solution to the Gaussian version of this optimisation problem.

Here we present an outline of the derivation  in \citet{DEOPORTANOVA199380} and \citet{VENKATRAMAN198637}. The course by \citet{LectureCCA} provides an overview of canonical correlation analysis, which is used in the derivation.
The oriented volume of the $p$-dimensional parallelepiped spanned by the columns of $\boldsymbol{\Sigma_{xx}(\beta)}$ is minimised as the analogue of an error bar in the univariate case. \citet{10.1287/opre.33.3.661} proved that
\begin{equation}\label{eq:betaStarMV}
    \boldsymbol{\beta^{\star}} = \argmin_{\boldsymbol{\beta} \in \mathbb{R}^{p \times q}} \det \left(\boldsymbol{\Sigma_{xx}}(\boldsymbol{\beta})\right) = \boldsymbol{\Sigma_{yc}}\boldsymbol{\Sigma_{cc}^{-1}}\,.
\end{equation}
Combining equations \eqref{eq:betaStarMV} and \eqref{eq:varMV} gives the generalised variance reduction
\begin{equation} \label{eq:mvVarReduc}
\begin{aligned}
    \frac{\det \left(\boldsymbol{\Sigma_{xx}}(\boldsymbol{\beta^{\star}})\right)}{\det \left( \boldsymbol{\Sigma_{yy}} \right)} &= \frac{\det \left( \boldsymbol{\Sigma_{yy}} \left( \boldsymbol{I}_{p} - \boldsymbol{\Sigma_{yy}^{-1}} \boldsymbol{\Sigma_{yc}}\boldsymbol{\Sigma_{cc}^{-1}}\boldsymbol{\Sigma_{yc}^T} \right) \right)}{\det \left( \boldsymbol{\Sigma_{yy}} \right)} \\
    &= \prod_{n=1}^{s = \mathrm{rank}\left( \boldsymbol{\Sigma_{yc}} \right)} \left( 1 - \lambda_n^2 \right)\,,
    \end{aligned}
\end{equation}
where the scalars $\lambda_1^2 \geq \lambda_2^2 \geq \dots  \geq \lambda_s^2 \geq 0 $ are the eigenvalues of $ \boldsymbol{\Sigma_{yy}^{-1}} \boldsymbol{\Sigma_{yc}}\boldsymbol{\Sigma_{cc}^{-1}}\boldsymbol{\Sigma_{yc}^T}$ and whose square roots are the canonical correlations between $\boldsymbol{y}$ and $\boldsymbol{c}$.
More precisely, $\lambda_1$ is the maximum obtainable cross-correlation between any linear combinations $\boldsymbol{u_1^T}\boldsymbol{y}$ and $\boldsymbol{v_1^T}\boldsymbol{c}$,
\begin{equation}\label{eq:cca}
\displaystyle
    \lambda_1 = \argmax_{\boldsymbol{u_1} \in \mathbb{R}^p, \boldsymbol{v_1} \in \mathbb{R}^q} \frac{\boldsymbol{u_1^T}\boldsymbol{\Sigma_{yc}}\boldsymbol{v_1}}{\sqrt{\boldsymbol{u_1^T}\boldsymbol{\Sigma_{yy}}\boldsymbol{u_1}} \sqrt{\boldsymbol{v_1^T}\boldsymbol{\Sigma_{cc}}\boldsymbol{v_1}}}\,,
\end{equation}
and $\left\{\lambda_n ; n \leq s \right\}$ are found recursively with the constraint of uncorrelatedness between $\left\{ \boldsymbol{u_n^T y}, \boldsymbol{v_n^T c} \right\}$ and $\left\{ \boldsymbol{u_1^T y}, \boldsymbol{v_1^T c}, \dots, \boldsymbol{u_{n-1}^T y}, \boldsymbol{v_{n-1}^T c}  \right\}$.
At the end, we have two bases for the transformed vectors $\boldsymbol{u}=\left[ \boldsymbol{u_1^Ty}, \dots, \boldsymbol{u_s^Ty}\right]^{\boldsymbol{T}}$ and $\boldsymbol{v}=\left[ \boldsymbol{v_1^Tc}, \dots, \boldsymbol{v_s^Tc}\right]^{\boldsymbol{T}}$ in which their cross-covariance matrix is diagonal, i.e. $\boldsymbol{\Sigma_{uv}}=\mathrm{diag}\left( \lambda_1,\dots, \lambda_s \right)$.

\subsection{Estimation in practice}\label{sec:InPractice}
In this section, we examine practical implications of the control variates implementation when the optimal control matrix $\boldsymbol{\beta}$ (or coefficients) and the mean of the cheap estimator $\boldsymbol{\mu_c}$ are unknown. We will consider an online approach in order to  improve the estimates of \eqref{eq:betaStar} or \eqref{eq:betaStarMV} as simulations and surrogates are computed. Estimating $\boldsymbol{\mu_c}$ is done through an inexpensive pre-computation step that consists in running  fast surrogates.
From now on, to differentiate our use  of the control variates principle and its application to cosmological simulations from the theory presented above, we will refer to it as the CARPool technique.

For the purposes of this paper, we will take as our goal to produce low-variance estimates of expectation values of full simulation observables. When we discuss model error, it is therefore only relative to the full simulation. From an absolute point of view the accuracy of the full simulation depends on a number of factors such as particle number, force resolution, timestepping, inclusion of physical effects, \textit{et cetera.} The numerical examples of full simulations we give are not selected for their unmatched accuracy, but for the availability of a large ensemble that we can use to validate the CARPool results.

\subsubsection{Estimation of $\boldsymbol{\mu_c}$}
In the textbook control variates setting, the crude approximation $\boldsymbol{\mu_c}$ of $\boldsymbol{\mu}$ is assumed to be known. There is no reason for this to be the case in the context of cosmological simulations, thus we compute $\bar{\boldsymbol{\mu}}_{\boldsymbol{c}}$ with surrogate samples drawn on a separate set of seeds $\mathcal{S}_{M} =  \left\{ r_{1}, \dots, r_{M} \right\}$ ($\mathcal{S}_{N} \cap \mathcal{S}_{M} = \emptyset$, where $\mathcal{S}_{N}$ is the set of initial conditions of simulations). What is then the additional variance-covariance of the control variates estimate stemming from the estimation of $\boldsymbol{\mu_c}$?

First, write each cheap-estimator realisation as $\boldsymbol{c}=\boldsymbol{\mu_c} + \boldsymbol{\delta}$, with $\mathbb{E}\left[ \boldsymbol{\delta} \right] = \boldsymbol{0}_q$,
\begin{equation}\label{eq:errMuc}
\begin{aligned}
   \bar{\boldsymbol{\mu}}_{\boldsymbol{c}} &= \boldsymbol{\mu_c} + \frac{1}{M}\sum_{i=1}^M \boldsymbol{\delta}_i\,, \\
   \boldsymbol{\Sigma}_{\boldsymbol{\bar{\boldsymbol{\mu}}_{\boldsymbol{c}}}\boldsymbol{\bar{\boldsymbol{\mu}}_{\boldsymbol{c}}}} &= \boldsymbol{\Sigma}_{\boldsymbol{\bar{\delta}}\boldsymbol{\bar{\delta}}} = \frac{1}{M}\boldsymbol{\Sigma_{cc}}\,.
\end{aligned}
\end{equation}
Replacing $\boldsymbol{\mu_c}$ by $\bar{\boldsymbol{\mu}}_{\boldsymbol{c}}$ in equation~\eqref{eq:mvCV} and computing the covariance results in
\begin{equation}
\begin{aligned}
    \boldsymbol{x}(\boldsymbol{\beta}, \boldsymbol{\boldsymbol{\bar{\mu}}_{\boldsymbol{c}}}) &= \boldsymbol{y} - \boldsymbol{\beta} \left( \boldsymbol{c} - \boldsymbol{\mu_c} \right) + \boldsymbol{\beta}\boldsymbol{\bar{\delta}}\,, \\
   \boldsymbol{\Sigma_{xx}}(\boldsymbol{\beta},\boldsymbol{\bar{\boldsymbol{\mu}}_{\boldsymbol{c}}}) &=  \boldsymbol{\Sigma_{xx}}(\boldsymbol{\beta}) + \boldsymbol{\beta} \frac{\boldsymbol{\Sigma_{cc}}}{M}\boldsymbol{\beta}^{\boldsymbol{T}}\,,
    \end{aligned}
\end{equation}
with $\boldsymbol{\Sigma_{xx}}(\boldsymbol{\beta})$ from equation \eqref{eq:varMV}. The $\boldsymbol{\beta}\boldsymbol{\bar{\delta}}$ term above is statistically independent of the rest of the sum, since it is computed on a separate set of seeds.
As expected, additional uncertainty is brought by $\boldsymbol{\Sigma_{cc}}$ and scaled by the estimated control matrix. See Appendix \ref{app:bayesDer} for a Bayesian derivation of the combined uncertainty in the Gaussian case while taking into account possible prior information on $\boldsymbol{\mu}$ and/or $\boldsymbol{\mu_c}$.

\subsubsection{Estimation of the control matrix}\label{sec:estControl}
The matrices in equation \eqref{eq:betaStarMV} need to be estimated from data via the bias-corrected sample covariance matrix:
\begin{equation}\label{eq:empMat}
\begin{aligned}
    \boldsymbol{\widehat{\Sigma}_{yc}} &= \frac{1}{N-1} \sum_{i=1}^N \left( \boldsymbol{y}_i -\boldsymbol{\bar{y}} \right) \left( \boldsymbol{c}_i -\boldsymbol{\bar{c}} \right)^{\boldsymbol{T}}\,, \\
    \boldsymbol{\widehat{\Sigma}_{cc}} &= \frac{1}{N-1} \sum_{i=1}^N \left( \boldsymbol{c}_i -\boldsymbol{\bar{c}} \right) \left( \boldsymbol{c}_i -\boldsymbol{\bar{c}} \right)^{\boldsymbol{T}}\,.
    \end{aligned}
\end{equation}
The computational cost of $\boldsymbol{y}$ is the limiting factor for estimating $\boldsymbol{\Sigma_{yc}}$.
Therefore, the cross-covariance matrix is estimated online, as our primary motivation is to reduce the computation time: for instance, we certainly do not want to run more costly simulations in a precomputation step like we do for $\boldsymbol{\mu_c}$ with fast simulations. Simply put, $\boldsymbol{\widehat{\Sigma}_{yc}}$ is updated each time a new simulation pair is available. 

Note that for finite $N$, the inverse of $\boldsymbol{\widehat{\Sigma}_{cc}}$ in equation \eqref{eq:empMat} is not an unbiased estimator of the  precision matrix  $\boldsymbol{\Sigma_{cc}^{-1}}$ \citep{Hartlap_2006}. 
Moreover, $\boldsymbol{\widehat{\Sigma}_{cc}^{-1}}$ is not defined when $\boldsymbol{\widehat{\Sigma}_{cc}}$ is rank-deficient, which is guaranteed to happen when $N$ is smaller that $p$ . 
We have consequently replaced $\boldsymbol{\Sigma_{cc}^{-1}}$ by the Moore-Penrose pseudo-inverse -- always defined and unique -- $\boldsymbol{\Sigma_{cc}^{\dagger}}$ in equation \eqref{eq:betaStarMV} for the numerical analysis presented in Section \ref{sec:results} to be able to compute multivariate CARPool estimates even when $N<p$.

Since the singular value decomposition (SVD) exists for any complex or real matrix, we can write $\boldsymbol{\Sigma_{yc}} = \boldsymbol{UVW^{T}}$ and $\boldsymbol{\Sigma_{cc}} = \boldsymbol{OPQ^{T}}=\boldsymbol{OPO^{T}}$ by symmetry.
The optimal control matrix now gives $\boldsymbol{\beta^{\star}} = \boldsymbol{UVW^{T}OP^{-1}O^{T}}$.  The product $\boldsymbol{-P^{\frac{1}{2}}O^{T}}$ whitens the centered surrogate vector elements (principal component analysis whitening), $\boldsymbol{OP^{-\frac{1}{2}}}$ restretches the coefficients and returns them to the surrogate basis, and then $\boldsymbol{UVW^{T}}$ projects the scaled surrogate elements into the high-fidelity simulation basis and rescales them to match the costly simulation covariance. It follows that, when using ${\boldsymbol{\hat\beta}}$ in practice, the projections are done in bases specifically adapted to the $\boldsymbol{y}$ and $\boldsymbol{c}$ samples available.  With this argument, we justify why we use the same simulation/surrogate pairs to compute ${\boldsymbol{\hat\beta}}$ first (with the Moore-Penrose pseudo-inverse of the surrogate covariance replacing the precision matrix) and estimate the CARPool mean after that.

An online estimation of both ${\boldsymbol{\hat\beta}}$ and $\boldsymbol{\bar{x}}(\boldsymbol{\hat{\beta}})$, considering incoming $\left\{ \boldsymbol{y}_n,\boldsymbol{c}_n \right\}$ pairs computed on the same seed $r_n$, amounts to computing a collection of $N$ samples as functions of $\widehat{\boldsymbol{\beta}}$,
\begin{equation}\label{eq:mvComp}
     \boldsymbol{x}_n(\boldsymbol{\hat{\beta}}) = \boldsymbol{y}_n - \boldsymbol{\hat{\beta}} \left( \boldsymbol{c}_n - \boldsymbol{\bar{\mu}_c} \right)\,.
\end{equation}
We implement equation \eqref{eq:mvCV} by taking the sample mean of $N$ such variance-reduced samples,
\begin{equation}\label{eq:sampleCV}
    \boldsymbol{\bar{x}}(\boldsymbol{\hat{\beta}}) = \boldsymbol{\bar{y}} - \boldsymbol{\hat{\beta}} \left( \boldsymbol{\bar{c}} - \boldsymbol{\bar{\mu}_c} \right)\,.
\end{equation}
This way, equation \eqref{eq:sampleCV} can be computed each time a simulation/surrogate  pair is drawn from a seed in $\mathcal{S}_{N} =  \left\{ r_{1}, \dots, r_{N} \right\}$, after updating ${\boldsymbol{\hat\beta}}$ according to equation~\eqref{eq:empMat}.

\subsubsection{Multivariate versus univariate CARPool}
So far we have not assumed any special structure for $\boldsymbol{\beta}$. If, as in the classical control variates setting, the (potentially dense) covariances on the right-hand side of equation \eqref{eq:betaStarMV} are known \textit{a priori}, then  $\boldsymbol{\beta^{\star}}$ is the best solution because it exploits the mutual information between all elements of $\boldsymbol{y}$ and $\boldsymbol{c}$.

In practice, we will be using the online approach discussed in Section \ref{sec:estControl} for a very small number of simulations. 
If we are limited by a very small number of $\left\{ \boldsymbol{y}_n,\boldsymbol{c}_n \right\}$ pairs compared to the number of elements of the vectors, the estimate of $\boldsymbol{\beta^{\star}}$ can be unstable and possibly worsen the variance of equation~\eqref{eq:sampleCV}, though unbiasedness remains guaranteed.

We will demonstrate below that in the case of small number of simulations and a large number of statistics to estimate from the simulations, it is advantageous to impose structure on $\boldsymbol{\beta}$. In the simplest case, we can set the off-diagonal elements to zero. This amounts to treating each vector element separately and results in a decoupled problem with a separate solution \eqref{eq:betaStar} for each vector element.

The univariate setting of \ref{sec:unicv} applied individually to each vector element (\textit{bin}) will be referred to as ``diagonal $\boldsymbol{\beta}$'' or $\boldsymbol{\beta^{\mathrm{diag}}}$, as it amounts to fixing the non-diagonal elements of $\boldsymbol{\Sigma_{cc}}$ and $\boldsymbol{\Sigma_{yc}}$ to zero in equation \eqref{eq:betaStarMV} and only estimating the diagonal elements:
\begin{equation}
\boldsymbol{\beta^{\mathrm{diag}}} =
\begin{pmatrix}
\frac{cov(y_1,c_1)}{\sigma_{c_{1}}^2} \\
 & \frac{cov(y_2,c_2)}{\sigma_{c_{2}}^2}  & &  \text{\huge0} \\
 \text{\huge0} &  &    \ddots       \\
 &   &   & \frac{cov(y_p,c_p)}{\sigma_{c_{p}}^2}
\end{pmatrix}\,.
\end{equation}
The intent of this paper is to show the potential of control variates for cosmological simulations; to this end, we will compare the following unbiased estimators:
\begin{itemize}[labelwidth=*]
    \item \texttt{GADGET}, where we compute the sample mean $\boldsymbol{\bar{y}}$ from $N$-body simulations only.
    \item Multivariate CARPool described by equation \eqref{eq:mvCV}, where we estimate the control matrix $\boldsymbol{\beta}$ online using equations \eqref{eq:empMat}, and denote it by $\boldsymbol{\beta^{\star}}$.
    \item Univariate CARPool, where we use the empirical counterpart of equation \eqref{eq:betaStar} as the control coefficient for each element of a vector: we estimate $\boldsymbol{\beta^{\mathrm{diag}}}$.
\end{itemize}
Other, intermediate choices between fully dense and diagonal $\boldsymbol{\beta}$ are possible and may be advantageous in some circumstances. We will leave an exploration of these to future work, and simply note here that this freedom to tune $\boldsymbol{\beta}$ does not affect the mean of the CARPool estimate.

\section{Cosmological simulations} \label{sec:3sec}
This section describes the simulation methods that we use to compute the statistics presented in Section~\ref{sec:results}.
The simulations assume a $\Lambda$ Cold Dark Matter ($\Lambda$CDM) cosmology  congruent with the {\it Planck} constraints provided by \citet{2018arXiv180706209P}: $\Omega_\mathrm{m}=0.3175$, $\Omega_\mathrm{b}=0.049$, $h=0.6711$, $n_s=0.9624$, $\sigma_8=0.834$, $w=-1.0$ and $M_{\nu}=0.0~\mathrm{eV}$.

\subsection{\textit{Quijote} simulations at the fiducial cosmology}
\citet{villaescusanavarro2019quijote} have publicly released data outputs from $N$-body cosmological simulations run with the full TreePM code \texttt{GADGET-III}, a development of the previous version \texttt{GADGET-II} by \citet{2005MNRAS.364.1105S}.\footnote{Instructions to access the data are given at \url{https://github.com/franciscovillaescusa/Quijote-simulations}.}  Available data and statistics include simulation snapshots, matter power spectra, matter bispectra and matter probability density functions. The sample mean of each statistic computed from all available realisations gives the unbiased estimator of $\mathbb{E} \left[ \boldsymbol{y} \right] = \boldsymbol{\mu}$. 
The fiducial cosmology data set contains 15,000 realisations; their characteristics are grouped in Table~\ref{table:yFeat}.
\begin{table}
    \centering
    \caption{Characteristics of \texttt{GADGET-III} simulations}
    \begin{tabular}{ | m{12em} | m{12em}| }
    \hline
      Characteristic/Parameter   & Value  \\
      \hline \hline
       Simulation box volume  & $\left( 1000~h^{-1} {\rm Mpc} \right)^3$ \\
       \hline
       Number of CDM particles & $N_\mathrm{p} = 512^3$ \\
       \hline
       Force mesh grid size & $N_\mathrm{m} = 1024$ \\
       \hline
       Starting redshift &  $z_i = 127$ \\
       \hline
       Initial conditions & Second-order Lagrangian perturbation theory (2LPT) \\
       \hline
       Redshift of data outputs & $z \in \left\{ 3.0, 2.0, 1.0, 0.5, 0.0\right\}$ \\
       \hline
    \end{tabular}
    \label{table:yFeat}
\end{table}

As discussed in section \ref{sec:InPractice}, the \textit{Quijote} simulations are selected because we have access to an extensive ensemble of simulations that we can use to validate the CARPool approach. In the following we will look at wavenumbers $k=\sim 1~h{\rm Mpc}^{-1}$ where the \textit{Quijote} simulations may not be fully resolved. This is not important for the purposes of this paper; we will consider the full simulation ensemble as the gold standard that we attempt to reproduce with a much smaller number of simulations plus fast surrogates.

In the next subsection, we present the chosen low-fidelity simulation code which provides an approximate statistic $\boldsymbol{c}$ for our numerical experiments.

\subsection{Choice of approximate simulation method}
Any fast solution can be used for $\boldsymbol{c}$, provided that it can be fed with the same initial conditions as of the \textit{Quijote} simulations.
To this end, the matter power spectrum from \texttt{CAMB} \citep{Lewis_2000} at $z=0$ is rescaled at the initial redshift $z_i = 127$ to generate the initial conditions, as in~\citet{villaescusanavarro2019quijote}.
In this work, we use the \texttt{L-PICOLA} code developed by \citet{Howlett_2015}, an MPI parallel implementation of the COLA method \citep{Tassev_2013}. The core idea of COLA is to add residual displacements computed with a PM $N$-body solver to the trajectory given by the first- and second-order LPT approximations. If $\boldsymbol{l}$ is the initial Lagrangian position of a particle and $\boldsymbol{x}$ is its Eulerian comoving coordinates, the evolution of the residual displacement field $\boldsymbol{\Psi_\mathrm{res}}$ appears by rewriting the equation of motion in a frame comoving with the LPT trajectory,
\begin{align}\label{eq:colaEOM}
    \partial_a^2 \boldsymbol{\Psi_\mathrm{res}} &= -\nabla_{\boldsymbol{x}} \Phi - \partial_a^2 \boldsymbol{\Psi_\mathrm{LPT}}\,,
\end{align}
where $a$ is the cosmological scale factor and
\begin{align}\label{eq:colaEOM2}
    \boldsymbol{\Psi_\mathrm{res}} &\equiv \boldsymbol{\Psi} - \boldsymbol{\Psi_\mathrm{LPT}}\nonumber\,, \\
    \boldsymbol{x}(\boldsymbol{l},a) &\equiv \boldsymbol{l} + \boldsymbol{\Psi}\left( \boldsymbol{l}, a \right)\nonumber \,.
\end{align}
Here, we have omitted the Hubble expansion rate and constants for simplicity, $\boldsymbol{\Psi_\mathrm{LPT}}$ is the displacement vector associated to $\boldsymbol{x_\mathrm{LPT}}$, the LPT approximation to the Eulerian position $\boldsymbol{x}$ of matter particles, and $\Phi$ is the gravitational potential obtained by solving the Poisson equation with $\nabla_{\boldsymbol{x}}$ the gradient operator in Eulerian comoving coordinates. Time integration is performed by discretising the derivative $\partial_a^2$ only on the left-hand side of equation \eqref{eq:colaEOM}, while the (second-order) LPT displacements are computed analytically and stored. \texttt{L-PICOLA} has its own initial conditions generator and uses a slightly modified version of the \texttt{2LPTic} code.\footnote{The parallelised version of the code is available at \url{http://cosmo.nyu.edu/roman/2LPT/}.}
To generate \texttt{L-PICOLA} snapshots and extract statistics, we set the free parameters as presented in Table~\ref{table:cFeat}. Justification for these choices, along with more details on COLA and the \texttt{L-PICOLA} implementation, can be found in Appendix \ref{app:cola}.
\begin{table}
    \centering
    \caption{Characteristics of \texttt{L-PICOLA} simulations}
    \begin{tabular}{ | m{12em} | m{12em}| }
    \hline
      Characteristic/Parameter   & Value  \\
      \hline \hline
       Number of timesteps & 20 (linearly spaced)\\
       \hline
       Modified timestepping from \citet{Tassev_2013} & $nLPT=+0.5$\\
       \hline
       Force mesh grid size & $N_\mathrm{m} = 512$ \\
       \hline
       Starting redshift &  $z_i = 127$ \\
       \hline
       Initial conditions & Second-order Lagrangian perturbation theory (2LPT) \\
       \hline
       Redshift of data outputs & $z \in \left\{1.0, 0.5, 0.0\right\}$ \\
       \hline
    \end{tabular}
    \label{table:cFeat}
\end{table}

\section{Application and results}\label{sec:results}
In this section, we apply the CARPool technique to three standard cosmological statistics: the matter power spectrum, the matter bispectrum, and the one-dimensional probability density function (PDF) of matter fractional overdensity. We seek to improve the precision of estimates of theoretical expectations of these quantities as computed by \texttt{GADGET-III}. To assess the actual improvement, we need the sample mean $\boldsymbol{\bar{y}}$ of the \textit{Quijote} simulations on the one hand, and the estimator \eqref{eq:sampleCV} on the other hand.

\begin{figure*}
    \centering
    \includegraphics[width=1.85\columnwidth, height = 0.95\textheight]{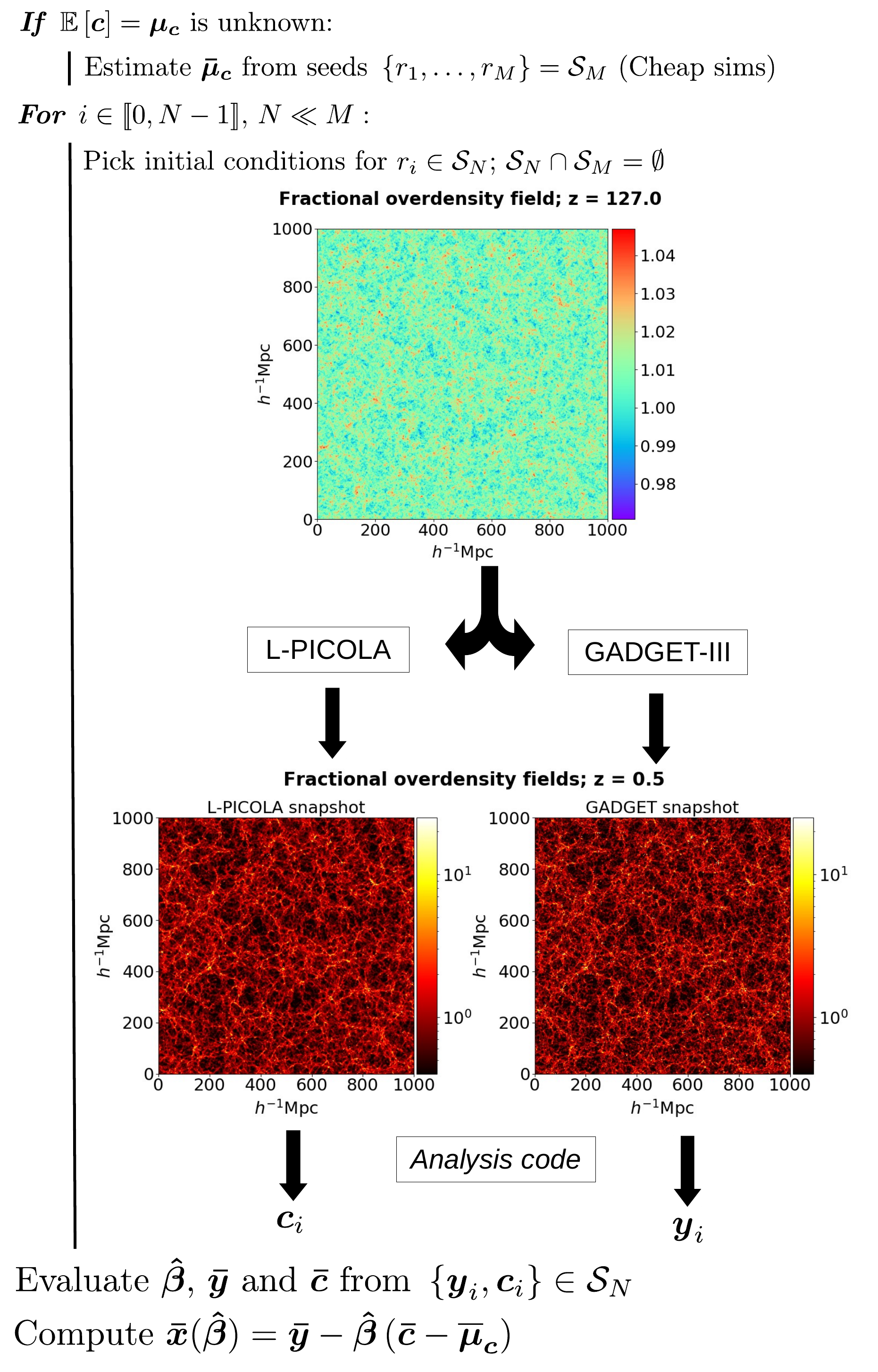}
    \caption{Flowchart of the practical application of CARPool to cosmological simulations. We highlight the estimation of $\boldsymbol{\mu_c}$ as a precomputation step using $M$ fast simulations. The larger the $M$, the less impacted the variance/covariance of the control variates estimator, as expressed in \eqref{eq:errMuc} and Appendix~\ref{app:bayesDer}. The fractional overdensity images are projected slices of $60~h^{-1}{\rm Mpc}$.}\label{fig:flow}
\end{figure*}

Additionally, unless stated otherwise, each test case has the following characteristics:
\begin{itemize}
    \item $N_\mathrm{max}=500 $ $\left\{ \boldsymbol{y}_i, \boldsymbol{c}_i \right\}$ simulation pairs are generated, and the cumulative sample mean $\boldsymbol{\bar{y}}$ (resp. $\boldsymbol{\bar x(\beta)}$) is computed for every other $5$ additional simulations (resp. simulation pairs).
    \item $M=1,500$ additional fast simulations are dedicated to the estimation of $\boldsymbol{\mu_c}$.
    \item The sample mean of 15,000 $N$-body simulations, accessible in the \textit{Quijote} database, is taken as the true $\boldsymbol{\mu}$.
    \item $p = q$ since we post-process \texttt{GADGET-III} and \texttt{L-PICOLA} snapshots with the same analysis codes (e.g. same vector size for $\boldsymbol{y}$ and $\boldsymbol{c}$).
    \item The analysis is performed at redshift $z=0.5$. The lower the redshift, the more non-linear (and hence more difficult) the structure formation problem. We pick the lowest redshift that is relevant for upcoming galaxy surveys. We expect CARPool to be even more efficient for higher redshifts.
    \item $\delta(\boldsymbol{x}) \equiv \rho(\boldsymbol{x})/\bar{\rho} - 1$ is the matter density contrast field; the first term  designates the matter fractional overdensity field computed with the Cloud-in-Cell (CiC) mass assignment scheme. $\boldsymbol{x}$ exceptionally denotes the three-dimensional comoving grid coordinates here.
    \item $N_\mathrm{grid}$ designates the density contrast grid size when post-processing snapshots.
    \item We use bias-corrected and accelerated (BCa) bootstrap,\footnote{Available at \url{https://github.com/cgevans/scikits-bootstrap}} with $B =$~5,000 samples with replacement, to compute the $95\%$ confidence intervals of the estimators. \citet{efron1994introduction} explain the computation.
\end{itemize}
The procedure of the method is illustrated in Figure~\ref{fig:flow}. The first step is to run $M$ fast surrogates to compute the approximate mean $\boldsymbol{\mu}_{\boldsymbol{c}}$. 
How large $M$ should be depends on the accuracy demanded by the user. Then, for each newly picked initial condition, both the expensive simulation code and the low-fidelity method are run to produce a snapshot pair. Only in this step do we need to run the high-fidelity simulation code $N$ times. The mean \eqref{eq:sampleCV} can be computed for each additional pair to track the estimate. In the next section, we assess the capacity of CARPool to use less than $10$ simulations and a set of fast surrogates to match the precision of a large number of $N$-body simulations. All the statistics are calculated from the snapshots with the Python 3 module \texttt{Pylians3}.\footnote{Available at \url{https://github.com/franciscovillaescusa/Pylians3}}

\subsection{Matter power spectrum}
This section is dedicated to estimating the power spectrum of matter density in real space at $z=0.5$, the lower end of the range covered by next-generation galaxy redshift surveys.
The density contrast $\delta(\boldsymbol{x})$ is computed from each snapshot with the grid size $N_\mathrm{grid} = 1024$. The publicly available power spectra range from $k_\mathrm{min}=$ \num{8.900e-3} $h {\rm Mpc^{-1}}$ to $k_\mathrm{max}=5.569$ $h {\rm Mpc^{-1}}$ and contain $886$ bins.
The following analysis is restricted to $k_\mathrm{max}=1.194$ $h {\rm Mpc^{-1}}$, which results in $190$ bins. We simplify our test case by compressing the power spectra into $p=95$ bins, using the appropriate re-weighting by the number of modes in each $k$ bin given in \texttt{Pylians3}. Univariate CARPool gives the best results since we are using the smallest possible number of costly $N$-body simulations; for this reason, power spectrum estimates using the multivariate framework are not shown here. 
As we discuss in appendix \ref{app:cola},  we intentionally run our fast surrogate (COLA) in a mode that produces a power spectrum that is highly biased compared to the full simulations, with a power deficit of more than $60\%$ on small scales.

\subsubsection{CARPool versus $N$-body estimates}\label{sec:estPk}
\begin{figure}
    \includegraphics[width=\columnwidth]{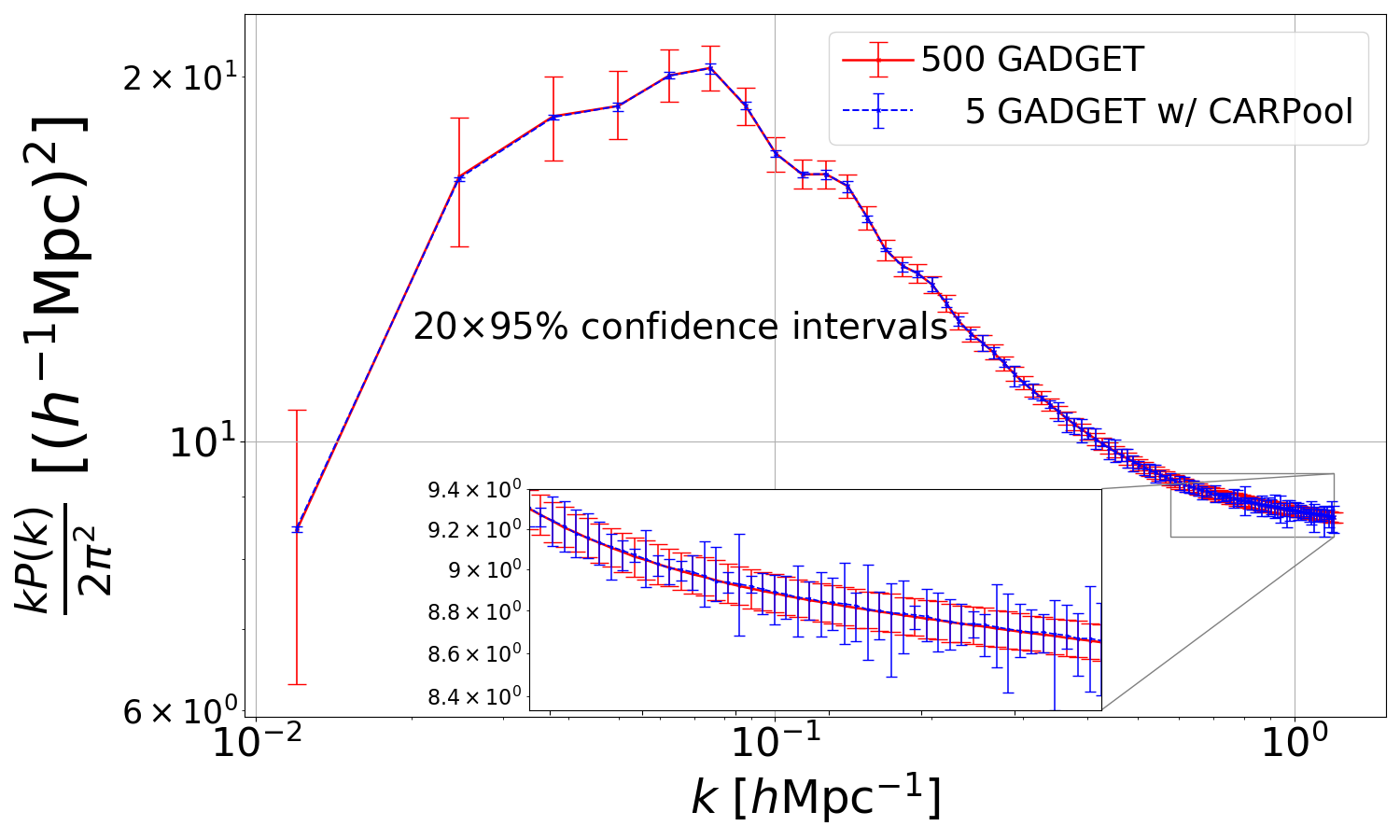}
    \caption{Estimated power spectrum with $500$ $N$-body simulations versus $5$ pairs of ``$N$-body + cheap'' simulations, from which $\widehat{\boldsymbol{\beta^\mathrm{diag}}}$ is derived. The estimated $95\%$ confidence intervals are computed with the BCa bootstrap. They are enlarged by a factor of 20 for better visibility.}
    \label{fig:pkEst5v500}
\end{figure}

Figure \ref{fig:pkEst5v500} shows the estimated power spectrum with $95\%$ confidence intervals enlarged by a factor of $20$ for better visibility. Only $5$ $N$-body simulations are needed to compute an unbiased estimate of the power spectrum with much higher precision than $500$ $N$-body runs on large scales and on the scale of Baryon Acoustic Oscillations (BAO). On small scales, confidence intervals are of comparable size.\footnote{While bootstrap is robust for estimating the $95\%$ error bars of a sample mean with $500$ simulation, it is not equally reliable with a very small number of realisations. This leads to large bin-to-bin variations of the estimated CARPool confidence intervals in Figure \ref{fig:pkEst5v500}.
An alternative, parametric computation of confidence intervals with very few samples can be found in Appendix \ref{app:collFigs}, using Student $t$-score values.}

\begin{figure}
    \includegraphics[width=0.49\textwidth]{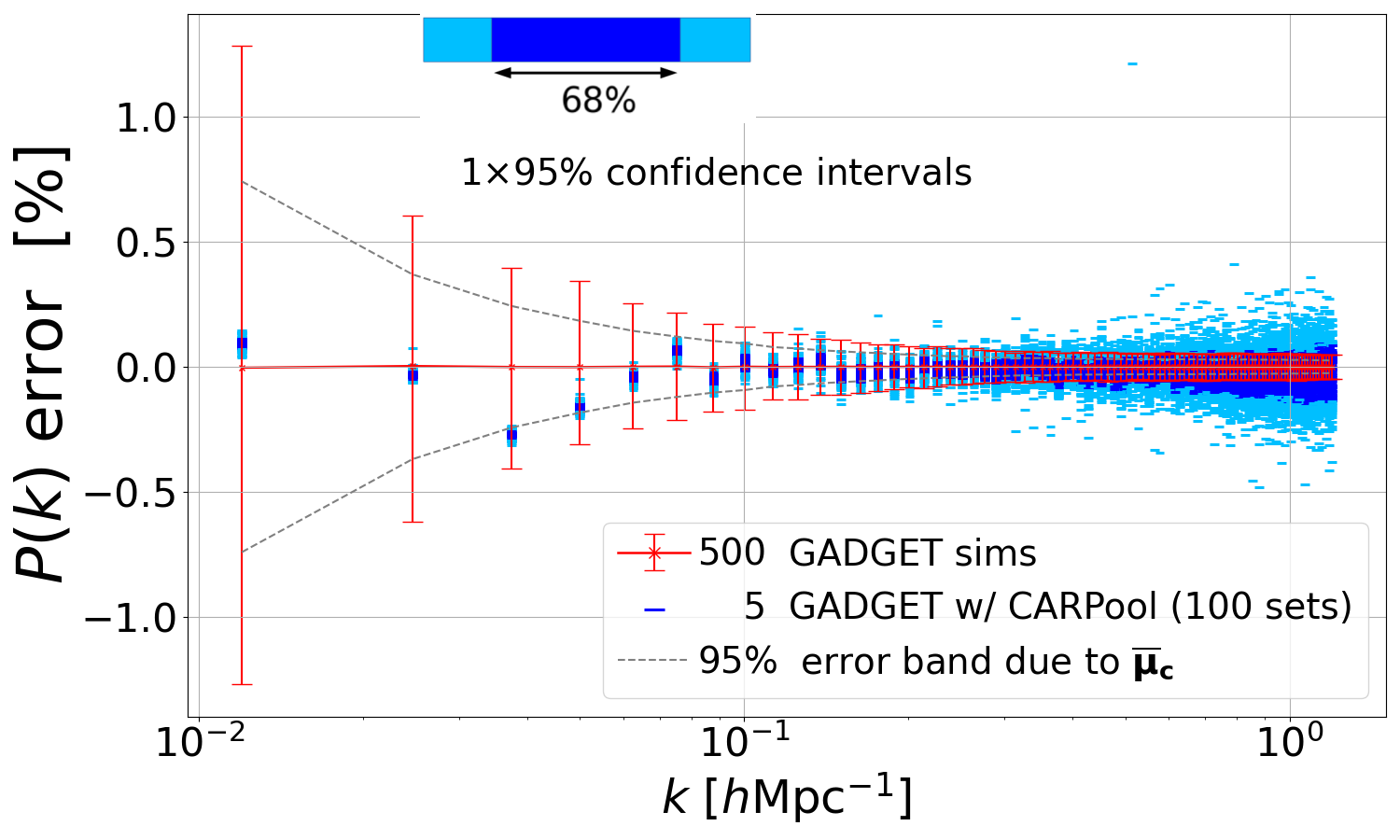}
    \includegraphics[width=0.49\textwidth]{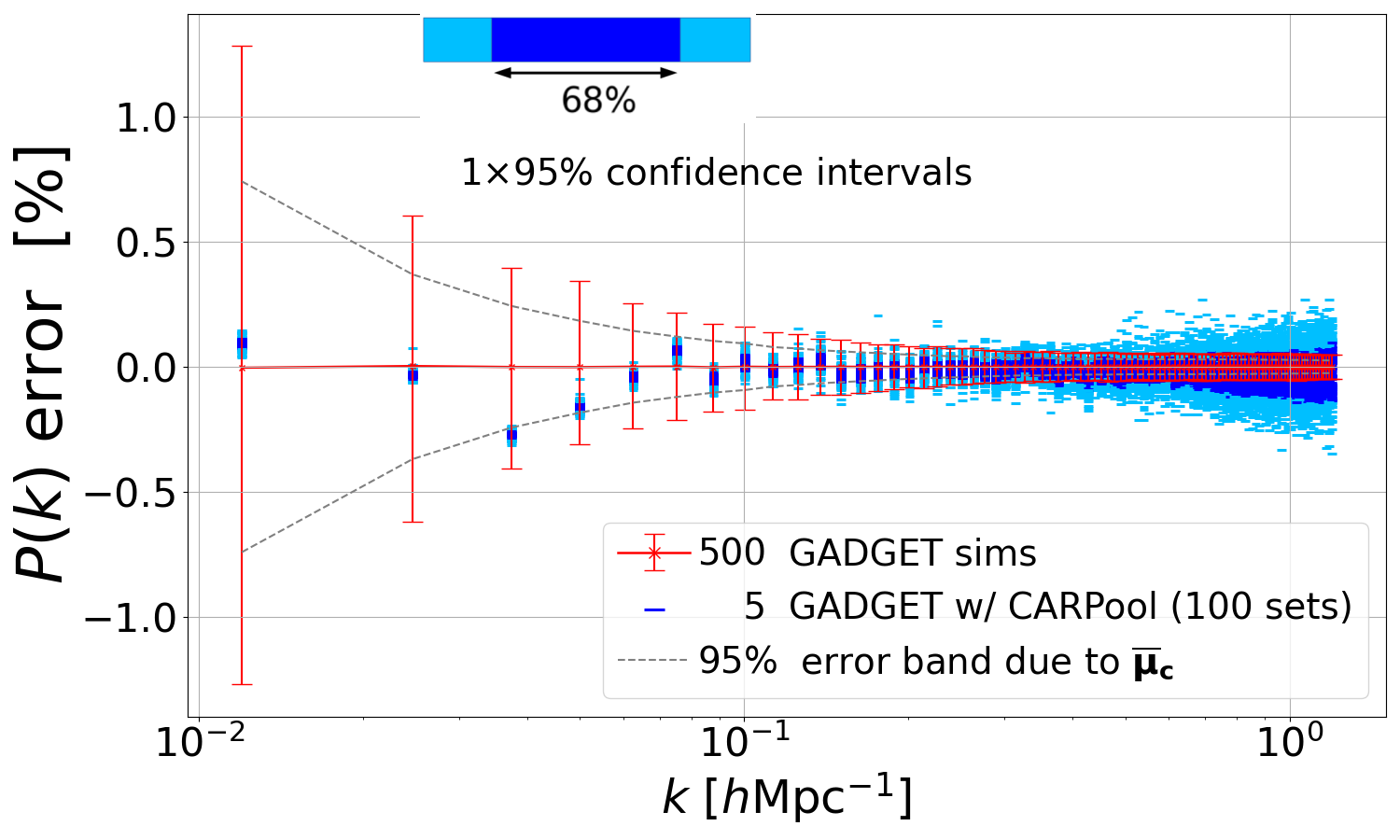}
    \caption{Estimated power spectrum percentage error with respect to 15,000 $N$-body runs: $500$ $N$-body simulations versus $100$ sets of $5$ pairs of ``$N$-body + cheap'' simulations. Each set uses a distinct $\widehat{\boldsymbol{\beta^\mathrm{diag}}}$, calculated with the same seeds used for $\boldsymbol{\bar{x}}$. The upper panel estimate uses $\widehat{\boldsymbol{\beta^\mathrm{diag}}}$ while the lower panel convolves the diagonal elements of $\widehat{\boldsymbol{\beta^\mathrm{diag}}}$ with a narrow top-hat window. Beta smoothing removes outliers and Gaussianises the tails by effectively increasing the number of degrees of freedom for each $\beta$ estimate. Both panels use the same random seeds. 
    The estimated $95\%$ confidence intervals are plotted for the $N$-body sample mean only, using BCa bootstrap. The dark blue symbols show the $68\%$ percentile of the CARPool estimates ordered by the absolute value of the percentage error; the rest appears in light blue symbols.}
    \label{fig:pkRatios5v500}
\end{figure}

We must verify that these results are not produced by a ``lucky'' set of $5$ simulation pairs. To this end, we compute $100$ CARPool means $\boldsymbol{\bar{x}(\widehat{\beta^\mathrm{diag})}}$ from distinct sets of $5$ random seeds. The CARPool estimates fall within a sub-percent accuracy relative to the sample mean from 15,000 $N$-body simulations, as illustrated by the upper panel of Figure~\ref{fig:pkRatios5v500}. The \texttt{GADGET} sample mean percentage error of $500$ simulations with respect to 15,000 simulations is plotted with $95\%$ confidence intervals. We stress here that every percentage error plot in this paper shows an error with respect to 15,000 $N$-body simulations. The mean of $500$ \texttt{GADGET} realisations is thus not at zero percent, though the difference is very small.

\paragraph*{Beta smoothing.} Since we use a very small number of simulations, the estimates of the diagonal elements of $\widehat{\boldsymbol{\beta^\mathrm{diag}}}$ are noisy. This leads to some heavy tailed distributions for the CARPool estimates. Using the freedom we have to modify $\boldsymbol{\beta}$ without affecting unbiasedness, we can exploit the fact that we expect neighboring bins to have similar optimal $\boldsymbol{\beta}$. Convolving the diagonal elements  with a $5$-bin-wide top-hat window slightly reduces the spread at small scales of CARPool estimates computed with only $5$ \texttt{GADGET} power spectra and  removes outliers. The comparison of the two panels in Figure \ref{fig:pkRatios5v500} illustrates this point. Using a $9$-bin-wide Hanning window for the smoothing yields similar results.
We call this technique beta smoothing and  use it with a 5-bin-wide top-hat window in what follows.

Both panels of Figure~\ref{fig:pkRatios5v500} show the symmetric $95\%$ confidence intervals of the surrogate mean with grey dashed lines. They represent the $95\%$ error band likely to stem from the estimation of $\boldsymbol{\mu_c}$, relatively to the mean of 15,000 \texttt{GADGET} simulations, hence the fact that, at large scales especially, the CARPool means concentrate slightly away from the null percentage error. Though the unbiased estimator in equation \eqref{eq:sampleCV} takes a precomputed cheap mean, the practitioner can decide to run more approximate simulations on the fly to improve the accuracy of $\boldsymbol{\bar{\mu}_c}$. Note that the CARPool means with $5$ $N$-body simulations still land withing the $95\%$ confidence intervals from $500$ \texttt{GADGET} simulations, even at large scales where the difference due to the surrogate mean is visible.

\begin{figure}
    \includegraphics[width=\columnwidth]{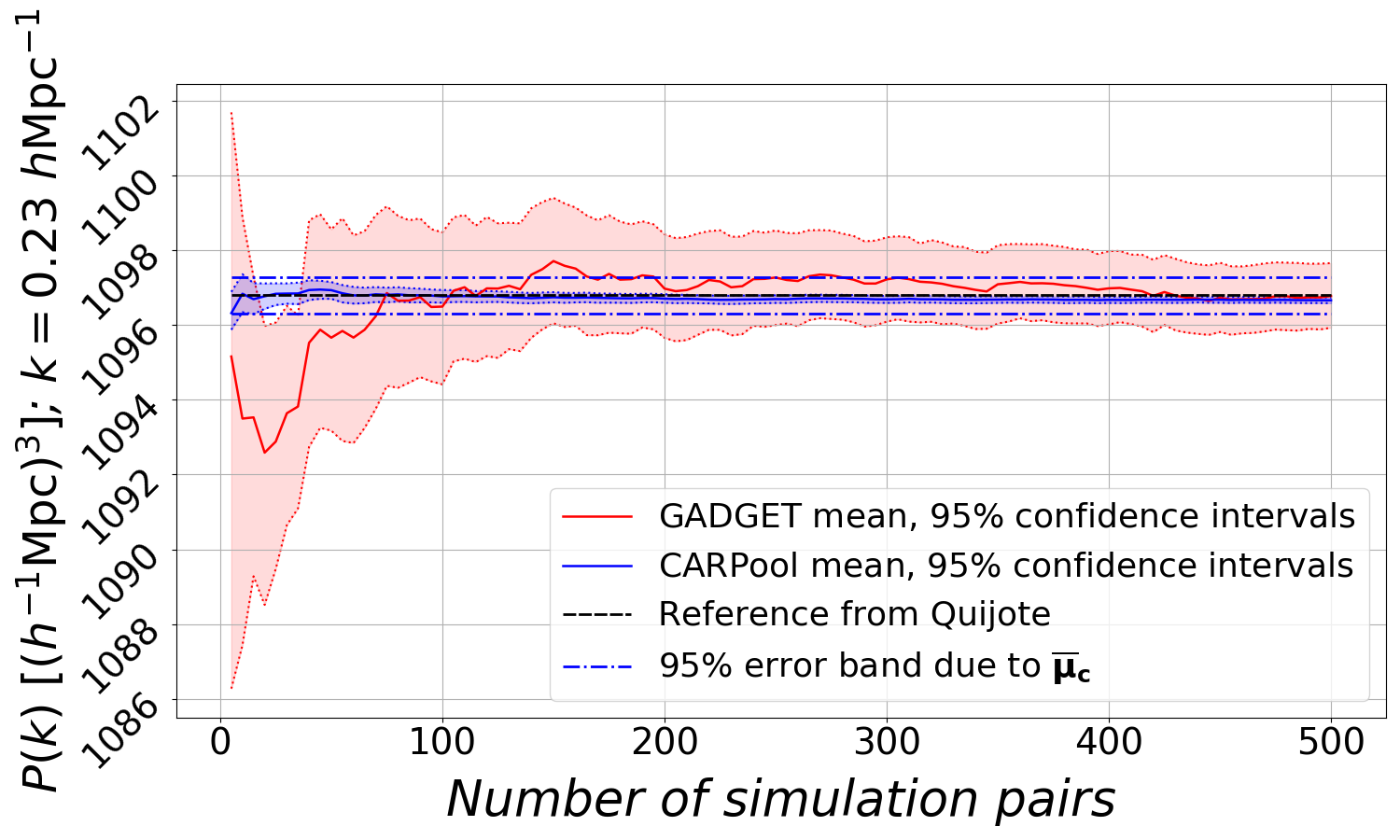}
    \caption{Convergence of a single $k$-bin at the BAO scale: the cumulative sample mean $\boldsymbol{\bar{y}}$ of $N$-body simulations versus the sample mean $\boldsymbol{\bar{x}(\widehat{\beta^\mathrm{diag}})}$. Confidence intervals take into account that   $\boldsymbol{\beta^\mathrm{diag}}$ is estimated from the same number of samples used to compute the CARPool estimate of $P(k)$.}
    \label{fig:pkBin}
\end{figure}

Figure \ref{fig:pkBin} exhibits the convergence of one power spectrum bin at the BAO scale as we add more simulations: the $95\%$ error band of the control variates estimate shrinks extremely fast compared to that of the $N$-body sample mean. 

\subsubsection{Empirical variance reduction}\label{sec:varPk}
\begin{figure*}
    \includegraphics[width=0.49\textwidth]{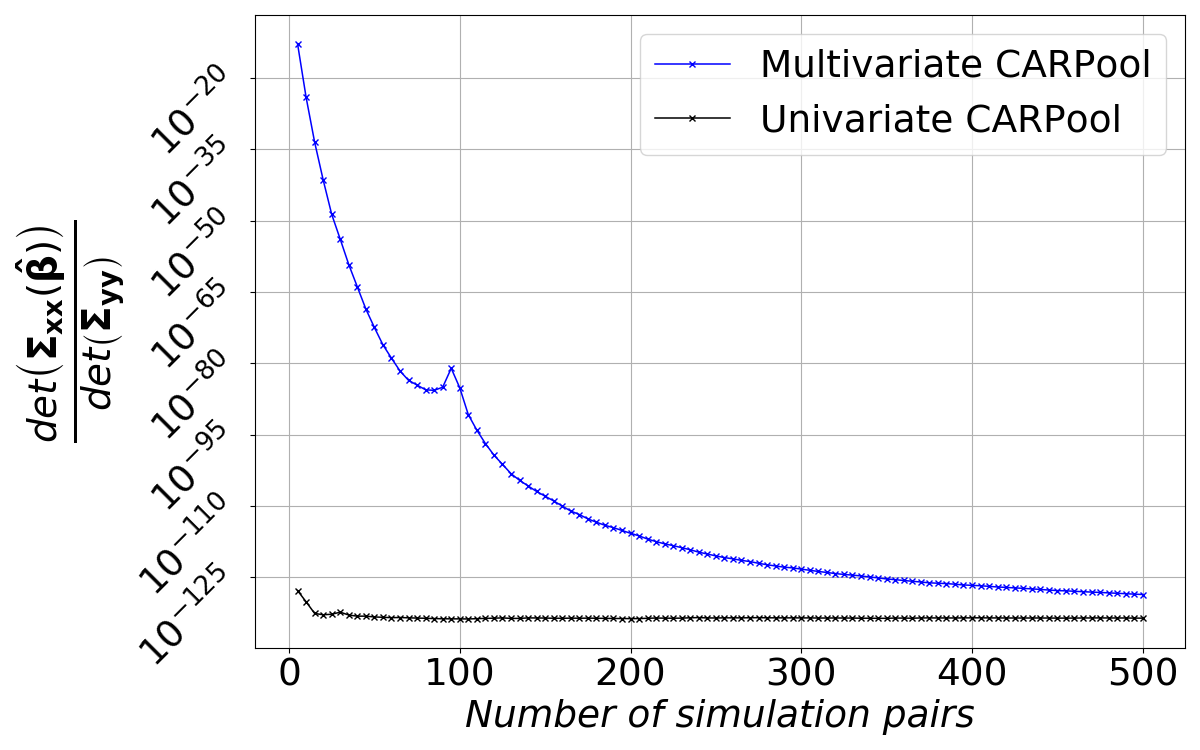}
    \includegraphics[width=0.49\textwidth]{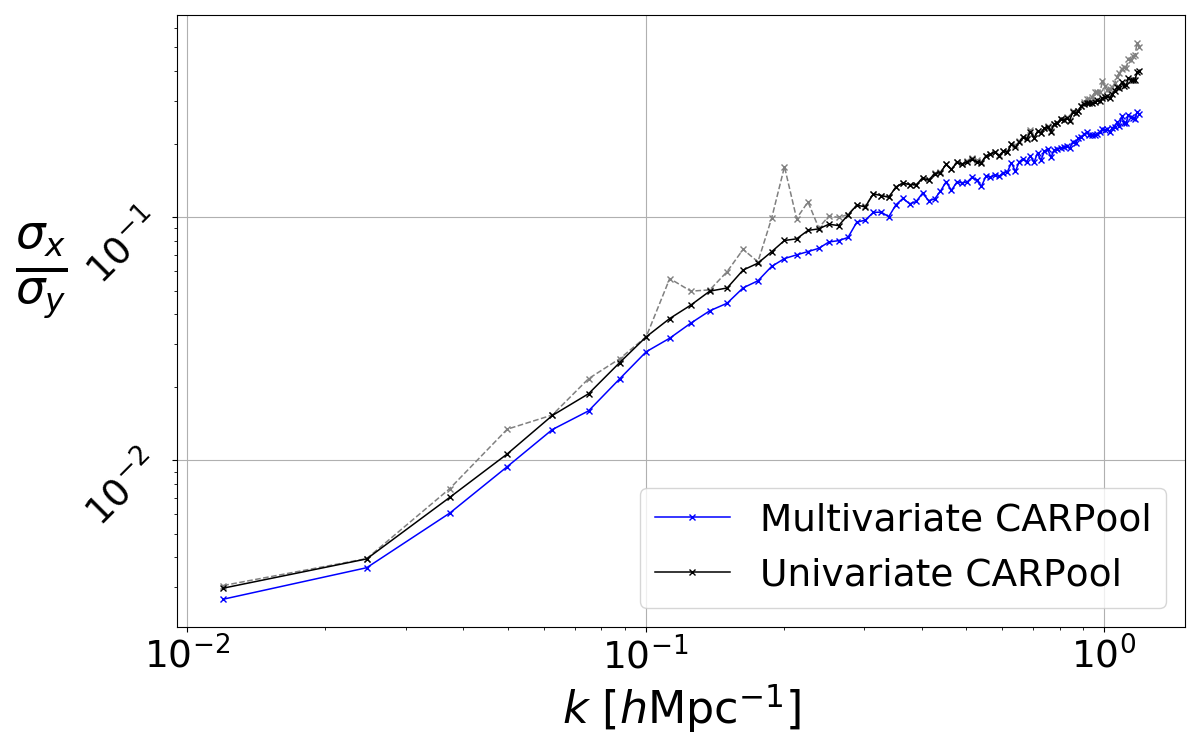}
    \caption{Left panel: Generalised variance ratio for the power spectrum up to $k_\mathrm{max} \approx 1.2~h {\rm Mpc^{-1}}$ as a function of the number of available simulations. Each $\widehat{\boldsymbol{\beta}}$ and $\widehat{\boldsymbol{\beta^\mathrm{diag}}}$ serves to generate 3,000 samples according to \eqref{eq:mvComp} to estimate the CARPool covariance matrix.
    Right panel: Standard deviation reduction for each power spectrum bin due to CARPool. The blue and black curves  use $\widehat{\boldsymbol{\beta}}$ and $\widehat{\boldsymbol{\beta^\mathrm{diag}}}$ estimated with $500$ samples. The dashed grey curve exhibits the actual standard deviation ratio when we have $5$ samples only to compute $\widehat{\boldsymbol{\beta^\mathrm{diag}}}$.
    $\boldsymbol{\Sigma_{yy}}$ is estimated using all 15,000 available power spectra from the \textit{Quijote} simulations.}
    \label{fig:empVarPk}
\end{figure*}

The left panel of Figure~\ref{fig:empVarPk} shows the empirical generalised variance reduction of the CARPool estimate compared to the standard estimate, as defined in equation \eqref{eq:mvVarReduc}. The vertical axis corresponds to the volume ratio of two parallelepipeds of dimension $p=95$, in other words the volume ratio of error ``boxes'' for two estimators. The determinant 
$\det\left(\boldsymbol{\widehat{\Sigma_{yy}}}\right)$ is fixed because we take all 15,000 $N$-body simulations available in \textit{Quijote} to compute the most accurate estimate of $\boldsymbol{\Sigma_{yy}}$ we have access to, whereas $\det\left(\boldsymbol{\Sigma_{xx}}(\boldsymbol{\hat{\beta}})\right)$ changes each time new simulation pairs are run. More precisely, for each data point in Figure~\ref{fig:empVarPk}, we take the control matrix estimate computed with $5k,k \in \llbracket 1,100 \rrbracket$ simulation pairs and generate 3,000 $\boldsymbol{x}$ samples according to \eqref{eq:mvComp} to obtain an estimator of $\boldsymbol{\Sigma_{xx}}$. For that, we use 3,000  \textit{Quijote} simulations and 3,000 additional \texttt{L-PICOLA} surrogates run with the corresponding seeds.

The simpler univariate scheme outperforms the estimation of the optimal $\boldsymbol{\beta^{\star}}$ for $N=5k,k \in \llbracket 1,100 \rrbracket$, corroborating the experiments of Section \ref{sec:estPk}. Furthermore, variance reduction granted by a sub-optimal diagonal $\boldsymbol{\beta^\mathrm{diag}}$ improves rapidly and reaches its apparent limit quickly. We suspect that the slight worsening of the variance reduction, when the number of available samples to estimate $\boldsymbol{\beta^{\star}}$ neighbors the vector size $p$, is linked to the eigenspectrum of $\boldsymbol{\Sigma_{c,c}^{\dagger}}$ and could be improved by projecting out the eigenmodes corresponding to the smallest, noisiest eigenvalues.  

We depict the scale-dependent performance of CARPool for the matter power spectrum in the right panel of Figure~\ref{fig:empVarPk}. The vertical axis is the variance reduction to expect from the optimal control coefficients (or matrix). Namely, we take the data points of the left panel for $500$ simulation/surrogate pairs, extract the diagonal of the covariance matrices and divide the arrays. The blue and black curves show the variance reduction with respect to the sample mean of $N$-body simulations using all 500 simulation/surrogate pairs to estimate the control matrix. 
In practice, we estimate $\boldsymbol{\beta}$ using only 5 simulation/surrogate pairs; does this noisy $\boldsymbol{\hat\beta}$  lead to significant inefficiency? The grey dashed curve shows the actual standard deviation reduction brought by the rough estimate of $\boldsymbol{\beta^\mathrm{diag}}$ using $5$ simulation pairs only, with which the results of Figures \ref{fig:pkEst5v500} and \ref{fig:pkRatios5v500} are computed. A few $k$-bins fluctuate high but the variance reduction remains close to optimal, especially considering that only 5 simulations were used, and we have not attempted any further regularisation except for beta smoothing.

\subsection{Matter bispectrum}
We compute the shot-noise corrected matter bispectrum in real space \citep{Hahn_2020,villaescusanavarro2019quijote}, using \texttt{pySpectrum}\footnote{Available at \url{https://github.com/changhoonhahn/pySpectrum}} with $N_\mathrm{grid}=360$ and bins of width $\Delta k= 3k_\mathrm{f}$ = \num{1.885e-2} $h {\rm Mpc^{-1}}$, where $k_\mathrm{f} = \frac{2\pi}{L}$ $h {\rm Mpc^{-1}}$ is the fundamental mode depending on the box size $L$.
As in the previous section, we present only the results using $\boldsymbol{\beta^\mathrm{diag}}$ instead of $\boldsymbol{\beta^{\star}}$.
We examine two distinct sets of bispectrum coefficients: in the first case we study the bispectrum for squeezed isosceles triangles as a function of opening angle only, averaging over scale; in the second case we compute equilateral triangles as a function of $k$.

\subsubsection{Squeezed isosceles triangles}
We start the analysis by regrouping isosceles triangles ($k_1=k_2$) and re-weighting the bispectrum monopoles for various $\frac{k_3}{k_1}$ ratios in ascending order. Only squeezed triangles are considered here: $\left(\frac{k_3}{k_1}\right)_\mathrm{max} = 0.20$ so that the dimension of $\boldsymbol{y}$ is $p=98$ (see Table \ref{table:notations}).

\begin{figure}
    \includegraphics[width=0.49\textwidth]{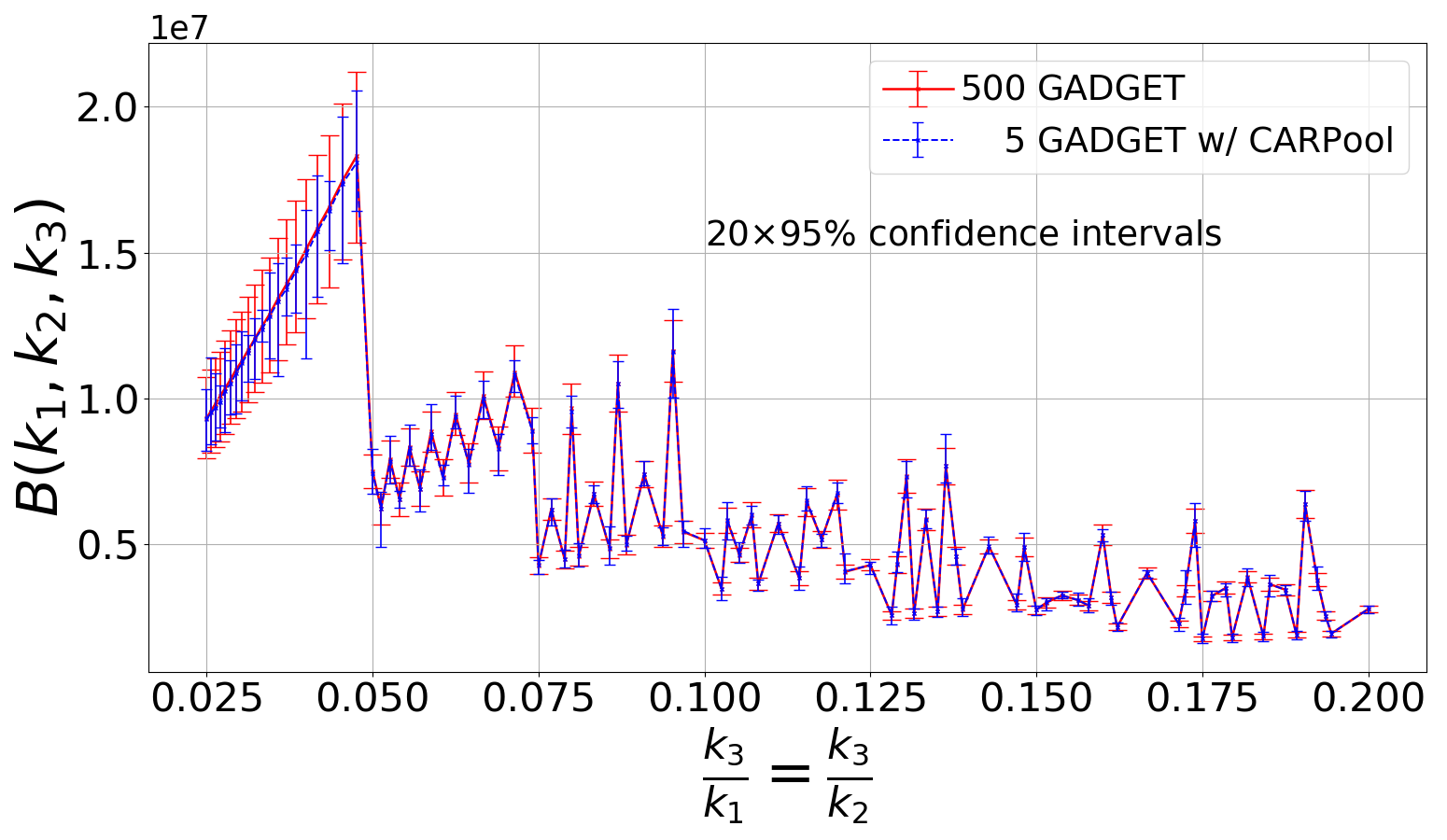}
    \includegraphics[width=0.49\textwidth]{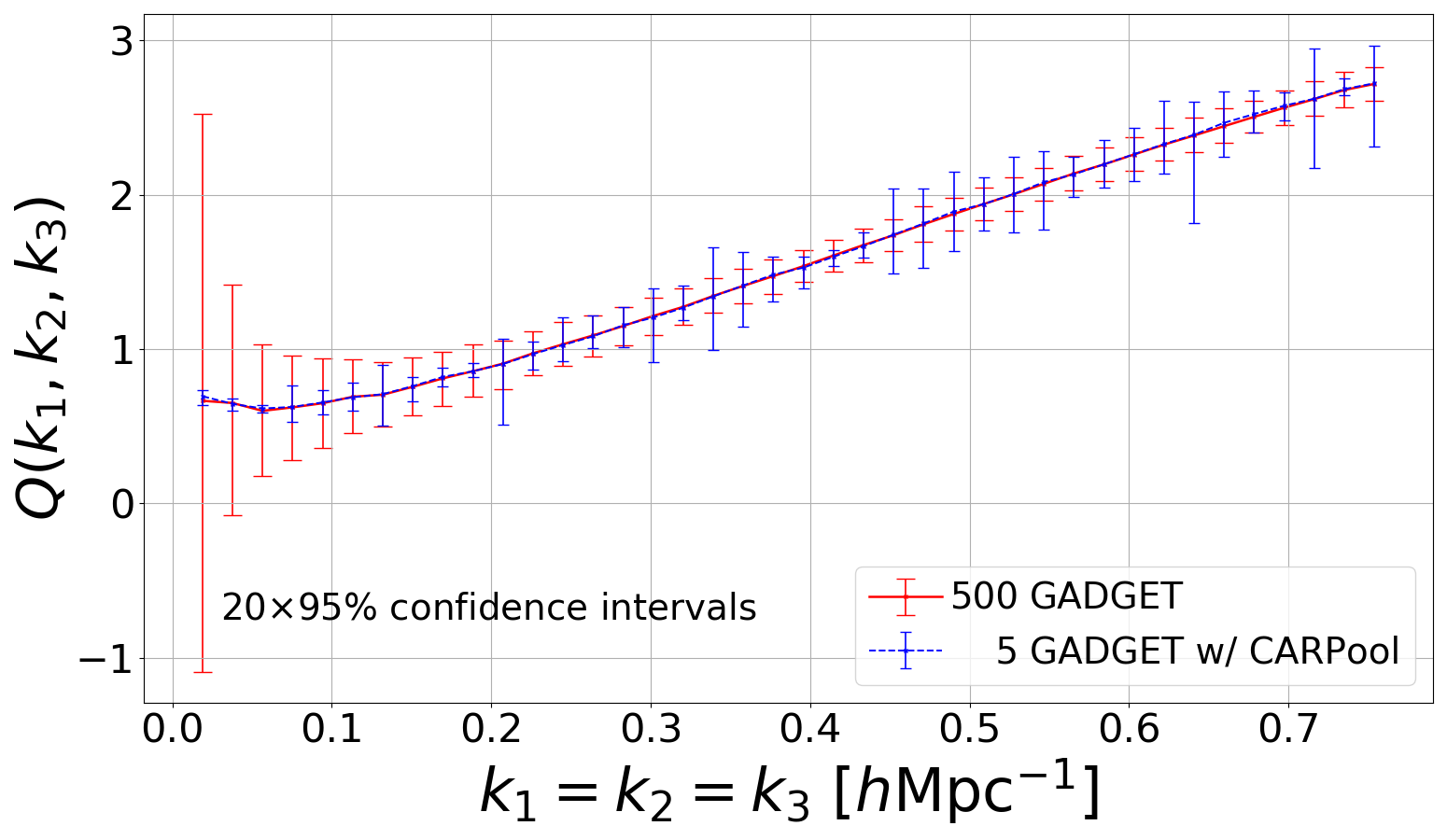}
    \caption{Upper panel: Estimated bispectrum for squeezed isosceles triangles with $500$ $N$-body simulations versus $5$ pairs of ``$N$-body + cheap'' simulations, from which the smoothed $\widehat{\boldsymbol{\beta^\mathrm{diag}}}$ is derived. The estimated $95\%$ confidence intervals are computed with the BCa bootstrap. They are enlarged by a factor of 20 for better visibility.
    Lower panel: As in the upper panel, but for the reduced bispectrum of equilateral triangles.\label{fig:bkQkEstSmF5v500}}
\end{figure}

\begin{figure}
    \includegraphics[width=0.49\textwidth]{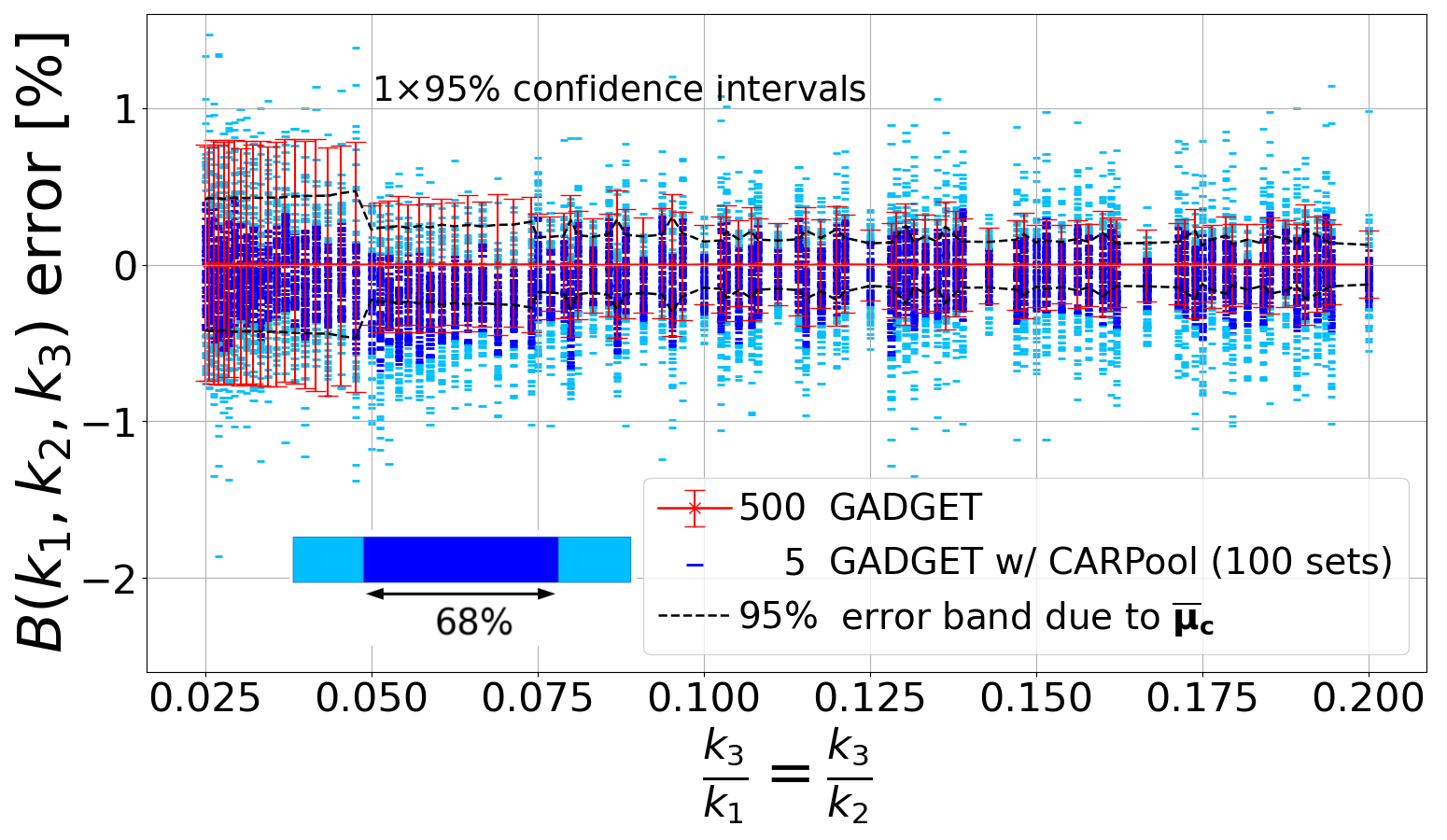}
    \includegraphics[width=0.49\textwidth]{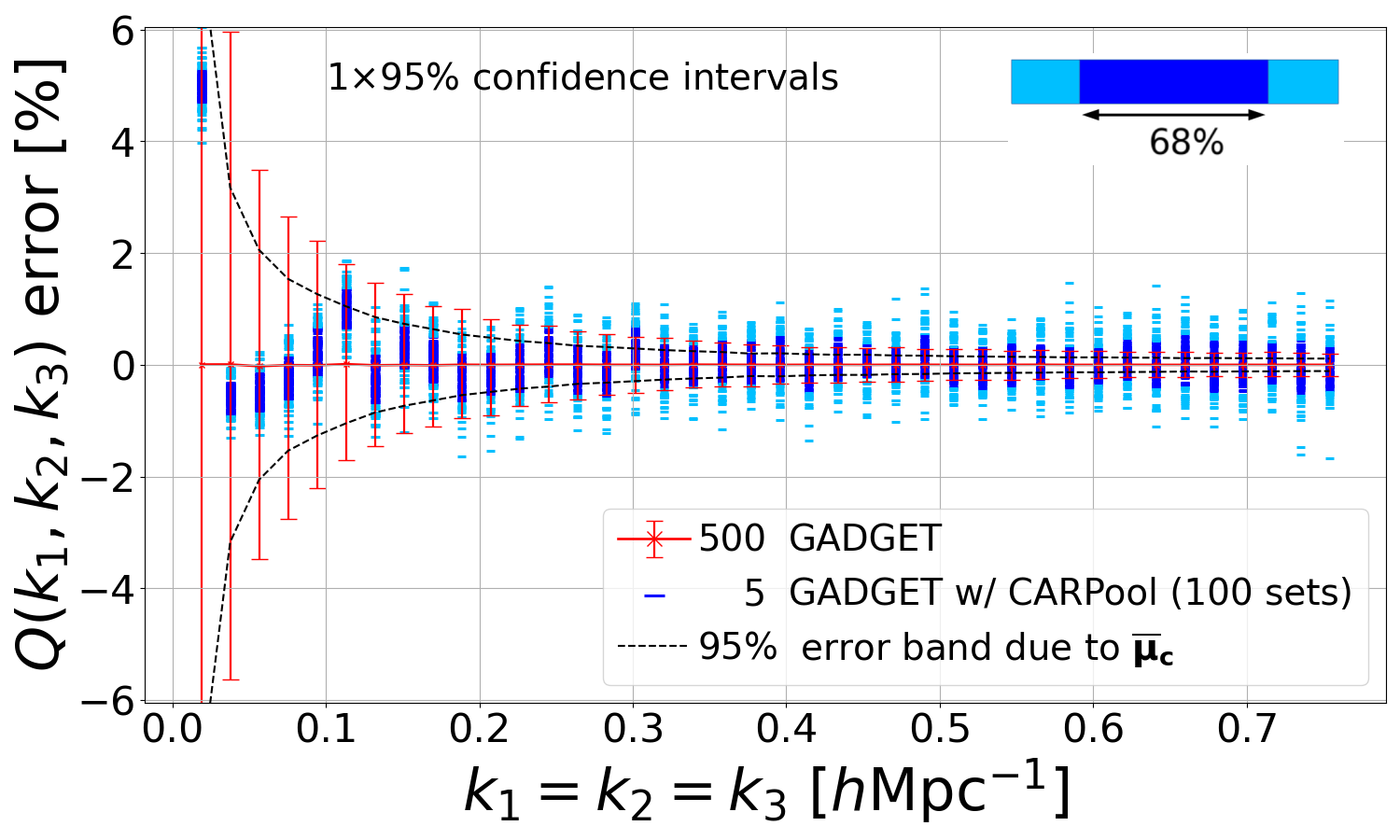}
    \caption{Upper panel: Estimated bispectra percentage error for squeezed isosceles triangles with respect to 15,000 $N$-body runs: $500$ $N$-body simulations versus $100$ sets of $5$ pairs of ``$N$-body + cheap'' simulations. Each set uses a distinct $\widehat{\boldsymbol{\beta^\mathrm{diag}}}$, calculated with the same seeds intervening in $\boldsymbol{\bar{x}}$ and smoothed by a $5$-bin-wide flat window. The estimated $95\%$ confidence intervals are plotted for the $N$-body sample mean only, using BCa bootstrap. The dark blue symbols show the $68\%$ percentile of the CARPool estimates ordered by the absolute value of the percentage error; light-blue symbols represent the rest.
    Lower panel: As in the upper panel, but for the reduced bispectrum of equilateral triangles.\label{fig:bkQkRatiosSmF5v500}}
\end{figure}

\paragraph*{CARPool versus $N$-body estimates.} On the order of $5$ samples are required to achieve a precision similar to that of the sample mean of 500 $N$-body simulations as we show in Figure \ref{fig:bkQkEstSmF5v500} (upper panel). Figure \ref{fig:bkQkRatiosSmF5v500} (upper panel) corroborates the claim by showing the percentage error of 100 CARPool means using $5$ costly simulations each. The reference is the mean of the 15,000 bispectra from the \textit{Quijote} simulations. As in the previous section, we show the $95\%$ error band due to estimation of the surrogate mean $\boldsymbol{\mu_c}$ with dashed curves.

\begin{figure*}
    \includegraphics[width=0.49\textwidth]{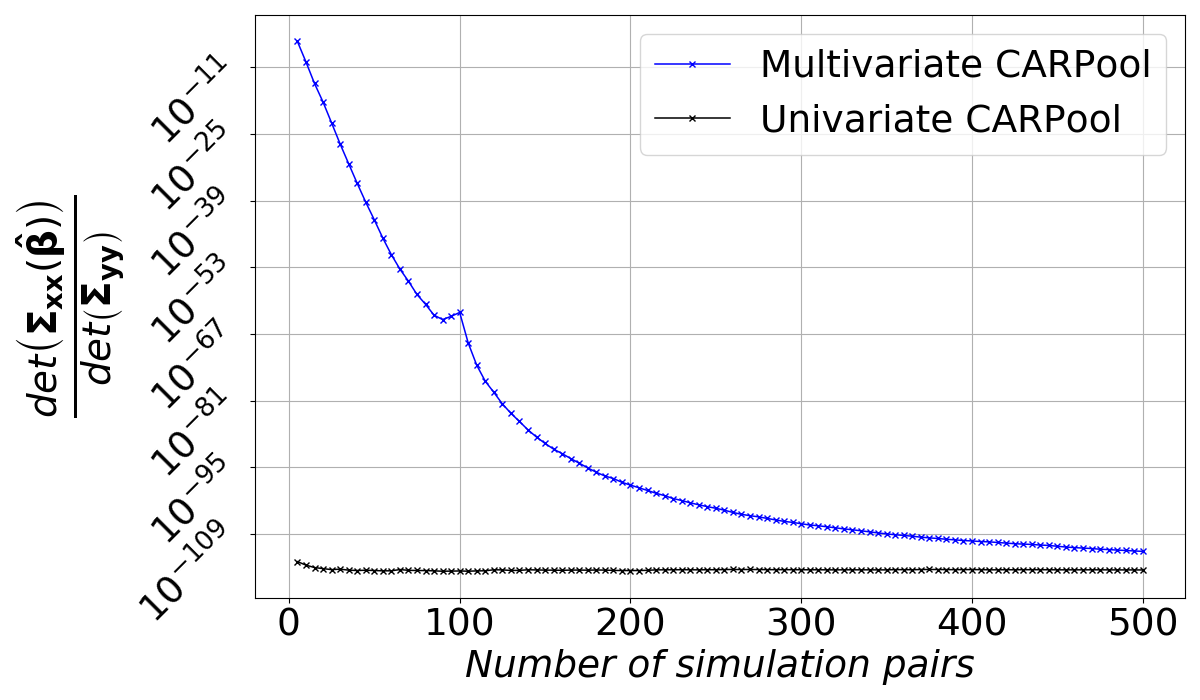}
    \includegraphics[width=0.47\textwidth]{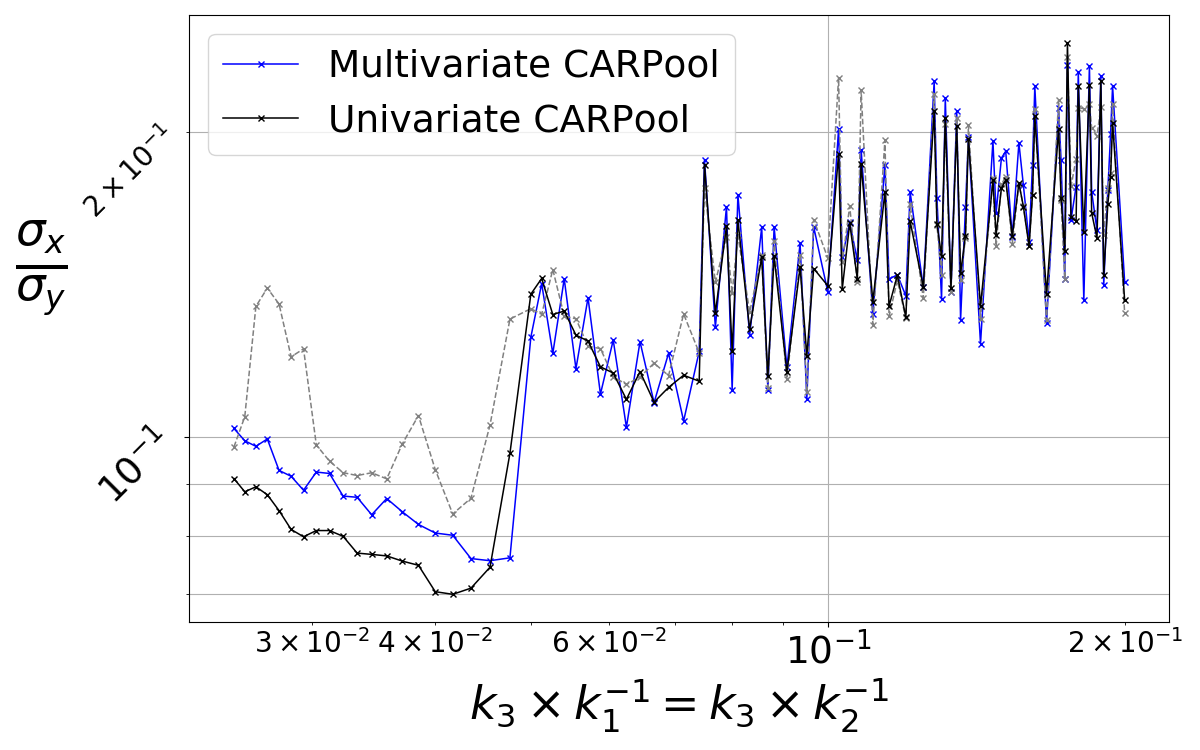}
    \includegraphics[width=0.49\textwidth]{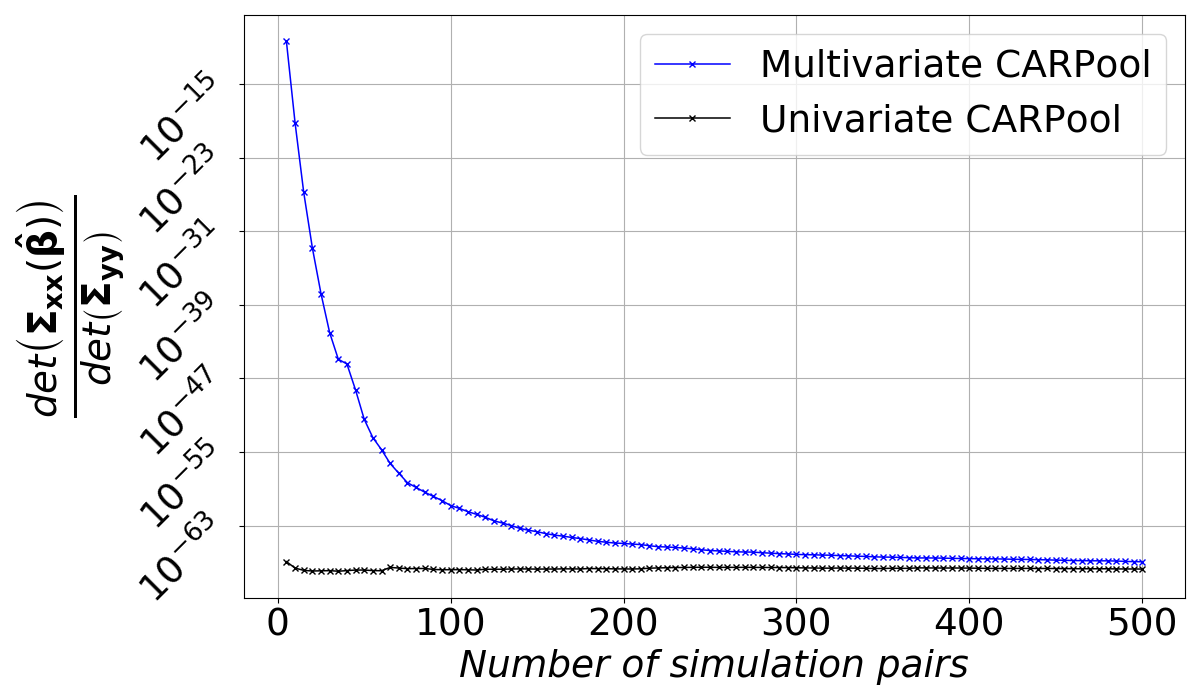}
    \includegraphics[width=0.47\textwidth]{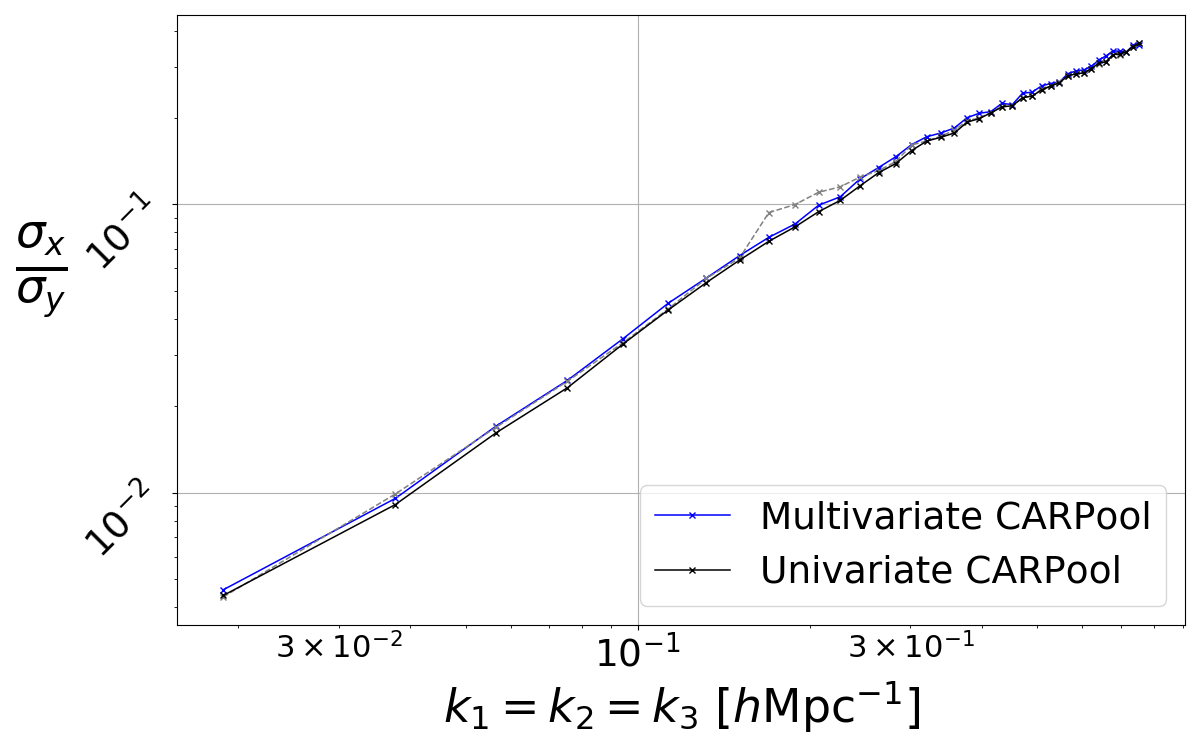}
    \caption{Upper left panel: Generalised variance ratio of bispectrum for squeezed isosceles triangles as a function of the number of available simulations. Each $\widehat{\boldsymbol{\beta}}$ and $\widehat{\boldsymbol{\beta^\mathrm{diag}}}$ serves to generate 1,800 samples according to \eqref{eq:mvComp} to estimate the CARPool covariance matrix.
    Upper right panel: Standard deviation reduction for each squeezed isosceles triangle to expect from CARPool. The blue and black curves respectively use $\widehat{\boldsymbol{\beta}}$ and $\widehat{\boldsymbol{\beta^\mathrm{diag}}}$ estimated with $500$ samples. The dashed grey curve exhibits the actual standard deviation ratio when we have $5$ samples only to compute $\widehat{\boldsymbol{\beta^\mathrm{diag}}}$.
    $\boldsymbol{\Sigma_{yy}}$ is estimated  with all 15,000 available bispectra from the \textit{Quijote} simulations. Lower panels: As in the upper panels, but for the reduced bispectrum of equilateral triangles.}
    \label{fig:varBkQk}
\end{figure*}

\paragraph*{Empirical variance reduction.} As for the power spectrum, the upper left panel of Figure~\ref{fig:varBkQk} shows that the generalised variance reduction is much more significant when separately estimating control coefficients for each triangle configuration. The right-hand side of the curve suggests an increasing improvement of the multivariate case, but in this range of numbers of required samples the variance reduction scheme loses its appeal.
We have used 1,800 additional simulations to compute the covariance matrices intervening in the generalised variance estimates.
In the upper right panel of the figure, the calculation of the standard deviation ratio for each triangle configuration follows the same logic as in Section~\ref{sec:varPk}. The grey dashed curve corresponds to the standard deviation reduction brought by control coefficients (i.e. the univariate CARPool framework) estimated with $5$ simulation/surrogate pairs only.

\subsubsection{Equilateral triangles}
Here, we analyse equilateral triangles with the modulus of $k_1=k_2=k_3$ varying up to $k_\mathrm{max} = 0.75$ $h {\rm Mpc^{-1}}$ ($p=40$). For better visibility, we show the reduced bispectrum monopole $Q(k_1,k_2,k_3)$.

\paragraph*{CARPool versus $N$-body estimates.} Similarly to the previous set of triangle configurations, we compare the precision of the CARPool estimator using $5$ $N$-body simulations with that of the sample mean from $500$ \texttt{GADGET} runs. Figure~\ref{fig:bkQkEstSmF5v500} (lower panel) exhibits the estimated reduced bispectrum with $5$ seeds, while Figure~\ref{fig:bkQkRatiosSmF5v500} (lower panel) shows the relative error of various CARPool sets with respect to the reference from $15,000$ $N$-body samples.

\paragraph*{Empirical variance reduction.} In Figure~\ref{fig:varBkQk} (lower panels), we observe a trend similar to that of the previous experiments: the univariate control coefficients are much better than the control matrix in terms of generalised variance reduction for a realistic number of full $N$-body simulations.

\subsection{Probability density function of smoothed matter fractional overdensity}
\begin{figure}
    \includegraphics[width=\columnwidth]{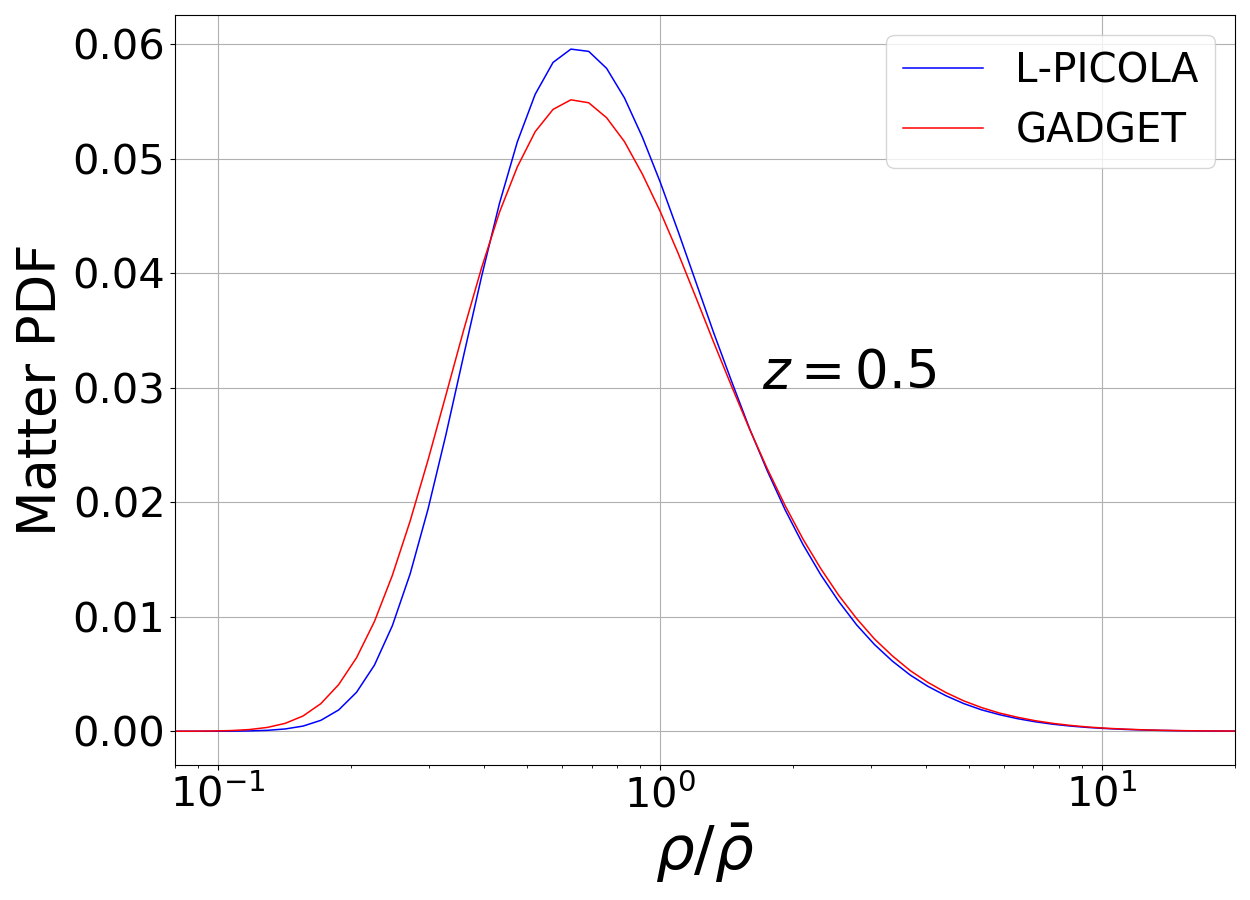}
    \caption{Probability density function of the smoothed matter fractional overdensity of \texttt{GADGET-III} and \texttt{L-PICOLA} snapshots at $z=0.5$ for the same initial conditions. The characteristics of \texttt{L-PICOLA} are provided in Table \ref{table:cFeat}.}
    \label{fig:pdfEx}
\end{figure}

The power spectrum and the bispectrum are Fourier-space statistics. How does CARPOOL fare on a purely direct-space statistic? In the \textit{Quijote} simulations, the probability density function of the matter fractional overdensity, or the \textit{matter PDF}, is computed on a grid with $N_\mathrm{grid}=512$, smoothed by a top-hat filter of radius $R$. There are $100$ histogram bins in the range $\rho/\bar{\rho} \in \left[ 10^{-2}, 10^{2}\right]$. We work with the $R=5~h^{-1} {\rm Mpc}$ case and restrict the estimation of the PDF to the interval $\rho/\bar{\rho} \in \left[ \num{8e-2}, \num{5e1}\right]$ that contains $p=70$ bins. Note that we intentionally do not do anything to improve the correspondence of the surrogate and simulation histograms, an example of which is displayed in Figure~\ref{fig:pdfEx}.

\subsubsection{Empirical variance reduction}
\begin{figure*}
    \includegraphics[width=0.49\textwidth]{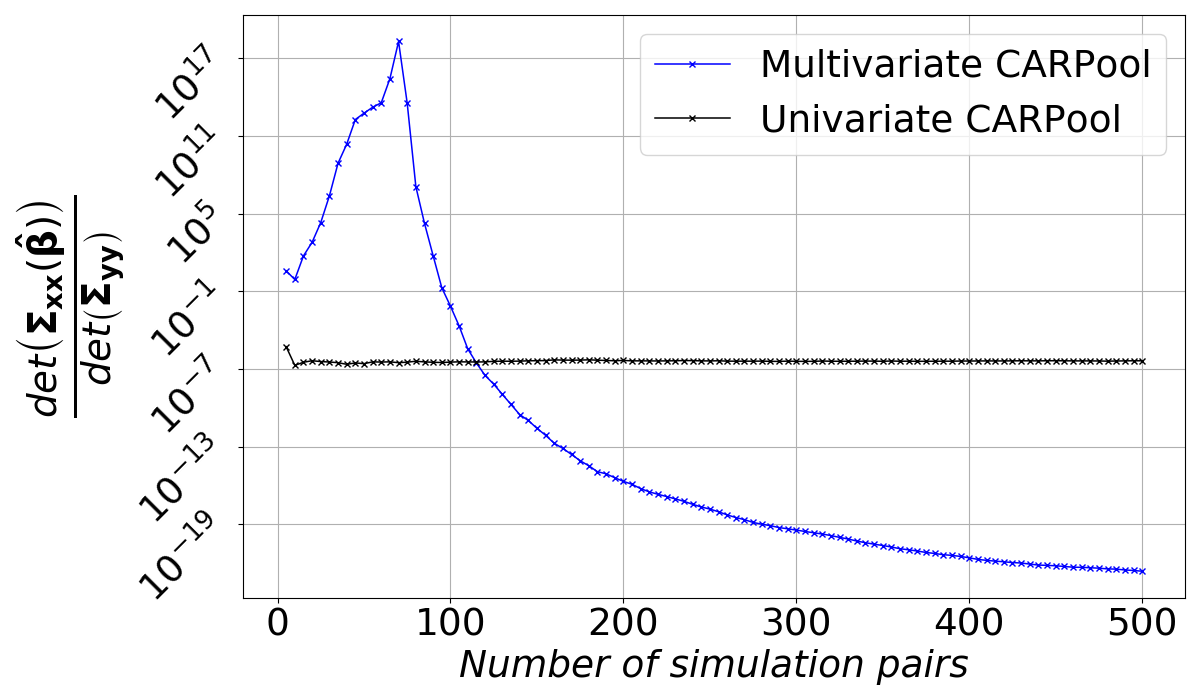}
    \includegraphics[width=0.47\textwidth]{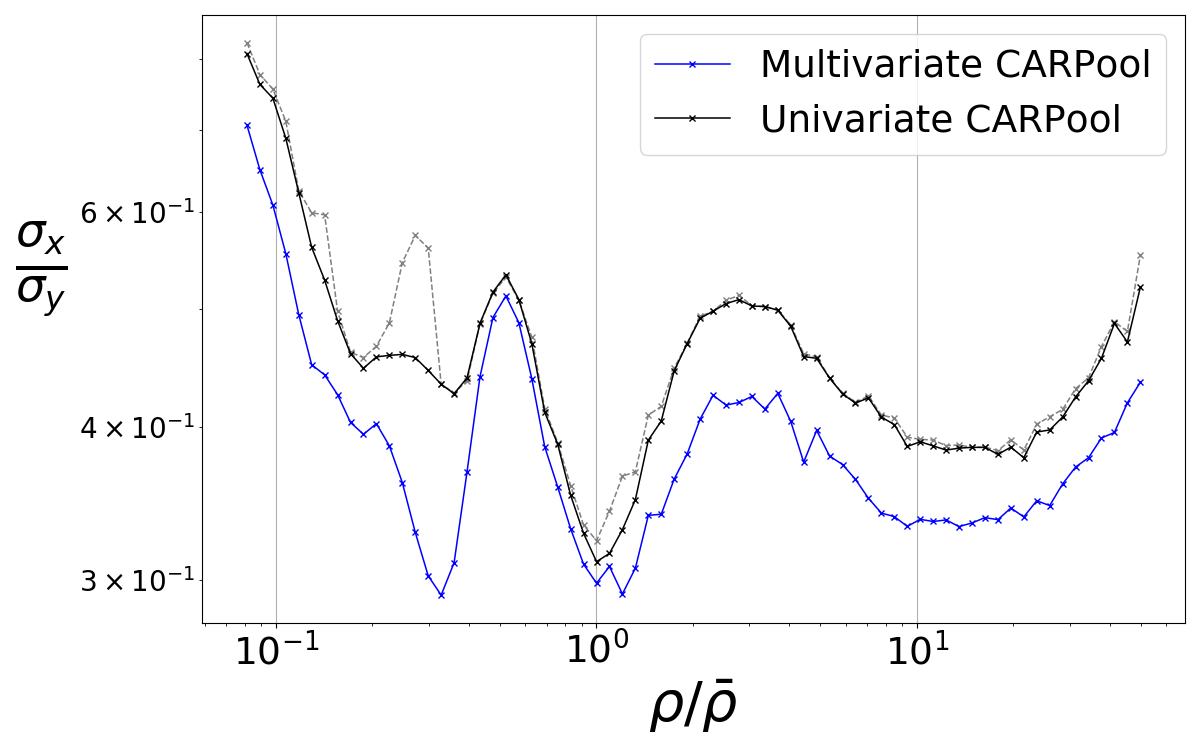}
    \caption{Left panel: Generalised variance ratio of the matter PDF as a function of the number of available simulations. Each $\widehat{\boldsymbol{\beta}}$ and $\widehat{\boldsymbol{\beta^\mathrm{diag}}}$ serves to generate 1,800 samples according to \eqref{eq:mvComp} to estimate the CARPool covariance matrix.
    Right panel: Standard deviation reduction for the PDF bin to expect from CARPool. The blue and black curves respectively use $\widehat{\boldsymbol{\beta}}$ and $\widehat{\boldsymbol{\beta^\mathrm{diag}}}$ estimated with $500$ samples. The dashed grey curve exhibits the actual standard deviation ratio when we have $10$ samples only to compute $\widehat{\boldsymbol{\beta^\mathrm{diag}}}$.
    $\boldsymbol{\Sigma_{yy}}$ is estimated  with all 15,000 available PDFs from the \textit{Quijote} simulations.}
    \label{fig:varPDF}
\end{figure*}

For the matter PDF, we show the empirical variance reduction results before the actual estimates: Figure~\ref{fig:varPDF} shows that  the variance reduction is much milder for the PDF than for the power spectrum or the bispectrum, both for the univariate and multivariate CARPool frameworks. While the multivariate case does eventually lead to  significant gains, CARPool needs $\mathcal{O}(100)$ simulations to learn how to map density contrast in COLA outputs to density contrast in \texttt{GADGET-III} simulations. While COLA places overdense structures close to the right position, their density contrast is typically underestimated, meaning a  level sets of the COLA output is informative about a different level set of the \texttt{GADGET-III} simulation. 

The right panel nonetheless proves that it is possible to reduce the variance of the one-point PDF with CARPool, unlike with paired-fixed fields \citep{Villaescusa_Navarro_2018}.
As for the bispectrum, we took the data outputs of 1,800 additional simulations to compute the covariance matrices intervening in the generalised variance and standard error estimates.

\subsubsection{CARPool versus $N$-body estimates}
\begin{figure}
    \includegraphics[width=0.49\textwidth]{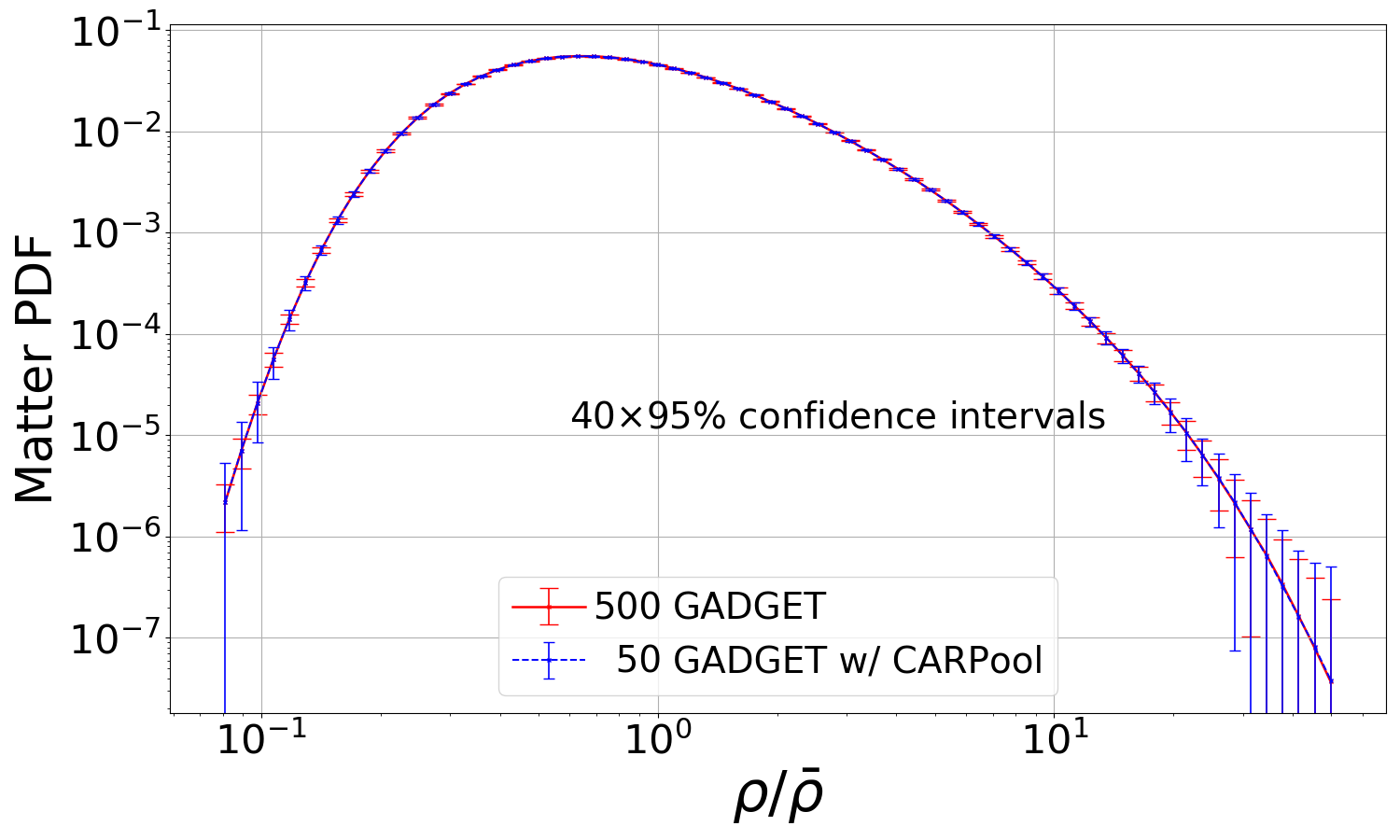}
    \includegraphics[width=0.49\textwidth]{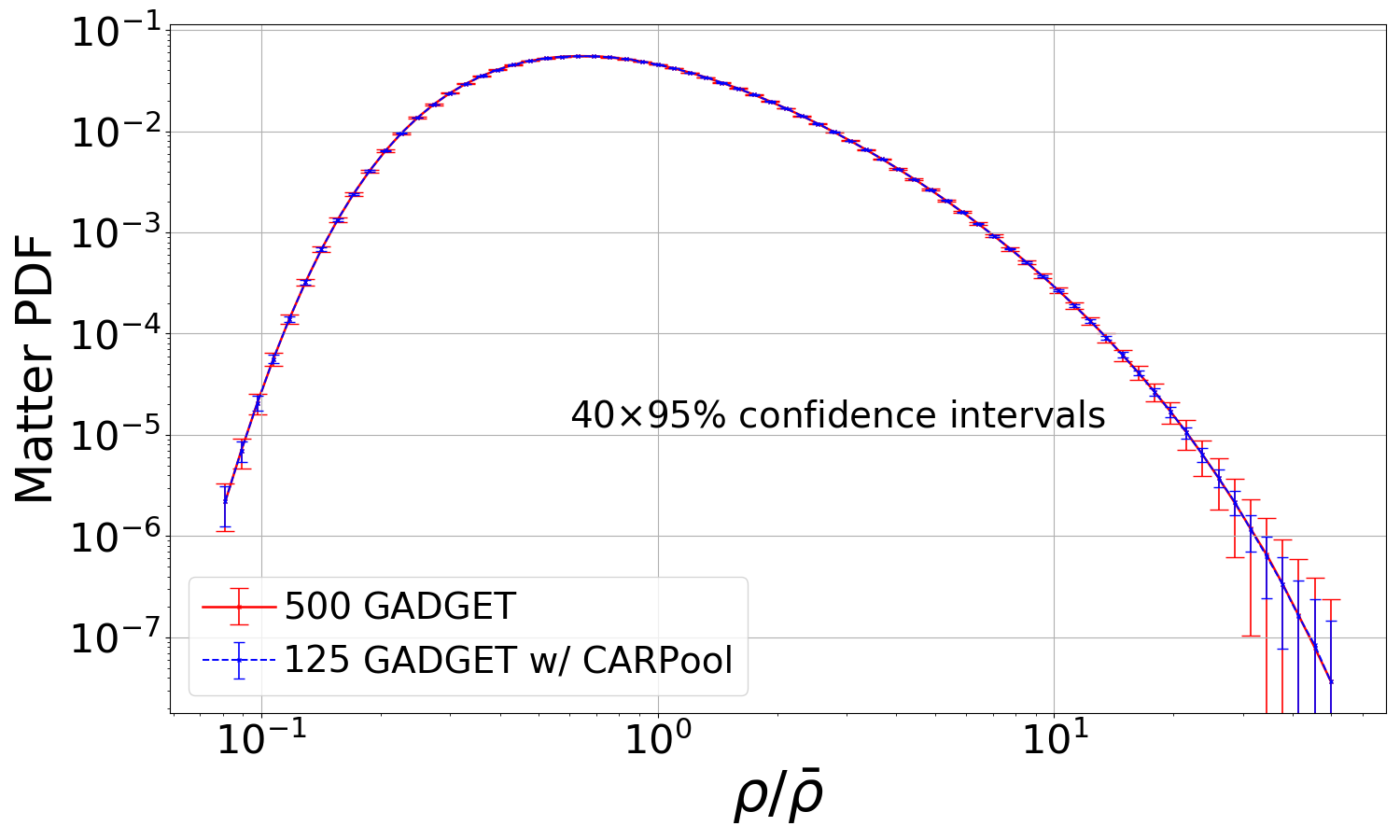}
    \caption{Estimated matter PDF with $500$ $N$-body simulations versus CARPool estimates. $\widehat{\boldsymbol{\beta^\mathrm{diag}}}$ is used in the upper panel whereas the full control matrix is computed in the lower panel. The estimated $95\%$ confidence intervals are computed with the BCa bootstrap. They are enlarged by a factor of $40$ for better visibility.}\label{fig:pdfEst}
\end{figure}

\begin{figure}
    \includegraphics[width=0.49\textwidth]{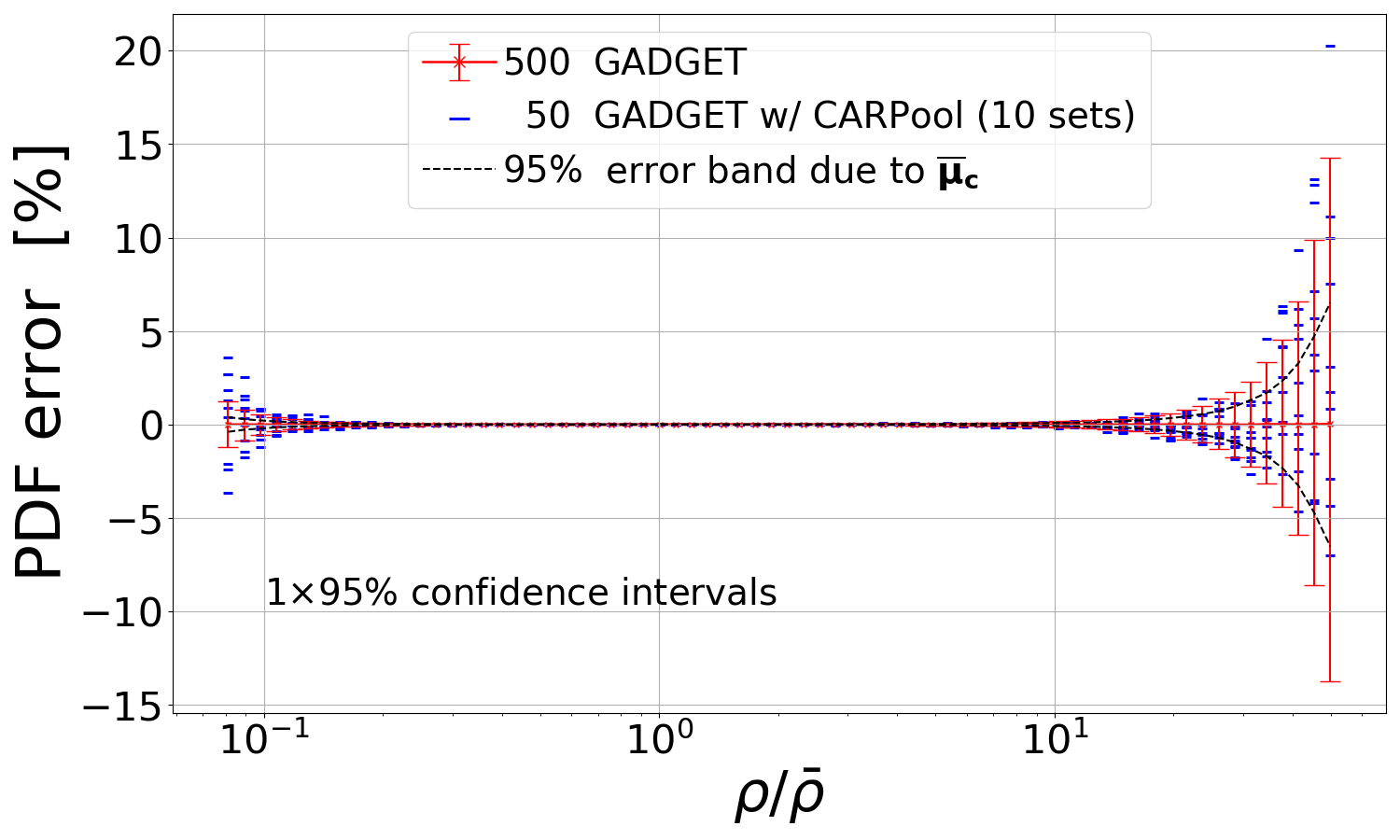}
    \includegraphics[width=0.49\textwidth]{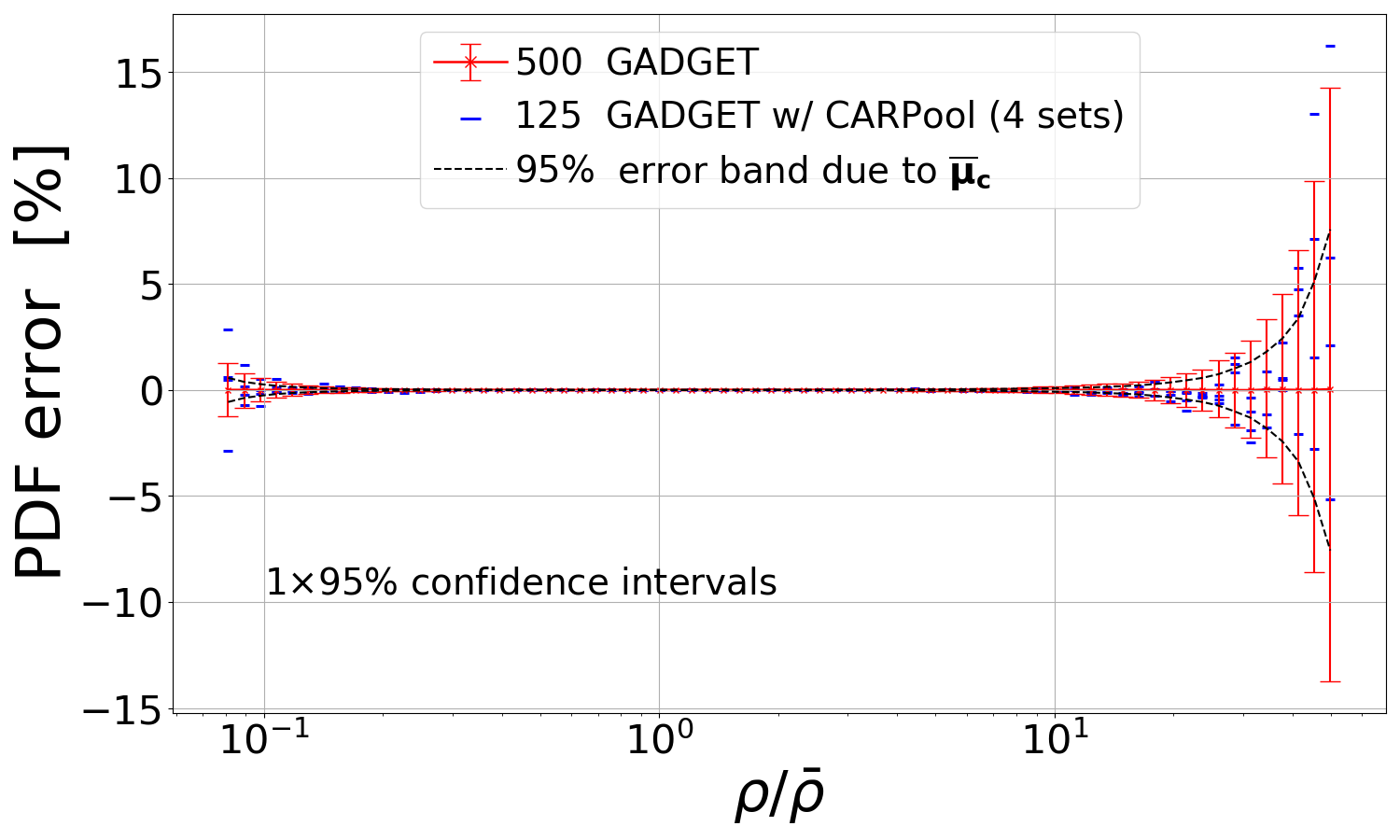}
    \caption{Estimated matter PDF percentage error with respect to 15,000 $N$-body runs: sample mean of 500 $N$-body simulations versus CARPool estimates. In the upper panel, $\widehat{\boldsymbol{\beta^\mathrm{diag}}}$ is used for each set and smoothed by a $5$-bin-wide flat window. In the lower panel, the full control matrix $\widehat{\boldsymbol{\beta}}$ is estimated for each group of seeds. The estimated $95\%$ confidence intervals are plotted for the $N$-body sample mean only, using BCa bootstrap.}\label{fig:pdfRatios}
\end{figure}

For the matter PDF we compare CARPool estimates in both the multivariate and univariate settings. Figures \ref{fig:pdfEst} and \ref{fig:pdfRatios} are paired and show the comparable performance at the tails of the estimated PDF for the smoothed $\widehat{\boldsymbol{\beta^\mathrm{diag}}}$ with $50$ samples on the one hand, and the dense $\widehat{\boldsymbol{\beta}}$ matrix  obtained with $125$  simulations on the other.
We can expect $\mathcal{O}(10^1)$ fewer $N$-body simulations to compute an accurate estimate of the PDF when applying the simple univariate CARPool technique ($50$ instead of $500$ here).
As discussed above, with enough simulations CARPool can learn the mapping between the density contrasts of COLA and  \texttt{GADGET} outputs. Therefore, the matter PDF is a case where the multivariate framework, which involves the estimation of $p \times p$ covariance matrices, shows improvement over the more straightforward univariate case once the number of available simulation pairs passes a threshold.

While we wanted to test the performance of CARPool with minimal tuning, we expect that  with some mild additional assumptions and tuning the univariate CARPool approach could be improved and similar gains to the multivariate case could be obtained with a smaller number of simulations. As an example, one could pre-process the COLA outputs to match the PDF (and power spectrum) of \texttt{GADGET-III} using the approach described in \citet{2013JCAP...11..048L} to guarantee a close correspondence between bins of density contrast.  In addition, a  regularising assumption would be to consider transformations from COLA to \texttt{GADGET-III} density contrasts that are smooth and monotonic.

\subsection{Summary of results}
Here we present a summary of the variance reduction observed in our numerical experiments. With $M=$1,500 additional fast simulations reserved for estimating the cheap mean $\boldsymbol{\bar{\mu}_c}$, and with percentage errors relative to the mean of $15,000$ full $N$-body runs available in \textit{Quijote}, we find:
\begin{itemize}[labelwidth=*,label=\textbullet, font = \color{black} \large]
    \item With only $5$ $N$-body simulations, the univariate CARPool technique recovers the $95$-bin power spectrum up to $k_\mathrm{max} \approx 1.2$ $h {\rm Mpc^{-1}}$ within the $0.5\%$ error band, when the control coefficients are smoothed.
    \item For the bispectrum of $98$ squeezed isosceles triangle configurations, the recovery is within $2\%$ when $5$ $N$-body simulations are available, and $1\%$ when we have $10$ of them, still with the smoothed $\widehat{\boldsymbol{\beta^\mathrm{diag}}}$.
    \item The bispectrum estimator of equilateral triangles on $40$ bins falls within the $2\%$ (resp. $1\%$) error band with $5$ simulations (resp. $10$) at large $k$, and performs better than the mean of $500$ \texttt{GADGET} simulations at large scales.
    \item The standard deviation of matter PDF bins can also be reduced with CARPool, by factors between 3 and 10, implying that the number of required costly simulations is lowered by an order of magnitude.
\end{itemize}

In Appendix \ref{app:collFigs}, we provide the power spectrum and bispectrum results when the CARPool means are computed with $10$ simulation/surrogate pairs instead of the $5$ pairs presented so far.

\section{Discussion and Conclusions}\label{sec:conclusions}

We presented Convergence Acceleration by Regression and Pooling (CARPool), a general scheme for  reducing variance on estimates of large-scale structure statistics. It operates on the idea of forming a combination (pooling) of a small number of accurate simulations  with a larger number of fast but approximate surrogates in such a way as to not introduce systematic error (zero bias) on the combination. The result is equivalent to having run a much larger number of accurate simulations. This apporach is particularly adapted to cosmological applications where our detailed physical understanding has resulted in a number of pertubative and non-perturbative methods to build fast surrogates for high-accuracy cosmological simulations.

To show the operation and promise of the technique, we computed high-accuracy and low-variance predictions for statistics of \texttt{GADGET-III} cosmological $N$-body simulations in the $\Lambda$CDM model at $z=0.5$. A large number of surrogates are available; for illustration we selected the approximate particle mesh solver \texttt{L-PICOLA}.

For three different examples of statistics, the matter power spectrum, the matter bispectrum, and the probability density function of the matter fractional overdensity, CARPool reduces variance by factors 10 to 100 even in the non-linear regime, and by much larger factors on large scales. Using only 5 \texttt{GADGET-III} simulations CARPool is able to compute Fourier-space two-point and three-point functions of the matter distribution at a precision comparable to 500 \texttt{GADGET-III} simulations. 

CARPool requires 1) inexpensive access to surrogate solutions, and 2) strong correlations of the fluctuations about the mean of the surrogate model with the fluctuations of the expensive and accurate simulations. 
By construction, CARPool estimates are unbiased compared to the full simulations no matter how biased the surrogates might be. In all our examples, we achieved substantial variance reductions even though the fast surrogate statistics were highly biased compared to the full simulations.

So far we have presented CARPool as a way to accelerate the convergence of ensemble averages of accurate simulations. An equivalent point of view would be to consider it a method to remove approximation error from ensembles of fast mocks by running a small number of full simulations. Such simulations often already exist, as in our case with the {\it Quijote} simulations, not least because strategies to produce fast surrogates are often tested against a small number of simulations.

In some cases there are opportunities to use CARPool almost for free: for instance, using linear theory from the initial conditions as a surrogate model  has the advantage that $\boldsymbol{\mu_c}$ (the mean linear theory power spectrum) is perfectly known \textit{a priori}. In addition, the de-correlation between linearly and non-linearly evolved perturbations is well-studied, and can be used to set $\boldsymbol{\beta}$. Even for just a single $N$-body simulation, and without the need to estimate $\boldsymbol{\mu_c}$ from an ensemble of surrogates, this would remove cosmic variance on the largest scales better than in our numerical experiments with \texttt{L-PICOLA}, which are limited by the uncertainty of the $\boldsymbol{\mu_c}$ estimate. 

Regardless of the details of the implementation, the reduction of sample variance on observables could be used to avoid having to run ensembles of simulations (or even surrogates) at the full survey volume. This would  simplify  simulation efforts for upcoming large surveys since memory limitations rather than computational time are currently the most severe bottleneck for full-survey simulations \citep{2017ComAC...4....2P}.

In comparison to other methods of variance reduction, CARPool has the main advantage of guaranteeing lack of model error (``bias'') compared to the full simulation.
``Fixing'' \citep{2016PhRvD..93j3519P,2016MNRAS.462L...1A} explicitly modifies the statistics of the generated simulation outputs; which observables are unbiased must be checked on a case-by-case basis, either through theoretical arguments or through explicit simulation \citep{Villaescusa_Navarro_2018}. \citet{2020MNRAS.496.3862K} argue that ``fixed'' field initialisation is unsuitable for simulation suites to estimate accurate covariance matrices, and they are pessimistic about the possibility of generating mock galaxy catalogues solely with this technique.

\citet{2016PhRvD..93j3519P} and \citet{2016MNRAS.462L...1A} also introduce and study the ``pairing'' technique. ``Pairing'' reduces variance for  $k$-space observables (such as the power spectrum) by a factor of $\mathcal{O}(1)$ by combining two simulations whose initial conditions only differ by an overall minus sign, that is they are \textit{phase-flipped}. This technique can be analysed simply in the  control variates framework of CARPool. Consider the phase-flipped simulation as the surrogate for the moment. The mean of an ensemble of phase-flipped simulations is identical to the mean of the unflipped simulations by symmetry. ``Pairing'' then amounts to taking  $\boldsymbol{\beta}=-1$ to cancel off contributions of odd-order terms in the initial conditions \citep{2016PhRvD..93j3519P,2016MNRAS.462L...1A} to reduce variance on the simulation output. Inserting this  $\boldsymbol{\beta}$ in the equation \eqref{eq:scalarCV} and taking the expectation shows that ``pairing'' is an unbiased estimator of the simulation mean.

Other opportunities of exploiting the control variates principle abound; related ideas have been used in the past. As an example, a very recent study \citep{2021MNRAS.500..259S} succeeds in reducing the variance of the quadrupole estimator of the two-point clustering statistic in redshift space. In this case, the variance reduction is achieved by combining different, correlated lines of sight through  the halo catalogue of the  Outer Rim simulation. Though not driven by a general theoretical framework that guarantees unbiasedness and optimal variance reduction, for the specific application at hand their approach  does not require pre-computation of fast surrogates and uses a control matrix set based on physical assumptions.

While we intentionally refrained from tuning CARPool for this first study, there are  opportunities to use physical insight to adapt it for  cosmological applications. For instance, the one-point remapping technique proposed by \citet{2013JCAP...11..048L}, which allows us to increase the cross-correlation between LPT-evolved density fields and full $N$-body simulations, could improve snapshots of a chosen surrogate for CARPool.

In future work we plan to explore intermediate forms of CARPool between the multivariate and univariate versions we study in this paper. Any given entry of $\boldsymbol{y}$ could be predicted by an optimal combination of a small subset of $\boldsymbol{c}$. In this case, the variance reduction could be improved compared to the univariate case while the reduced dimension of the control matrix would ensure a stable estimate using a moderate number of simulations. 

The CARPool setup can be applied to numerous ``$N$-body code plus surrogate'' couples for cosmology. It can be used to make high-resolution corrections to low-resolution simulations, while reducing variance. This will provide an alternative to the procedure suggested by \citet{2014MNRAS.440.1420R}, where the mass resolution effect is estimated by a polynomial fit of the matter power spectrum ratio, and the work of \citet{2015MNRAS.446.1756B}, where a linear transformation of the low-resolution power spectra preserving the mean and variance is smoothed by a polynomial fit. Furthermore, rather than using a single surrogate, taking advantage of multiple low-fidelity methods for variance reduction is also a possibility to explore, especially if the cost of running a large number of surrogates is non-neglible. For instance, taking the linear theory as a second surrogate in addition to \texttt{L-PICOLA} would have strongly reduced the number of \texttt{L-PICOLA} runs required to match the variance of the $\boldsymbol{\mu_c}$ estimate to the massively reduced variance of $\boldsymbol{y}-\boldsymbol{\beta}\left(\boldsymbol{c} - \boldsymbol{\mu_c} \right)$.
In this regard, the multi-fidelity Monte Carlo  scheme of \citet{doi:10.1137/15M1046472} and the approximate control variates framework of \citet{GORODETSKY2020109257} are recent techniques that reduce variance with multiple surrogates for a fixed computational budget. We can also combine CARPool with other techniques. For instance, if the paired-fixed fields initialisation of \citet{2016MNRAS.462L...1A} is found to be unbiased in practice for a particular statistic, then one can combine it with CARPool for further variance reduction.

The simplicity of the theory behind CARPool makes the method attractive for various applications both in and beyond cosmology, as long as the conditions given above are satisfied. Our results suggest that CARPool allows estimating the expectation values of any desired large-scale structure correlators with negligible variances from a small number of accurate simulations, thereby providing a useful complement to analytical approaches such as higher-order perturbation theory or effective field theory.  We are planning to explore a number of these applications in upcoming publications.

\section*{Acknowledgements}
We thank Martin Crocce, Janis Fluri, Cullan Howlett and Hans Arnold Winther for their advice on COLA, and Boris Leistedt for stimulating discussions. We are grateful to Pier-Stefano Corasaniti, Eiichiro Komatsu, Marius Millea, Andrew Pontzen, Yann Rasera and Matias Zaldarriaga for stimulating comments on an earlier version of the manuscript. N.C. acknowledges funding from LabEx ENS-ICFP (PSL). B.D.W. acknowledges support by the ANR BIG4 project, grant ANR-16-CE23-0002 of the French Agence Nationale de la Recherche;  and  the Labex ILP (reference ANR-10-LABX-63) part of the Idex SUPER, and received financial state aid managed by the Agence Nationale de la Recherche, as part of the programme Investissements d'avenir under the reference ANR-11-IDEX-0004-02.
The Flatiron Institute is supported by the Simons Foundation. Y.A. is supported by LabEx ENS-ICFP: ANR-10-LABX-0010/ANR-10-IDEX-0001-02 PSL*. F.V.N acknowledges funding from the WFIRST program through NNG26PJ30C and NNN12AA01C.

\section*{Data availability}
The data underlying this article are available through \textit{globus.org}, and instructions can be found at \url{https://github.com/franciscovillaescusa/Quijote-simulations}. Additionally, a \texttt{Python3} package and code examples are provided at \url{https://github.com/CompiledAtBirth/pyCARPool} to reproduce some results presented in this study.

\bibliographystyle{mnras}
\bibliography{CARPool.bib}

\appendix

\section{Analytical derivation: a Bayesian approach}\label{app:bayesDer}
There is an elegant Bayesian derivation of the optimal form of the control variates estimator for the Gaussian case. The result coincides with the minimum variance estimator even in the non-Gaussian case. As in the derivation by \cite{10.1287/opre.33.3.661}, the covariance matrices of the full simulations $\boldsymbol{y}$ and of the fast simulations $\boldsymbol{c}$ are assumed to be known. In the main text, we use non-parametric approaches to estimate uncertainties since $\boldsymbol{\beta}$ is not known \textit{a priori} but estimated from the same simulations that we use to estimate $\boldsymbol{\mu_y}$.

For notational simplicity, we will use $\boldsymbol{y}$ for the empirical mean of the brute-force simulations, $\boldsymbol{c}$ for the empirical mean of cheap simulations, and $\boldsymbol{t}$ for the target, the unknown mean of $\boldsymbol{y}$. These quantities can be related in a linear model,
\begin{align}
    \boldsymbol{y}&=\boldsymbol{t}+\boldsymbol{\epsilon_y}\,,\\
    \boldsymbol{c}&=\boldsymbol{m}+\boldsymbol{\epsilon_c\,}.
\end{align}
We model the quantities on the right-hand side as
\begin{align}
\boldsymbol{t} &\sim N(\boldsymbol{\mu_y},\boldsymbol{\Sigma_{tt}})\,,\\
\boldsymbol{\epsilon_y}&\sim N(\boldsymbol{0}_p,\boldsymbol{\Sigma_{yy}}/N)\,,\\
\boldsymbol{\epsilon_c}&\sim N(\boldsymbol{0}_p,\boldsymbol{\Sigma_{cc}}/N)\,,\\
\boldsymbol{m}&\sim N(\boldsymbol{\mu_{c}},\boldsymbol{\Sigma_{mm}})\,,
\end{align}
which express, respectively, any prior information on $\boldsymbol{t}$ from previous runs, the noise terms for $\boldsymbol{y}$ and $\boldsymbol{c}$ after averaging over $N$ simulations, and prior information on $\boldsymbol{m}$ from a separate run of fast simulations of $\boldsymbol{c}$. In addition, the basis of our methods is to exploit correlation between the Monte Carlo noise $\boldsymbol{y}$ and $\boldsymbol{c}$, so $\mathrm{cov}(\boldsymbol{y},\boldsymbol{c})\equiv \boldsymbol{\Sigma}_{\boldsymbol{yc}}/N$.

Gathering these together in a single vector gives $\boldsymbol{z}=( \boldsymbol{t},\boldsymbol{y},\boldsymbol{c},\boldsymbol{m})^T$ with multivariate normal density $ p(\boldsymbol{z})=p( \boldsymbol{t},\boldsymbol{y},\boldsymbol{c},\boldsymbol{m})$. This joint vector $\boldsymbol{z}$ is a multivariate Gaussian $N(\boldsymbol{\mu},\boldsymbol{\Sigma})$, where
\begin{align}
    \boldsymbol{\mu}&=\begin{pmatrix}
     \boldsymbol{\mu_y}\\\boldsymbol{\mu_y}\\\boldsymbol{\mu_c}\\\boldsymbol{\mu_c}
    \end{pmatrix}
\end{align}
and
\begin{align}
    C
         &=\begin{pmatrix}
     \boldsymbol{\Sigma_{tt}} & \boldsymbol{\Sigma_{tt}}            & \boldsymbol{0}_{p,p}                 & \boldsymbol{0}_{p,p}  \\
     \boldsymbol{\Sigma_{tt}} & \boldsymbol{\Sigma_{tt}}+\boldsymbol{\Sigma_{yy}}/N & \boldsymbol{\Sigma_{yc}}/N     & \boldsymbol{0}_{p,p}  \\
     \boldsymbol{0}_{p,p}          & \boldsymbol{\Sigma_{yc}}^T/N     & \boldsymbol{\Sigma_{mm}}+\boldsymbol{\Sigma_{cc}}/N & \boldsymbol{\Sigma_{mm}}\\
     \boldsymbol{0}_{p,p}          & \boldsymbol{0}_{p,p}                     & \boldsymbol{\Sigma_{mm}}          & \boldsymbol{\Sigma_{mm}}
         \end{pmatrix}\,.
\end{align}
The diagonal covariances are the block marginals, representing prior information; e.g., $\boldsymbol{\Sigma_{mm}}$ expresses the uncertainty in $\boldsymbol{m}$ obtained from a prior, independent simulation set of the fast surrogate. For that reason $\boldsymbol{\Sigma_{ym}}=\boldsymbol{\Sigma_{tm}}=\boldsymbol{\Sigma_{tc}}=\boldsymbol{0}_{p,p}$.

We are interested in the posterior $p(\boldsymbol{t}|\boldsymbol{y},\boldsymbol{c})$; this expresses the information we have about our target $\boldsymbol{t}$ when we have obtained the set of correlated sample pairs $(\boldsymbol{y},\boldsymbol{c})$. Based on our assumptions, we know the posterior $p(\boldsymbol{t}|\boldsymbol{y},\boldsymbol{c})$ to be Gaussian with mean 
\begin{align}
    \boldsymbol{\mu_{t|y,c}}=\boldsymbol{\mu_y}+&\mqty(\boldsymbol{\Sigma_{tt}} &\boldsymbol{0}_{p,p})
    \mqty(\boldsymbol{\Sigma_{tt}}+\boldsymbol{\Sigma_{yy}}/N & \boldsymbol{\Sigma_{yc}}/N \\
    \boldsymbol{\Sigma_{yc}}^T/N     & \boldsymbol{\Sigma_{mm}}+\boldsymbol{\Sigma_{cc}}/N )^{-1}
    \mqty(\boldsymbol{y}-\boldsymbol{\mu_y}\\\boldsymbol{c}-\boldsymbol{\mu_c})\nonumber\\
    =\boldsymbol{\mu_y}+&\boldsymbol{\Sigma_{tt}}\qty[\boldsymbol{\Sigma_{tt}}+\frac1N\qty(\boldsymbol{\Sigma_{yy}}-\boldsymbol{\Sigma_{yc}}\qty(N\boldsymbol{\Sigma_{mm}}+\boldsymbol{\Sigma_{cc}})^{-1}\boldsymbol{\Sigma_{yc}}^T)]^{-1}\nonumber\\
&\quad\qty((\boldsymbol{y}-\boldsymbol{\mu_y})-\boldsymbol{\Sigma_{yc}}\qty(N\boldsymbol{\Sigma_m}+\boldsymbol{\Sigma_{cc}})^{-1}(\boldsymbol{c}-\boldsymbol{\mu_c}))
\end{align}
and covariance 
\begin{align}
    \boldsymbol{\Sigma_{t|y,c}}&=  \boldsymbol{\Sigma_{tt}} -\mqty(\boldsymbol{\Sigma_{tt}} &\boldsymbol{0}_{p,p})
    \mqty(\boldsymbol{\Sigma_{tt}}+\boldsymbol{\Sigma_{yy}}/N & \boldsymbol{\Sigma_{yc}}/N \\
    \boldsymbol{\Sigma_{yc}}^T/N     & \boldsymbol{\Sigma_{mm}}+\boldsymbol{\Sigma_{cc}}/N )^{-1}
    \mqty(\boldsymbol{\Sigma_{tt}} \\\boldsymbol{0}_{p,p})\nonumber\\
   &=\boldsymbol{\Sigma_{tt}}-\boldsymbol{\Sigma_{tt}} \qty[\boldsymbol{\Sigma_{tt}}+\frac1N\qty(\boldsymbol{\Sigma_{yy}}-\boldsymbol{\Sigma_{yc}}\qty(N\boldsymbol{\Sigma_{mm}}+\boldsymbol{\Sigma_{cc}})^{-1}\boldsymbol{\Sigma_{yc}}^T)]^{-1}\boldsymbol{\Sigma_{tt}}\nonumber\\
   &= \qty[\boldsymbol{\Sigma_{tt}}^{-1}+ N\qty(\boldsymbol{\Sigma_{yy}}- \boldsymbol{\Sigma_{yc}}\qty( N\boldsymbol{\Sigma_{mm}}+\boldsymbol{\Sigma_{cc}} )^{-1}
    \boldsymbol{\Sigma_{yc}}^T )^{-1}]^{-1}\,.
   \label{eq:fullresult}
\end{align}
These results generalise the earlier equations by 1) including $\boldsymbol{\Sigma_{mm}}$, the error estimate of $\boldsymbol{\mu_c}$ from a prior run of fast simulations, 2) allowing for information from previous runs to be included by specifying prior mean $\boldsymbol{\mu_t}$ and prior covariance $\boldsymbol{\Sigma_{tt}}$, and 3) giving analytical uncertainty estimates for the accelerated estimates.

To make contact with  equation \eqref{eq:sampleCV} we will consider special cases of this expression. Without prior information on $\boldsymbol{\mu_y}$ (i.e. $\boldsymbol{\Sigma_{tt} }\rightarrow\infty$) we obtain
\begin{equation}
    \boldsymbol{\mu_{t|y,c}}=\boldsymbol{y}-\frac1N\boldsymbol{\Sigma_{yc}}\qty(\boldsymbol{\Sigma_{mm}}+\frac1N\boldsymbol{\Sigma_{cc}})^{-1}(\boldsymbol{c}-\boldsymbol{\mu_c}) 
    \qq{(no prior on $\boldsymbol{y}$)}
\end{equation}
and 
\begin{equation}
   \boldsymbol{\Sigma_{t|y,c}}=
    + \frac1N\qty(\boldsymbol{\Sigma_{yy}}- \boldsymbol{\Sigma_{yc}}\qty( N\boldsymbol{\Sigma_{mm}}+\boldsymbol{\Sigma_{cc}})^{-1}
    \boldsymbol{\Sigma_{yc}}^T )
    \qq{(no prior on $\boldsymbol{y}$)\,.}
\end{equation}

For the case where the error on $\boldsymbol{m}$ can be neglected (i.e. $\boldsymbol{\Sigma_{mm}}\rightarrow \boldsymbol{0}_{p,p}$) but prior information is included, we obtain
\begin{align}
    &\boldsymbol{\mu}_{\boldsymbol{t}|\boldsymbol{y},\boldsymbol{c},\boldsymbol{\Sigma_{mm}}\rightarrow\boldsymbol{0}_p}\nonumber\\ &\quad=\boldsymbol{\mu_y}+\boldsymbol{\Sigma_{tt}}\qty(\boldsymbol{\Sigma_{tt}}+ \boldsymbol{\Sigma}_{\boldsymbol{t}|\boldsymbol{y},\boldsymbol{c},\boldsymbol{\Sigma_{mm}}\rightarrow\boldsymbol{0}_{p,p}})^{-1}\nonumber\\
&\qquad\qty((\boldsymbol{y}-\boldsymbol{\mu_y})-\boldsymbol{\Sigma_{yc}\Sigma_{cc}}^{-1}(\boldsymbol{c}-\boldsymbol{\mu_c}))
 \qq{($\boldsymbol{\mu_c}$ known)}
\end{align}
and
\begin{align}
   &\boldsymbol{\Sigma}_{\boldsymbol{t}|\boldsymbol{y},\boldsymbol{c},\boldsymbol{\Sigma_m}\rightarrow\boldsymbol{0}_{p,p}}\nonumber\\ &\quad=
   \qty[\boldsymbol{\Sigma_{tt}}^{-1}+ N\qty(\boldsymbol{\Sigma_{yy}}- \boldsymbol{\Sigma_{yc}\Sigma_c}^{-1}
    \boldsymbol{\Sigma_{yc}}^T )^{-1}]^{-1}
    \qq{($\boldsymbol{\mu_c}$ known)\,.}
\end{align}

In the absence of prior information and assuming that $\boldsymbol{\mu_c}$ is perfectly known (i.e. $\boldsymbol{\Sigma_{tt}}\rightarrow\infty$ and $\boldsymbol{\Sigma_{mm}}\rightarrow \boldsymbol{0}_{p,p} $), equation \eqref{eq:fullresult} simplifies to match the result of equation \eqref{eq:sampleCV} from \citet{10.1287/opre.33.3.661},
\begin{align}
    &\boldsymbol{\mu}_{\boldsymbol{t}|\boldsymbol{y},\boldsymbol{c}}= \boldsymbol{y}-\boldsymbol{\Sigma_{yc}\Sigma_{cc}}^{-1}(\boldsymbol{c}-\boldsymbol{\mu_c}))
 \qq{($\boldsymbol{\mu_c}$ known, no prior on $\boldsymbol{y}$)}
\end{align}
and 
\begin{equation}
   \boldsymbol{\Sigma}_{\boldsymbol{t}|\boldsymbol{y},\boldsymbol{c}}=
    \frac1N\qty(\boldsymbol{\Sigma_{yy}}-\boldsymbol{\Sigma_{yc}}\qty(\boldsymbol{\Sigma_{cc}} )^{-1}
    \boldsymbol{\Sigma_{yc}}^T) 
    \qq{($\boldsymbol{\mu_c}$ known, no prior on $\boldsymbol{y}$)\,.}
\end{equation}

\section{Additional insight on results and confidence intervals}\label{app:collFigs}
We start with a reminder about confidence intervals. The ``one-sigma rule of thumb'' (same for two and three) is a direct application of the {\it central limit theorem} (CLT) when estimating a random variable with the sample mean of $N$ realisations,
\begin{equation}
    \bar{y} \pm \gamma \frac{\widehat{\sigma_y}}{\sqrt{N}}\,, \label{eq:confInt}
\end{equation}
where $\gamma$ is the z-score---e.g. from a normal distribution---associated to a given confidence band. The $95\%$ symmetric confidence intervals correspond to $\gamma \approx 1.96$, hence the name ``two-sigma rule.'' With a very small number of samples, the CLT is not really ``working,'' so it is common practice to penalise the confidence intervals by taking $\gamma$ from a $t$-score table, i.e. from a Student distribution with $N-1$ degrees of freedom, which has fatter tails. For instance, for $N=10$, $\gamma \approx 2.26$ for the $95\%$ confidence band.
\begin{figure}
    \includegraphics[width=\columnwidth]{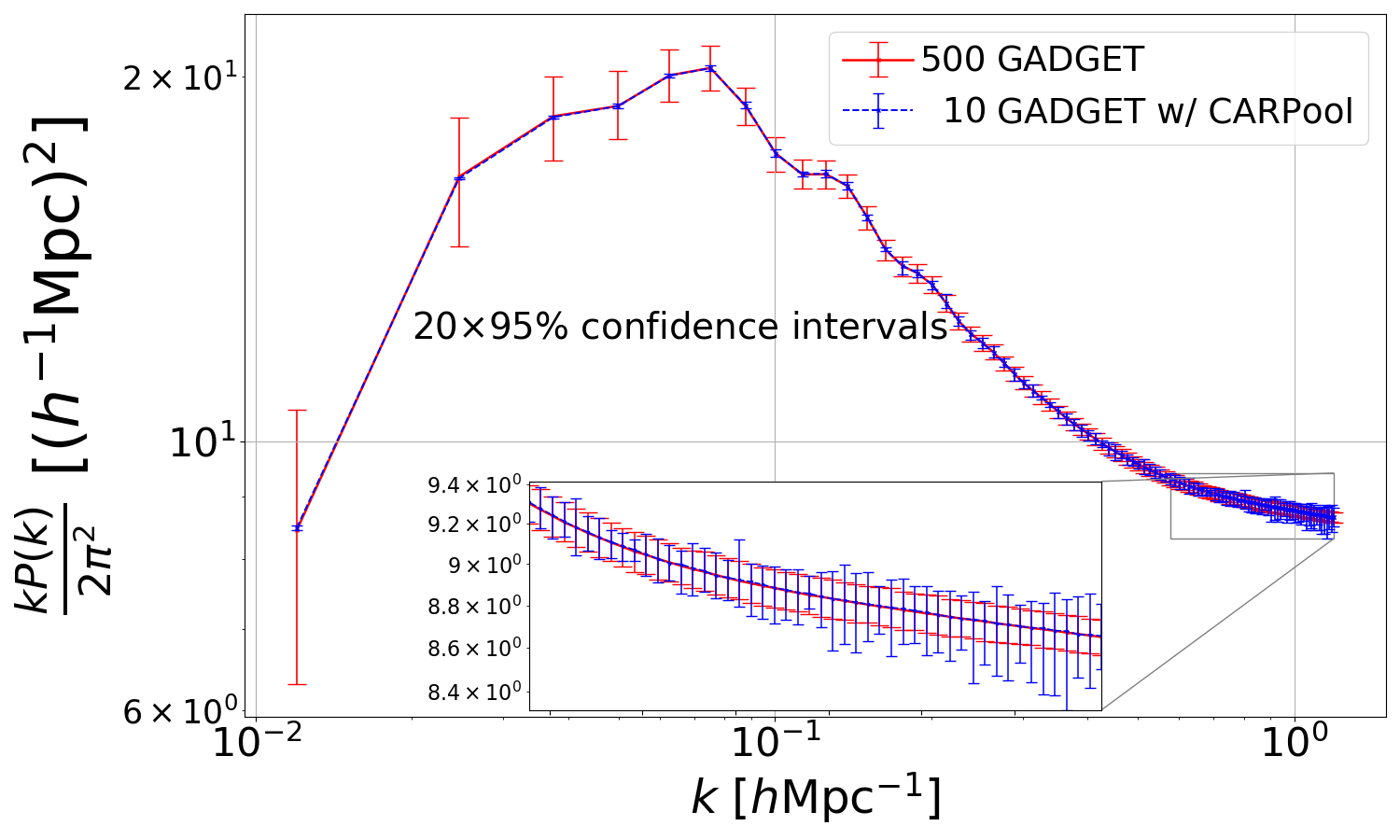}
    \caption{As in Figure \ref{fig:pkEst5v500}, but with $10$ $N$-body simulations used for the CARPool estimate. \label{fig:pkEst10v500}}
\end{figure}

\begin{figure}
    \includegraphics[width=0.49\textwidth]{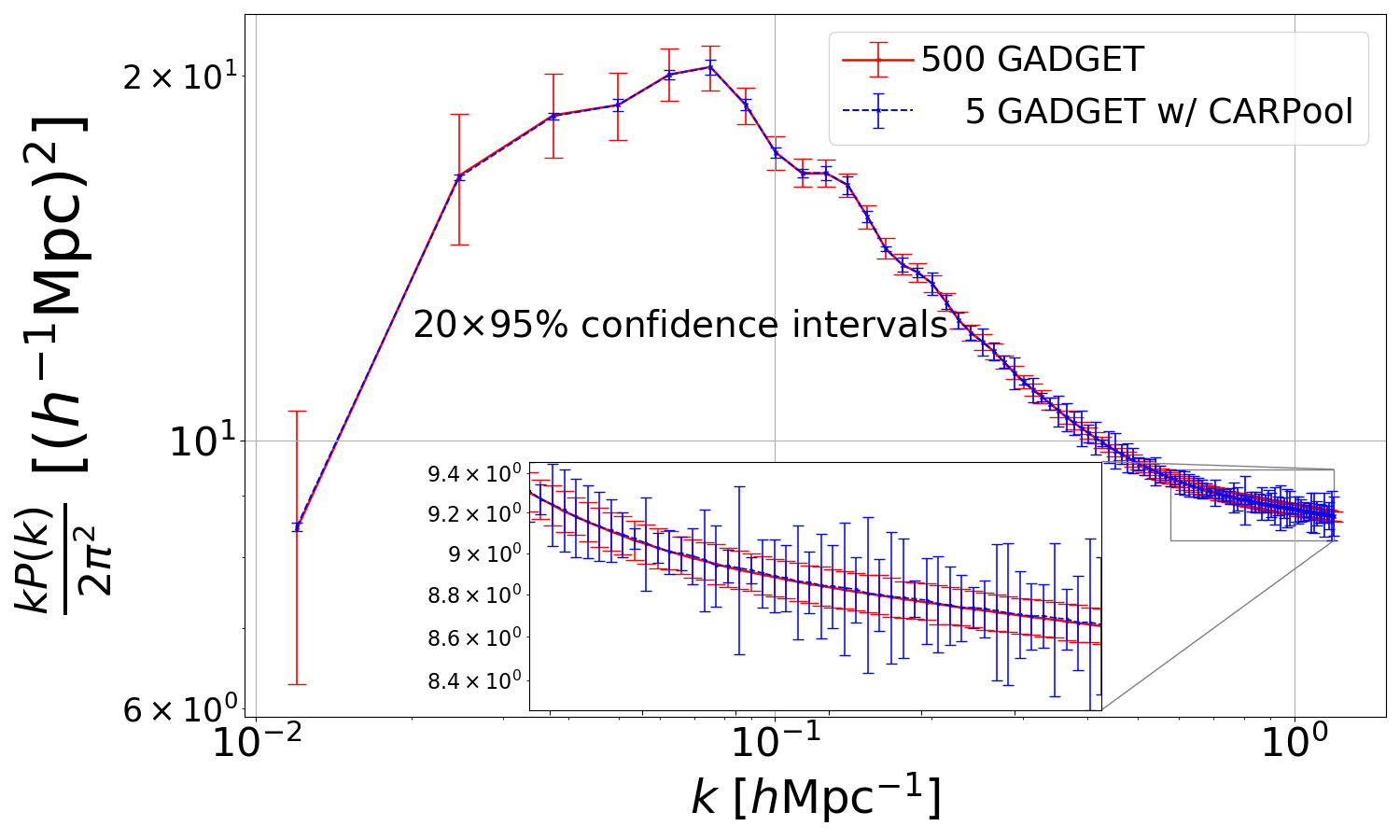}
    \includegraphics[width=0.49\textwidth]{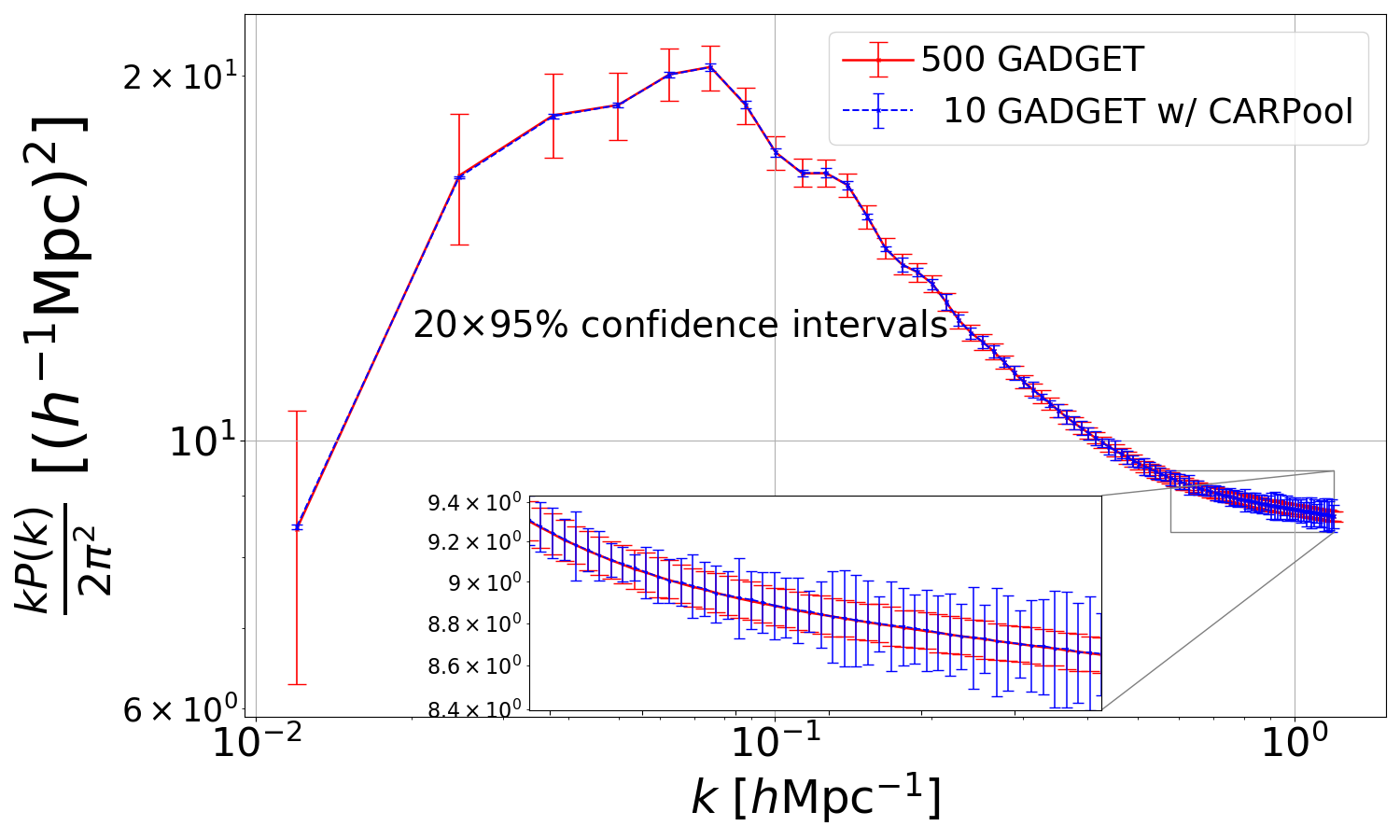}
    \caption{The upper panel shows the same data as in Figure \ref{fig:pkEst5v500} and the lower panel is paired with Figure \ref{fig:pkEst10v500}, except that the confidence intervals come from $t$-score values with $4$ and $9$ degrees of freedom, respectively.\label{fig:pkEstT5-10v500}}
\end{figure}

\begin{figure}
    \includegraphics[width=\columnwidth]{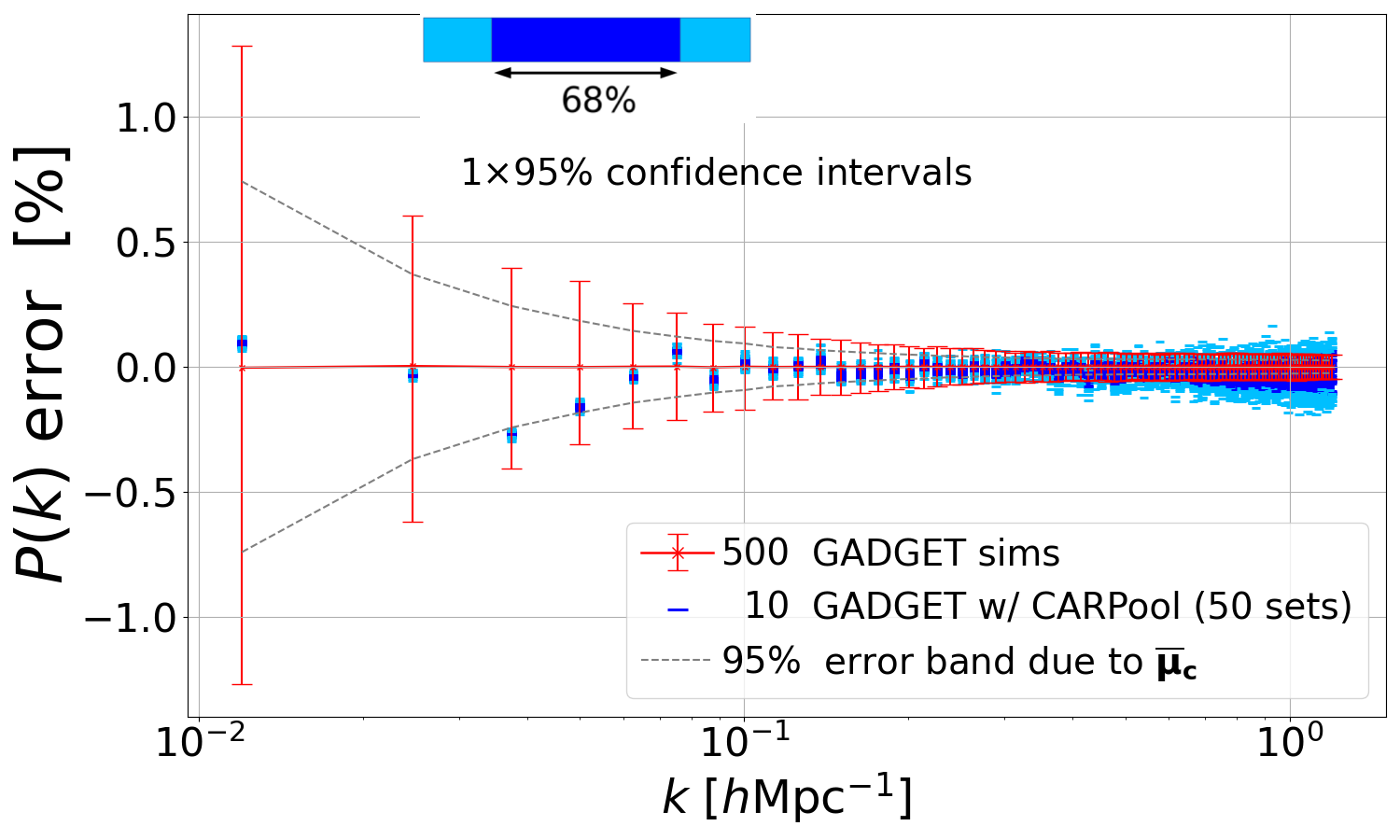}
    \caption{As in the lower panel of Figure \ref{fig:pkRatios5v500}, but with $10$ $N$-body simulations used for the CARPool estimate\label{fig:pkRatios10v500}}
\end{figure}

\begin{figure}
    \includegraphics[width=\columnwidth]{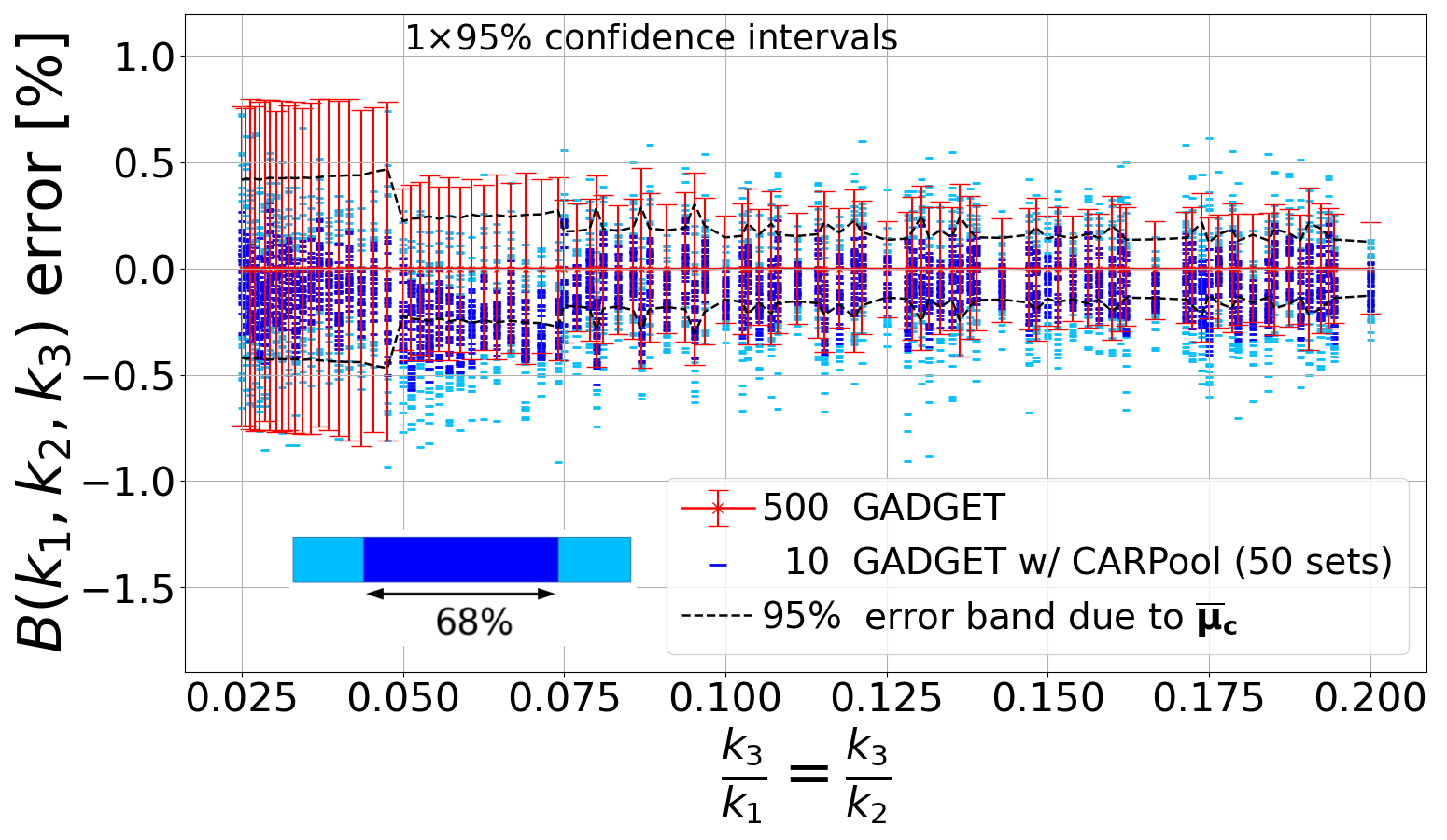}
    \includegraphics[width=\columnwidth]{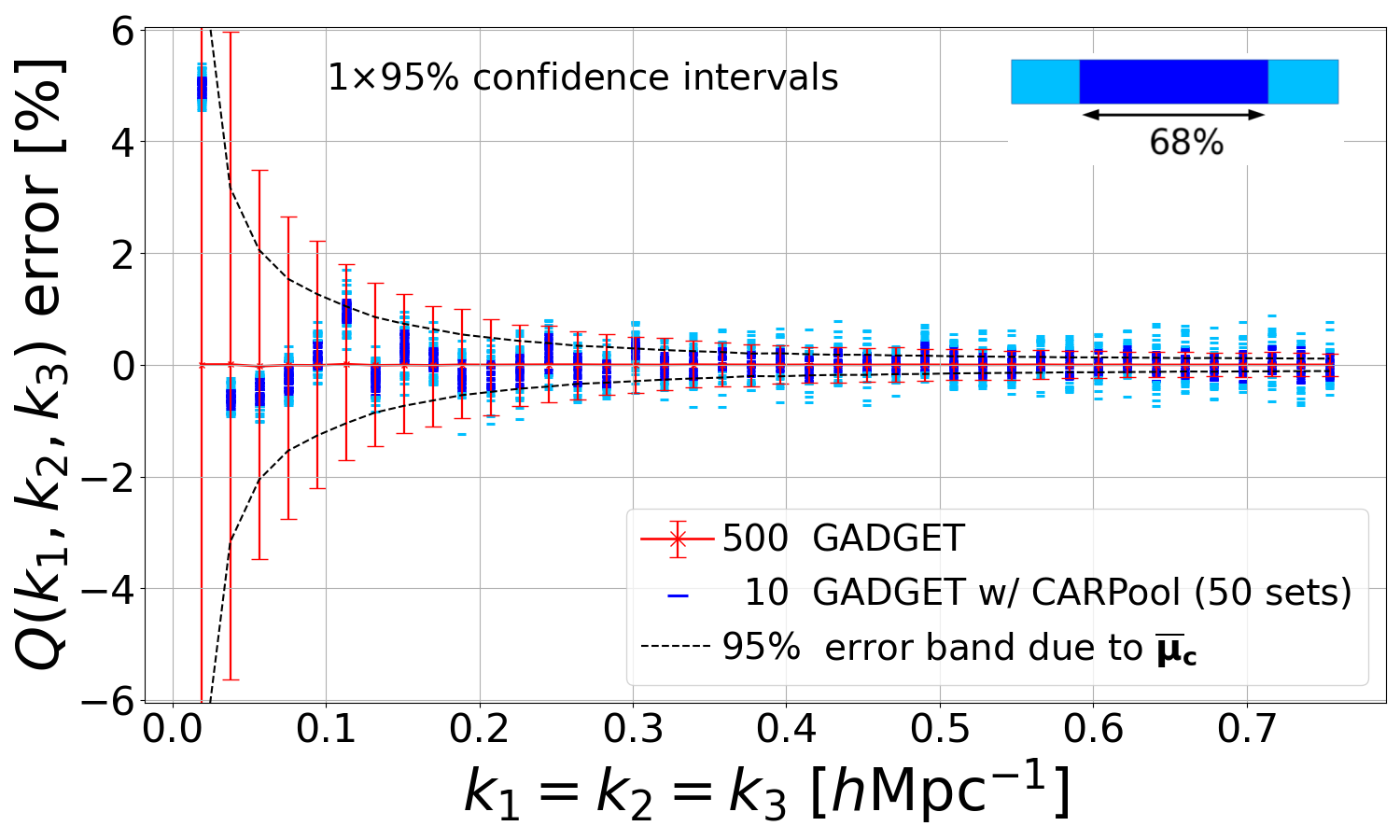}
    \caption{As in Figure \ref{fig:bkQkRatiosSmF5v500}, but with $10$ $N$-body simulations used for the CARPool estimate.\label{fig:bkQkRatiosSmF10v500}}
\end{figure}

Because the trustworthiness of confidence intervals for a sample mean with very few realisations is debatable, we provide here, by way of an example for the power spectrum only, Figure \ref{fig:pkEst10v500} with bootstrap confidence intervals of $10$ CARPool samples and Figure \ref{fig:pkEstT5-10v500} for CARPool with $5$ and $10$ $N$-body simulations but with $t$-score intervals accordingly to equation \eqref{eq:confInt}. The latter figure is to compare with Figures \ref{fig:pkEst5v500} and \ref{fig:pkEst10v500} (exact same data except for the blue CARPool confidence intervals). We have agreement between the paired plots, and we notice that the symmetric confidence intervals from $t$-score tend to be larger.
Additionally, for the two- and three-point clustering statistics, we present in Figures \ref{fig:pkRatios10v500} (power spectrum) and \ref{fig:bkQkRatiosSmF10v500} (bispectrum) the percentage error of CARPool means with $10$ simulations that are not shown in the main part of the paper.

\begin{figure*}
    \includegraphics[width=0.49\textwidth]{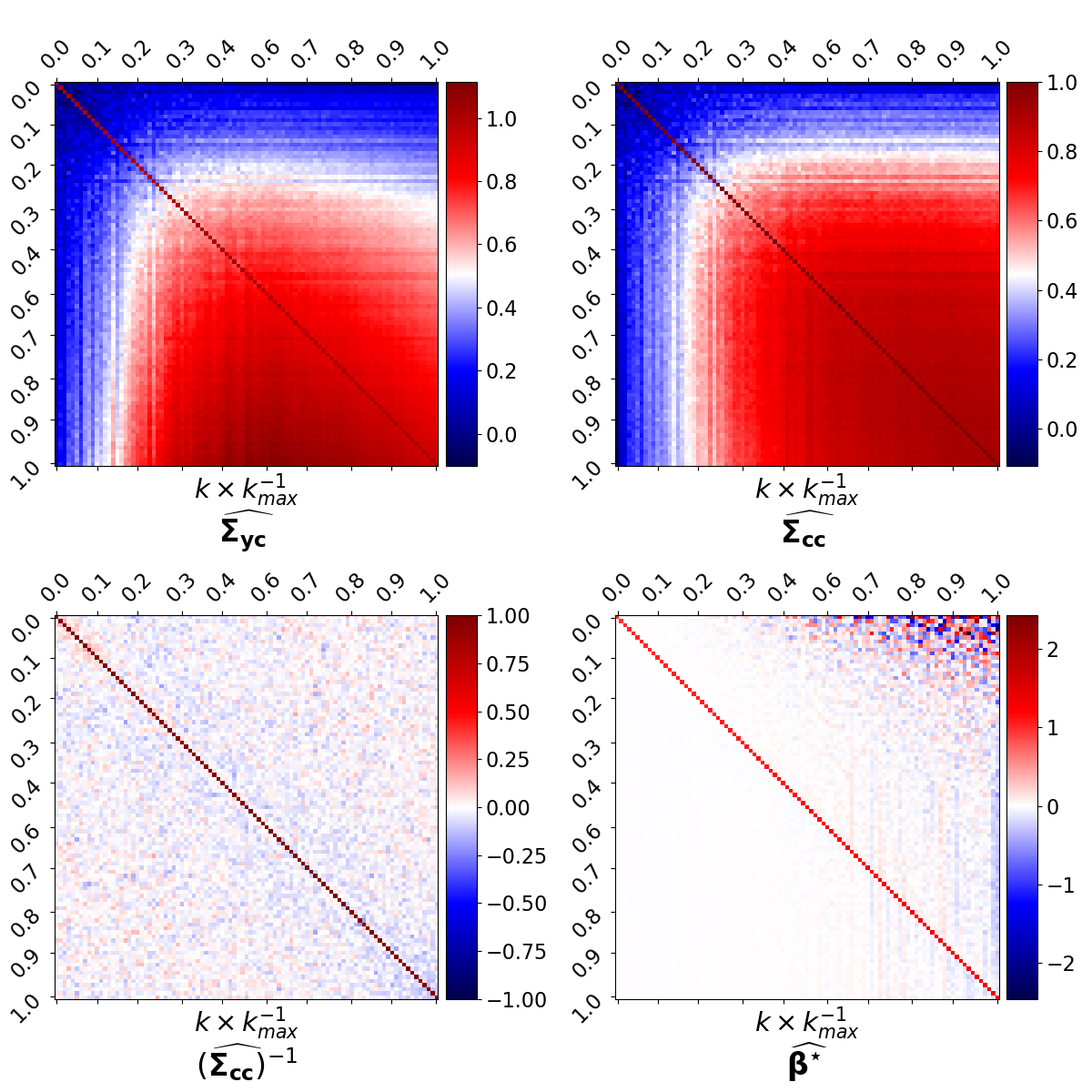}
    \includegraphics[width=0.49\textwidth]{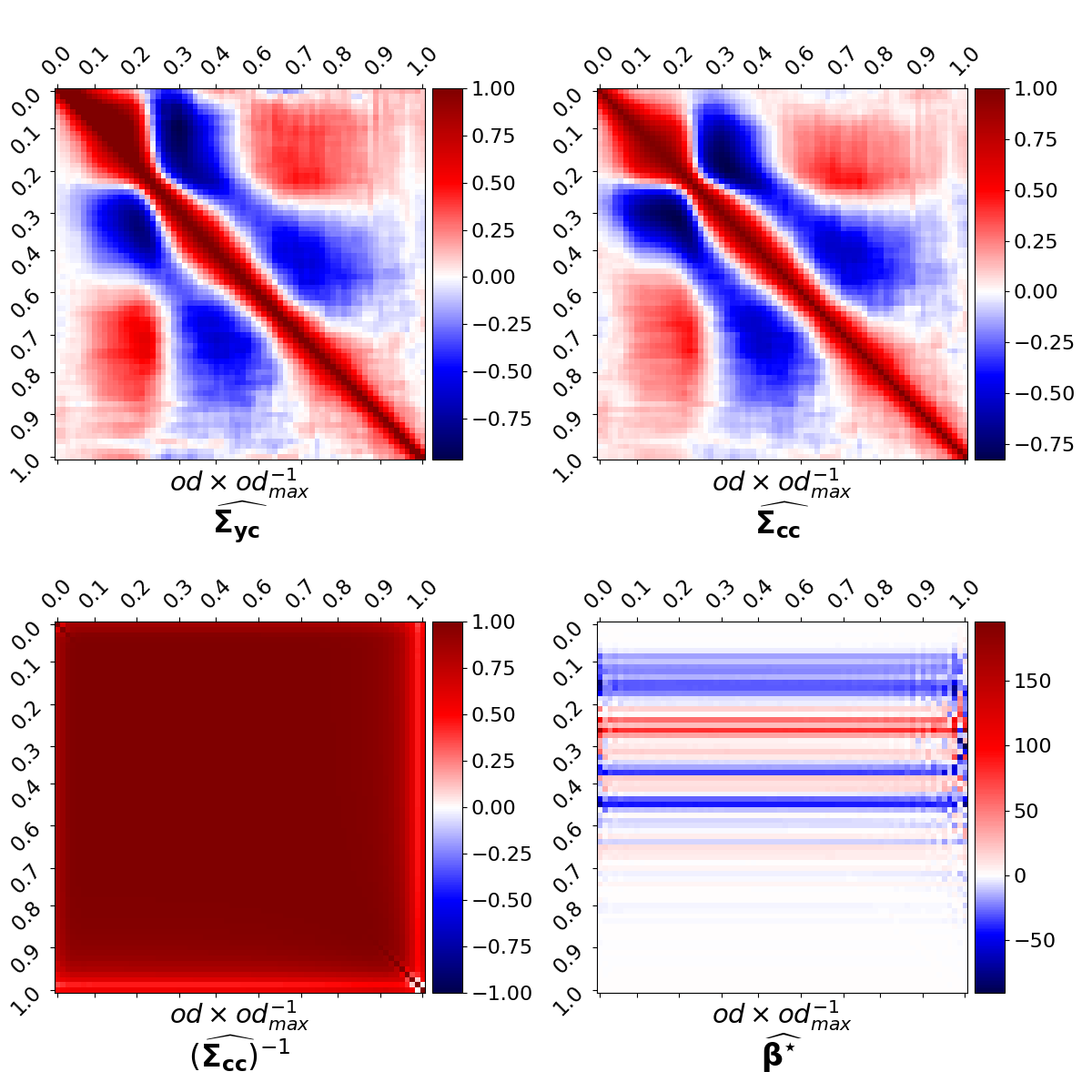}
    \caption{Estimated matrices intervening in $\boldsymbol{\beta^{\star}}$ for the matter power spectrum (left) and the matter PDF (right). The cross-covariance, covariance and precision matrices are normalised, i.e. we display $\boldsymbol{D}
    ^{-1}\boldsymbol{\widehat{\Sigma}}\boldsymbol{D}
    ^{-1}$ with $\boldsymbol{D} = \sqrt{\mathrm{diag}\left( {\boldsymbol{\widehat{\Sigma}}}\right)}$.
    ``od'' denotes the fractional overdensity bin $\rho/\bar{\rho}$. For better visibility, the diverging color scale is not forced to be centered at $0.0$ for the $\boldsymbol{\Sigma_{yc}}$ and $\boldsymbol{\Sigma_{cc}}$ estimates in the upper left corner (power spectrum). All matrices are estimated using $500$ simulation pairs, and represent the ``close to optimal'' $\boldsymbol{\beta^{\star}}$ towards which the control matrix estimator tends in the multivariate setting.} 
    \label{fig:controlMat}
\end{figure*}

We provide also in Figure \ref{fig:controlMat} an overview of the optimal control matrix $\boldsymbol{\beta^{\star}}$ from equation \eqref{eq:betaStarMV} for the matter power spectrum and matter PDF test cases.

\section{COLA timestepping and cross-correlation coefficients}\label{app:cola}
We briefly explain our choice of timestepping strategy to generate a collection of low-fidelity snapshots at $z=0.5$. In COLA, the cosmological scale factor $a$ is used to discretise the time derivative of the left-hand side of the COLA equation of motion \eqref{eq:colaEOM},
\begin{equation}
    \begin{aligned}
    \boldsymbol{v}_{i+\frac{1}{2}} &= \boldsymbol{v}_{i-\frac{1}{2}} - \Delta a_1 \partial_a^2 \boldsymbol{\Psi_\mathrm{res}}\,, \\
    \boldsymbol{r}_{i+\frac{1}{2}} &= \boldsymbol{r}_i + \boldsymbol{v}_{i+\frac{1}{2}}\Delta a_2 + \Delta D_1 \boldsymbol{\Psi_1} + \Delta D_2 \boldsymbol{\Psi_2}\,.
    \end{aligned}
\end{equation}
$\Delta D_l = D_{l, i+1} - D_{l,i}$ with $l \in \left\{1,2 \right\}$ are the changes of linear (or Zel'dovich) and second-order growth factors between the timesteps, normalised such that $D_1(a=1)=D_2(a=1)=1$. $\boldsymbol{\Psi_1}$ and $\boldsymbol{\Psi_2}$ are, respectively, the linear (or Zel'dovich) and second-order LPT (or 2LPT) displacement fields at $a=1$. We have enabled the timestepping scheme from \citet{Tassev_2013} in which the time intervals $\Delta a_i, i \in \left\{ 1,2\right\}$ are given by
\begin{equation}
    \begin{aligned}
    \Delta a_1 &= \frac{H_0}{nLPT} \frac{a_{i + \frac{1}{2}}^{nLPT} -a_{i - \frac{1}{2}}^{nLPT} }{a_i^{nLPT-1}}\,, \\
    \Delta a_2 &= \frac{H_0}{a_{i + \frac{1}{2}}^{nLPT}} \int_{a_i}^{a_{i+1}} \frac{a^{nLPT-3}}{H(a)} da\,.
    \end{aligned}
\end{equation}
Here, $nLPT$ is an additional free parameter which should be tuned experimentally for every simulation setting, as \citet{Tassev_2013}, \citet{Howlett_2015} and \citet{Izard_2016} already stressed. The Kick-and-Drift/Leapfrog algorithm of \citet{1997astro.ph.10043Q} can also be used in \texttt{L-PICOLA}.

Before generating our ensemble of fast surrogates, we tested the sensitivity of the cross-correlation coefficients $\zeta_{yc}$ between the full $N$-body dark matter density contrast field $\boldsymbol{\delta_y}$ and $\boldsymbol{\delta_c}$ produced by \texttt{L-PICOLA},
\begin{equation}\label{eq:crossP}
    \zeta_{yc} = \frac{\mathbb{E}\left[ \delta_{y}(\boldsymbol{k}) \delta_{c}(\boldsymbol{k})^{\ast} \right]}{\sqrt{\mathbb{E} \left[ \mid \delta_{y}(\boldsymbol{k})\mid^2 \right] \mathbb{E} \left[ \mid \delta_{c}(\boldsymbol{k})\mid^2 \right]}} = \frac{P_{yc}(\boldsymbol{k})}{\sqrt{P_y(\boldsymbol{k})P_c(\boldsymbol{k})}}\,,
\end{equation}
to the choice of timestepping.

\begin{figure*}
    \includegraphics[width=0.49\textwidth]{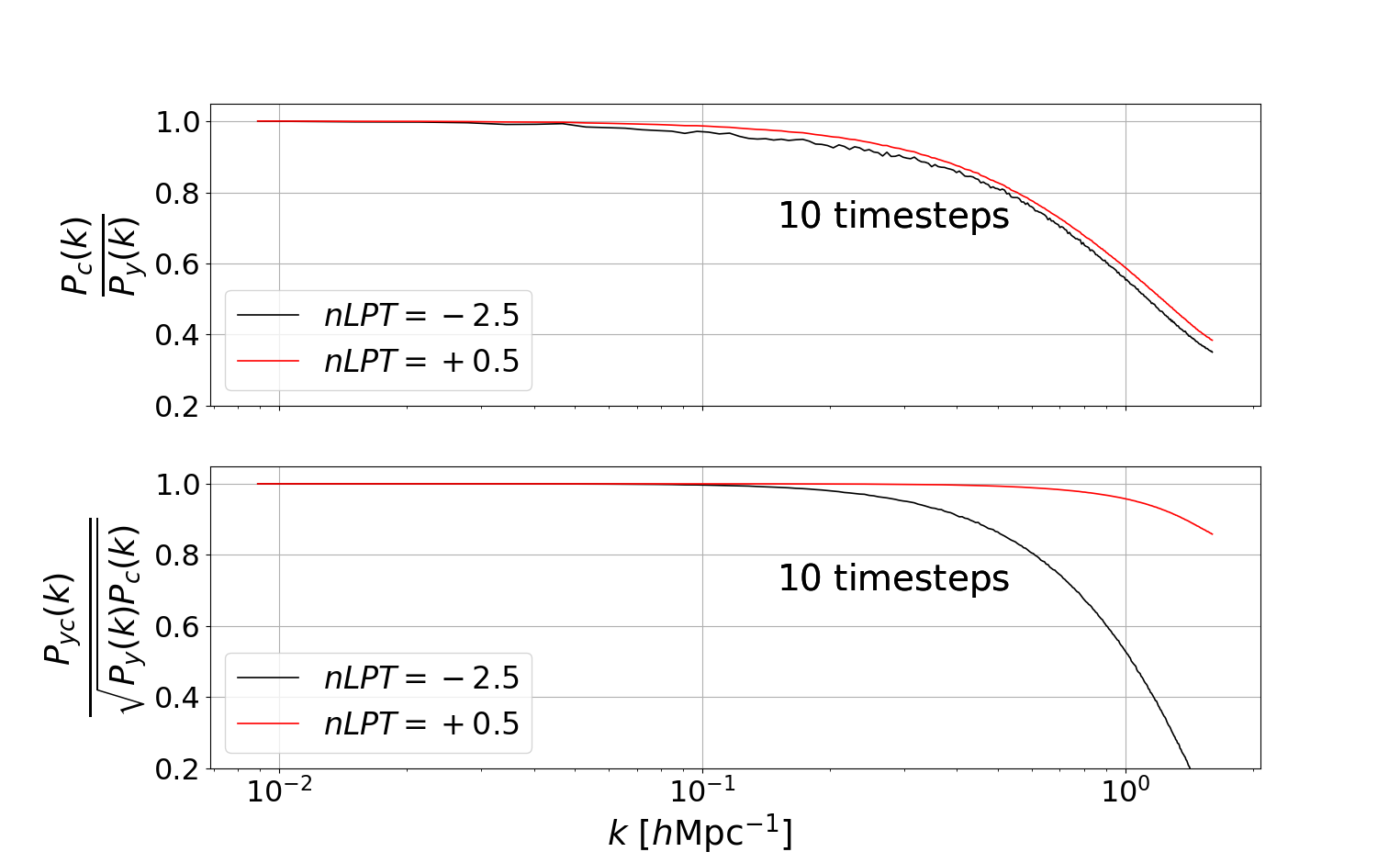}
    \includegraphics[width=0.49\textwidth]{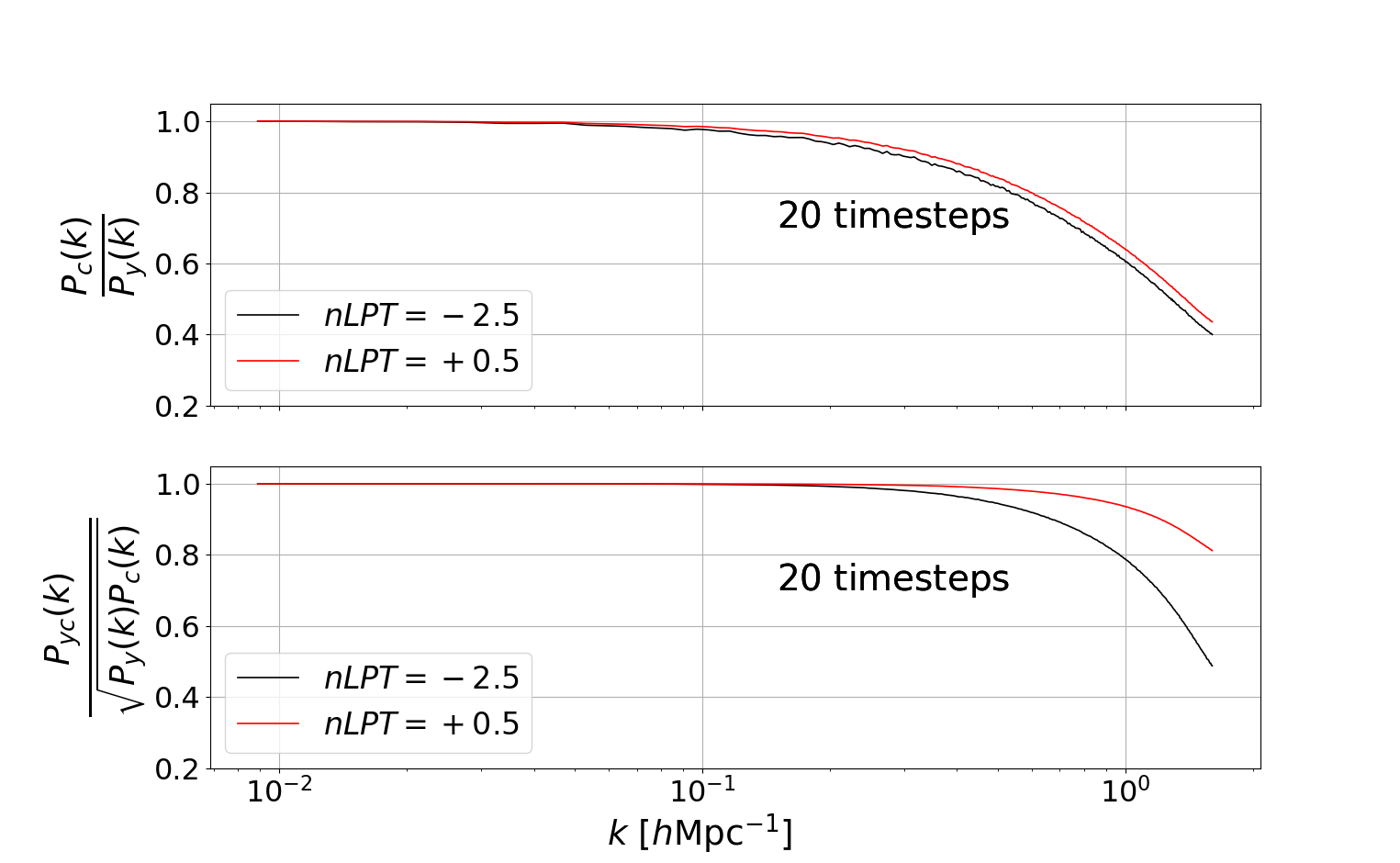}
    \caption{Power spectrum recovery ratio (top) and cross power spectrum coefficients (bottom) at $z=0.5$ between a specific \texttt{L-PICOLA} snapshot computed with $10$ (left) and $20$ (right) linearly-spaced timesteps and the corresponding $N$-body snapshot derived from the same initial conditions at $z_{i} = 127$.}
    \label{fig:colaPerf}
\end{figure*}

The numerator in \eqref{eq:crossP} is the cross power spectrum between the two aforementioned density contrast fields. $\delta(\boldsymbol{k})$ is the Fourier transform of the real-space density contrast $\delta(\boldsymbol{x})$. Note that these coefficients serve as a proxy for the correlation between the COLA and \texttt{GADGET} snapshots, but do not provide an estimation of the canonical cross-correlations of \eqref{eq:cca} between the statistics $\boldsymbol{y}$ and $\boldsymbol{c}$ computed from these snapshots. Having tested different schemes, we concluded that choosing linearly-spaced timesteps yields a better cross-correlation than with logarithmic ones, and that the fewer the timesteps, the more influential the modified timestepping parameter $nLPT$ in terms of cross-correlation coefficients (in the case of this study, with a very high starting redshift of $z_i=127$). Figure \ref{fig:colaPerf} shows an example with $10$ and $20$ linearly-spaced timesteps and $nLPT \in \left\{-2.5, +0.5 \right\}$ (the fiducial value and our experimentally ``best'' value, respectively). Although $\zeta_{yc}(k = 1.0~h {\rm Mpc^{-1}}) \approx 0.96$ with $10$ timesteps  exceeds  $\zeta_{yc}(k = 1.0~h {\rm Mpc^{-1}}) \approx 0.94$ with $20$ timesteps for $nLPT=+0.5$, we still chose to generate our \texttt{L-PICOLA} snapshots with $20$ timesteps between $z_i=127$ and $z=0.0$, again, to avoid tuning \texttt{L-PICOLA} for any one particular statistic. In any case, even with 20 timesteps the \texttt{L-PICOLA} surrogates are much faster than full \texttt{GADGET-III} simulations.

\bsp
\label{lastpage}
\end{document}